\documentclass[a4paper,11pt]{article}
\pdfoutput=1 
\usepackage[T1]{fontenc} 
\usepackage{lmodern} 
\usepackage{jcappub}
\usepackage{amsmath}
\usepackage{amssymb}
\usepackage{bm}
\usepackage{graphicx}
\usepackage{enumitem}
\usepackage{geometry}
\usepackage{xspace}
\usepackage{pdflscape}
\usepackage{enumerate}
\usepackage{xspace} 
\usepackage{soul}
\usepackage{accents}

\usepackage{mathtools}

\makeatletter
\gdef\@fpheader{}
\g@addto@macro\bfseries{\boldmath}
\makeatother

\newcommand{\ie}{\textsl{i.e.~}}

\newcommand{\eg}{\textsl{e.g.~}}

\newcommand{\dd}{\mathrm{d}}
\newcommand{\ee}{e}

\newcommand{\uin}{\mathrm{in}}

\newcommand{\uend}{\mathrm{end}}

\newcommand{\cs}{c_{_\mathrm{S}}}

\newcommand{\mpl}{m_\usssPl}

\newcommand{\efolds}{$e$-folds}
\newcommand{\efold}{$e$-fold}

\newcommand{\beq}{\begin{equation}}
\newcommand{\eeq}{\end{equation}}
\newcommand{\bea}{\begin{equation}\begin{aligned}}
\newcommand{\eea}{\end{aligned}\end{equation}}

\newlength{\wsingfig}
\newlength{\wdblefig}
\newlength{\wquadfig}
\newlength{\wtriplefig}
\newcommand{\Eq}[1]{Eq.~(\ref{#1})}
\newcommand{\Eqs}[1]{Eqs.~(\ref{#1})}
\newcommand{\Fig}[1]{Fig.~{\ref{#1}}}

\newcommand{\Refa}[1]{Ref.~{\cite{#1}}}
\newcommand{\Refs}[1]{Refs.~{\cite{#1}}}
\newcommand{\Sec}[1]{Sec.~\ref{#1}}

\newcommand{\App}[1]{Appendix~\ref{#1}}

\usepackage{xcolor}
\usepackage[normalem]{ulem} 
\usepackage{cancel} 

\newcommand{\eq}[1]{Eq.~(\ref{#1})}
\newcommand{\fig}[1]{Fig.~\ref{#1}}

\def\mpl{M_{\mathrm{Pl}}}
\def\N{\mathcal{N}}
\def\R{\mathcal{R}}
\def\adi{\mathcal{A}}
\def\grow{\mathcal{G}}
\def\decay{\mathcal{D}}
\def\Rk{\mathcal{R}_{k}}
\def\init{\mathrm{in}}
\def\Rh{\mathcal{R}_{\textrm{h}}}
\def\RhDot{\dot{\mathcal{R}}_{\textrm{h}}}
\def\iso{\mathcal{I}}

\def\d{{\mathrm d}}

\newcommand{\dbtilde}[1]{\accentset{\approx}{#1}}

\def\figurewidth{0.85}
\def\halffigurewidth{0.48}

\def\efold{e-fold\xspace}
\def\efolds{e-folds\xspace}

\subheader{}

\title{The separate-universe approach and sudden transitions during inflation}

\author[a]{Joseph H.~P. Jackson,}
\author[a,b]{Hooshyar Assadullahi,}
\author[a]{Andrew D.~Gow.}
\author[a]{Kazuya Koyama,}
\author[c,a]{Vincent Vennin,}
\author[a]{David Wands}

\affiliation[a]{Institute of Cosmology \& Gravitation, University of Portsmouth, Dennis Sciama Building, Burnaby Road, Portsmouth, PO1 3FX, United Kingdom}
\affiliation[b]{School of Mathematics and Physics, University of Portsmouth, Lion Gate Building, Lion Terrace, Portsmouth, PO1 3HF, United Kingdom}
\affiliation[c]{Laboratoire de Physique de l'École Normale Supérieure, ENS, CNRS, Université PSL, Sorbonne Université, Université Paris Cité, F-75005 Paris, France}

\emailAdd{joseph.jackson@port.ac.uk}
\emailAdd{hooshyar.assadullahi@port.ac.uk}
\emailAdd{andrew.gow@port.ac.uk}
\emailAdd{kazuya.koyama@port.ac.uk}
\emailAdd{vincent.vennin@ens.fr}
\emailAdd{david.wands@port.ac.uk}

\date{today}

\begin{document}

\sloppy

\abstract{
The separate-universe approach gives an intuitive way to understand the evolution of cosmological perturbations in the long-wavelength limit. It uses solutions of the spatially-homogeneous equations of motion to model the evolution of the inhomogeneous universe on large scales. We show that the separate-universe approach fails on a finite range of super-Hubble scales at a sudden transition from slow roll to ultra-slow roll during inflation in the very early universe. Such transitions are a feature of inflation models giving a large enhancement in the primordial power spectrum on small scales, necessary to produce primordial black holes after inflation. We show that the separate-universe approach still works in a piece-wise fashion, before and after the transition, but spatial gradients on finite scales require a discontinuity in the homogeneous solution at the transition. We discuss the implications for the $\delta N$ formalism and stochastic inflation, which employ the separate-universe approximation.
}



\maketitle

\section{Introduction}

Inflation is a period of rapid, accelerated expansion in the very early universe~\cite{Starobinsky:1980te, Sato:1980yn, Guth:1980zm, Linde:1981mu, Albrecht:1982wi, Linde:1983gd}, which explains a number of observational properties of our universe, including the origin of large-scale structure. Initial quantum field fluctuations, $\delta\phi$, are expanded to from small to super-Hubble scales by the accelerated expansion~\cite{Mukhanov:1981xt, Mukhanov:1982nu, Starobinsky:1982ee, Guth:1982ec, Hawking:1982cz, Bardeen:1983qw}, where they give rise to primordial curvature perturbations, $\R$~\cite{Lyth:2009zz,Baumann:2022mni}. The almost scale-invariant power spectrum of large-scale perturbations, $\mathcal{P}_{\R}$, predicted by canonical slow-roll inflation using linear perturbation theory is in excellent agreement with observations of the cosmic microwave background (CMB)~\cite{Ade:2015xua, Ade:2015lrj, Planck2018}.

We are yet to directly observe the primordial power spectrum on smaller scales, which left the Hubble radius closer to the end of inflation. A large enhancement of the perturbations compared to those observed on CMB scales would be required in order to produce observable signals~\cite{Chluba:2015bqa}, such as a stochastic gravitational-wave background~\cite{Caprini:2018mtu, Domenech:2021ztg} or primordial black holes (PBHs)~\cite{Ozsoy:2023ryl}. The former has been highlighted by the recent pulsar timing array detection of a stochastic gravitational-wave background~\cite{NANOGrav:2023gor, Antoniadis:2023rey, Xu:2023wog}, which could be of primordial origin~\cite{NANOGrav:2023hvm}. The latter is of particular interest to explain some or all of the dark matter in our universe~\cite{Carr:2016drx, Carr:2020gox, Carr:2020xqk, Green:2020jor}.

To produce a large enhancement in the power spectrum, the usual slow-roll behaviour needs to be interrupted, typically with a period of ultra-slow-roll (USR) inflation~\cite{Byrnes:2018txb, Cole:2022xqc}. As this requires breaking away from the slow-roll attractor solution, a localised feature in the potential is needed~\cite{Karam:2022nym}, which induces a sudden transition. The transition produces a large effective mass-squared for a limited time period in the Sasaki--Mukhanov mode equation~\cite{Sasaki:1986hm,Mukhanov:1988jd}. The enhanced power spectrum can also amplify non-linearities in $\R$. This gives the possibility of loop corrections to the power spectrum dominating over the tree-level result on large scales~\cite{Kristiano:2022maq, Ota:2022hvh, Choudhury:2023vuj, Motohashi:2023syh, Firouzjahi:2023ahg, Franciolini:2023lgy, Fumagalli:2023loc, Fumagalli:2023hpa, Tada:2023rgp}, potentially leading to a breakdown of the perturbative framework.

One approach which enables us to go beyond a purely perturbative description is the $\delta N$ formalism~\cite{Starobinsky:1982ee, Starobinsky:1986fxa, Sasaki1996, Sasaki:1998ug, Lyth:2004gb}. Here we identify the primordial curvature perturbation, $\R$,  with $\delta N$, the difference between $N(\phi + \delta\phi)$, the local integrated logarithmic expansion (or \efolds) from an initial spatially-flat hypersurface to the end of inflation hypersurface, and its mean value $N(\phi)$, \ie
\begin{equation}
    \label{eq:delta_N_formalsim}
    \delta N = N(\phi + \delta\phi) - N(\phi) \, .
\end{equation}
It is possible to treat long-wavelength (typically, super-Hubble scale) fluctuations as classical perturbations of a homogeneous and isotropic cosmology -- the so-called separate-universe approach~\cite{Wands:2000dp, Lyth:2003im, Lyth:2005fi, Artigas:2021zdk}. The homogeneous perturbation can be obtained as the long-wavelength limit in a gradient expansion~\cite{Salopek:1990jq,Rigopoulos:2003ak}. This enables us to find the local integrated expansion, $N(\phi + \delta\phi)$, using the fully non-linear equations of General Relativity for a homogeneous and isotropic cosmology, evolving the local fields and expansion rate, while neglecting spatial gradients. In the classical $\delta N$ approach we use the classical solutions for $N(\phi)$ to obtain $\delta N$ from \eq{eq:delta_N_formalsim},
which can give a highly non-Gaussian probability distribution for $\delta N$~\cite{Biagetti:2018pjj, Firouzjahi:2020jrj, Cai:2022erk, Pi:2022ysn, Hooshangi:2023kss}, even starting from Gaussian field perturbations, $\delta\phi$. In the stochastic $\delta N$ approach~\cite{Enqvist:2008kt, Fujita:2013cna, Vennin:2015hra} we also include the cumulative effect of quantum field fluctuations all along the trajectory, leading to a stochastic description for $\N(\phi)$, where $\N$ now indicates a stochastic variable. 

However, we will show that a sudden transition from slow-roll to ultra-slow-roll evolution during inflation leads to non-adiabatic behaviour. This gives rise to particle production on sub-Hubble scales and non-adiabatic perturbations on super-Hubble scales, sourced by gradient terms~\cite{Leach:2001zf}. While the separate-universe description applies both before and after the transition, the key role played by gradient terms implies that the naïve separate-universe approach breaks down at the transition itself. This necessitates a re-appraisal of how we apply the classical and stochastic $\delta N$ formalisms.

In this paper we first introduce the mathematical framework to identify the long-wavelength limit of the perturbed fields, which we refer to as homogeneous solutions, and then discuss their use in the separate-universe approach in \Sec{sec:seperate_universe_and_delta_N}. 
In \Sec{sec:starobinsky} we study the behaviour of the homogeneous perturbations in a simple model of an instantaneous transition using the piece-wise linear potential of Starobinsky~\cite{Starobinsky:1992ts}. Here, there is a non-adibatic transition from slow roll to ultra-slow roll, followed by an adiabatic transition back to slow roll. Using analytical and numerical solutions, we explicitly show that the sudden transition leads to discontinuities in the homogeneous solutions, sourced by gradient terms at the transition. 
We then demonstrate in \Sec{sec:swagat} that the same behaviour is seen in an example of a smooth potential with a localised feature, which induces a single sudden transition, using numerical methods to find the homogeneous component of the perturbation post-transition. 
Finally, we discuss in \Sec{sec:discussion} how the stochastic $\delta N$ approach might be modified to account for non-adiabatic effects in homogeneous perturbations at a transition from slow roll to ultra-slow roll. 
We conclude in \Sec{sec:conclusions}. Supporting calculations and another example of a single and sudden transition in a smooth potential are presented in the appendices. 
Throughout we work in natural units in which $c=\hbar=8\pi G=1$.

\section{Separate-universe approach}
\label{sec:seperate_universe_and_delta_N}

\subsection{Homogeneous background}

The background dynamics of inflation driven by a spatially-homogeneous canonical single scalar field, $\phi(t)$, with potential energy $V(\phi)$, in a homogeneous and isotropic spacetime with scale factor $a(t)$, are given by the Klein--Gordon equation
\begin{equation}
    \label{eq:klein_gordon}
    \Ddot{\phi} + 3H \dot{\phi} + \frac{\d V(\phi)}{\d \phi} = 0 \, ,
\end{equation}
where an over-dot denotes a derivative with respect to cosmic time, $t$. The Hubble expansion rate, $H = \dot{a}/a$, is given by the Friedmann equation
\begin{equation}
\label{eq:Friedmann}
    3 H^2 = V(\phi) + \frac12 \dot\phi^2 \,. 
\end{equation}

The inflationary expansion can be described by the Hubble-flow parameters~\cite{Schwarz:2001vv, Leach:2002ar}
\begin{equation}
\label{eq:hubble_flow_parameters}
    \epsilon_{i+1} = \frac{\d \ln \epsilon_{i}}{\d N} \, , \quad \epsilon_0 = \frac{H_\init}{H} \, ,
\end{equation}
where $H_\init$ is the Hubble expansion rate at some initial time. Inflation occurs whenever $\epsilon_1 < 1$. Slow-roll inflation requires $|\epsilon_i|\ll 1\ \forall i\geq 1$, corresponding to $\ddot{\phi}$ being sub-dominant in \eq{eq:klein_gordon}. The ultra-slow-roll limit of inflation requires $\epsilon_1\ll1$ and $\epsilon_2 \simeq -6$, corresponding to $\d V(\phi)/\d \phi$ being subdominant in \eq{eq:klein_gordon}~\cite{Dimopoulos:2017ged}.
Typically, ultra-slow roll corresponds to a transient phase, followed by a return to a slow-roll (or constant-roll) attractor~\cite{Pattison:2018bct}. Since slow roll is always an attractor whenever it exists, a deviation from the slow-roll attractor requires a feature in the potential, causing a sudden transition to ultra-slow roll.

\subsection{Linear perturbations}

Inhomogeneous perturbations about the homogeneous and isotropic background can be described by the Sasaki--Mukhanov variable~\cite{Sasaki:1986hm,Mukhanov:1988jd}, $v\equiv a\delta\phi$, where $\delta\phi(t,\vec{x})$ corresponds to scalar field perturbations in the spatially-flat gauge~\cite{Malik:2008im}.

Linear perturbations can be decomposed into independent wavemodes, $v_k$, with comoving wavenumber, $k$, which obey the Sasaki--Mukhanov mode equation~\cite{Sasaki:1986hm,Mukhanov:1988jd}
\begin{equation}
    \label{eq:sasaki_mukhanov_equation}
    v_k'' + \left( k^2 + \mu^2 \right)v_k = 0 \, ,
\end{equation}
where a prime denotes a derivative with respect to comoving time, $\eta\equiv\int \d t/a$, and $\mu^2\equiv -z''/z$, where we define $z = a\dot{\phi}/H = \mathrm{sign}(\dot{\phi}) \sqrt{2a^2 \epsilon_1}$.  
The time-dependent mass-squared term in \eq{eq:sasaki_mukhanov_equation} can be written in terms of the Hubble-flow parameters \eqref{eq:hubble_flow_parameters} as
\begin{equation}
\label{eq:z_prime_prime_by_z}
    \mu^2 =  (aH)^2 \left( - 2 + \epsilon_1 - \frac{3}{2} \epsilon_2 +  \frac{1}{2} \epsilon_1\epsilon_2 - \frac{1}{4} \epsilon_2^2 - \frac{1}{2} \epsilon_2\epsilon_3 \right) \,.
\end{equation}

If we assume that the scalar field starts in the Bunch--Davies vacuum state on small scales during inflation ($k^2\gg |\mu^2|$) then this sets the initial condition for the mode function
\begin{equation}
\label{eq:bunch_davies}
    v_{k}^{\mathrm{BD}} = \frac{e^{-ik\eta}}{ \sqrt{2k}} \,.
\end{equation}
In our numerical solutions we will assume that each mode $v_k$ is in the Bunch--Davies vacuum state at an initial time $-k \eta_{\init} = 10^3$. The subsequent evolution for a given mode is then given by \Eq{eq:sasaki_mukhanov_equation} and depends solely on the time-dependence of $\mu^2$ given by \Eq{eq:z_prime_prime_by_z}.

\begin{figure}
\begin{center}
        \includegraphics[width=\figurewidth\textwidth]{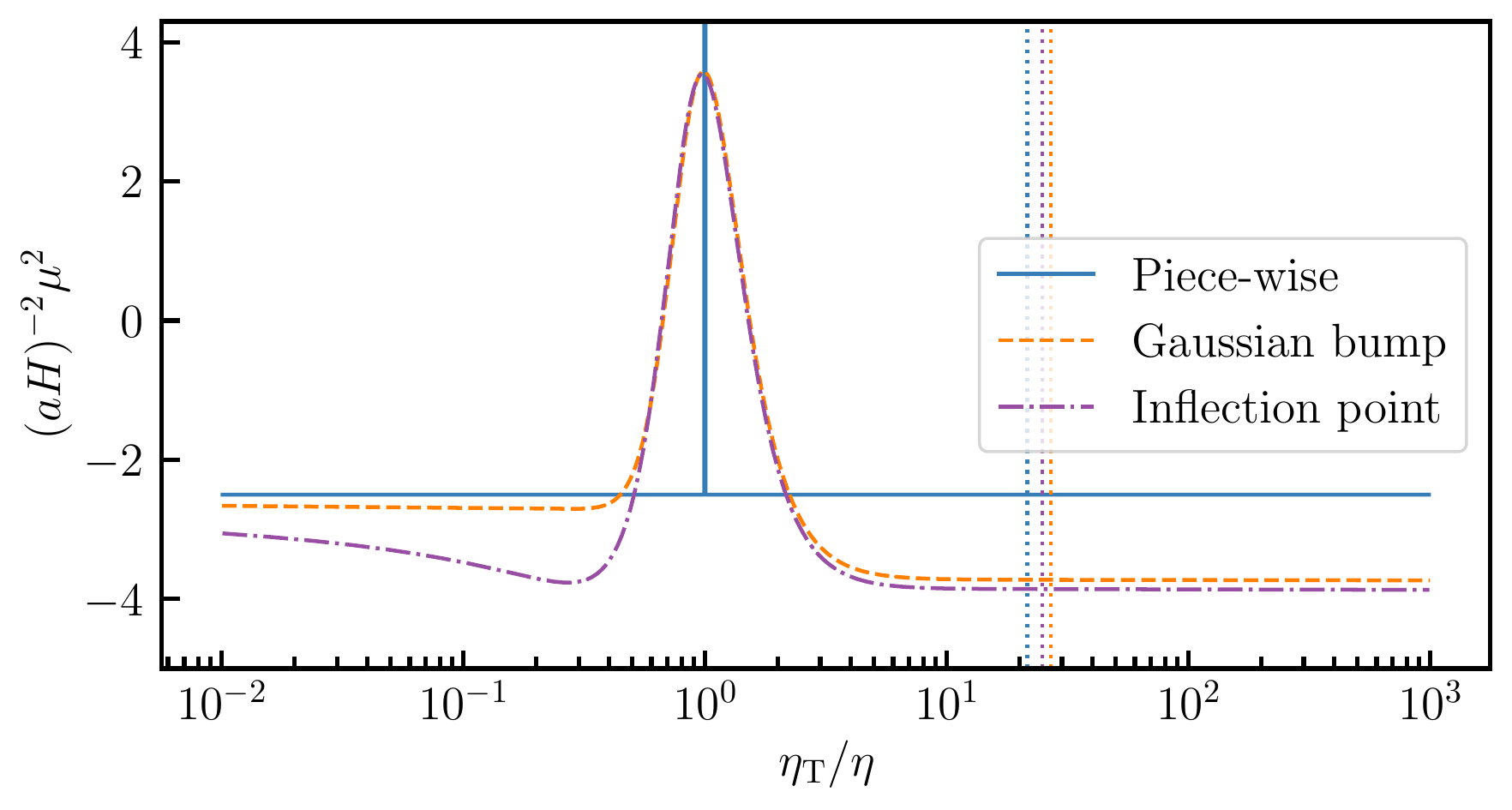}
        \caption{The evolution of the effective mass-squared, $\mu^2$, in the Sasaki--Mukhanov mode equation~\eqref{eq:z_prime_prime_by_z} relative to the comoving Hubble rate, near a sudden transition in the three inflation models investigated in this paper. The piece-wise linear potential is studied in \Sec{sec:starobinsky}, the Gaussian bump model in \Sec{sec:swagat} and the inflection-point model in Appendix~\ref{app:eemeli}. For each potential the transition time, $\eta_{\mathrm{T}}$, is defined to be the first time for which $\epsilon_2 < -3$. The dotted vertical lines indicate the time when ultra-slow roll ends, defined to be when $\epsilon_2>-3$. The similarity in the behaviour of $\mu^2$ close to its maximum is due to model parameters being tuned to produce a PBH abundance consistent with PBHs being all of the dark matter.}
\label{fig:z_double_prime}
\end{center}
\end{figure}

Note that so far we have placed no restriction on the values that the Hubble-flow parameters may take in \Eq{eq:z_prime_prime_by_z}.
We will consider models of inflation with a localised feature in the potential causing a sudden transition to ultra-slow roll, coinciding with a large mass-squared, 
$\mu^2> (aH)^2$, in \Eq{eq:z_prime_prime_by_z}. This behaviour is illustrated in \fig{fig:z_double_prime} for three specific inflation models which we will investigate in this paper. We will define a sudden transition to be a period of less than one \efold ($\Delta N\lesssim 1$) where $\mu^2$ is large and positive.

In addition to the Sasaki--Mukhanov variable, $v$, we will be interested in the comoving curvature perturbation, $\R=v/z$~\cite{Malik:2008im}. For general matter content, with density $\rho$ and pressure $P$, the time-dependence of $\R$ is given by
\begin{equation}
\label{eq:Rdot:general}
    \dot{\R} 
    =  \frac{H}{\rho+P} \delta P_{\mathrm{com}} \,,
\end{equation}
where the comoving pressure perturbation can be split into adiabatic and non-adiabatic components, $\delta P_{\mathrm{com}} = \cs^2 \delta\rho_{\mathrm{com}} + \delta P_{\mathrm{nad}}$, and $\cs^2=\dot{P}/\dot{\rho}$ is the adiabatic sound speed. The comoving density perturbation is related via the Einstein constraint equations to the divergence of the Bardeen metric potential~\cite{Malik:2008im} and thus the time-dependence of $\R$ can be written as 
\begin{equation}
    \dot{\R} = 
    \frac{H}{\rho+P} \left( \frac{2\cs^2}{a^2}\nabla^2\Psi + \delta P_{\mathrm{nad}} \right) \, .
\end{equation}
Therefore, in the absence of non-adiabatic pressure perturbations ($\delta P_{\mathrm{nad}}=0$), the comoving curvature is constant on sufficiently large scales where we can neglect spatial gradients of the Bardeen potential $\nabla^2\Psi\to0$~\cite{Weinberg:2003sw}.
This makes $\R$ convenient to track the evolution of adiabatic density perturbations on large scales during inflation, and through the end of inflation, as the inflaton energy is transferred to the particles of the standard model, reheating the primordial universe~\cite{Kofman:1994rk, Bassett:2005xm}.

During single-field inflation the comoving density perturbation in the scalar field can also be related to the non-adiabatic pressure perturbation~\cite{Gordon:2000hv}
and, noting that the adiabatic sound speed for a single field is given by $\cs^2=1-(2V'/3H\dot\phi)$, we find\footnote{See Ref.~\cite{Romano:2015vxz} for a general discussion of the time-dependence of the comoving curvature perturbation.}
\begin{equation}
\label{eq:Rdot:deltaPnad}
    \dot{\R}
   = -\frac{3H^2}{2\dot{V}} \delta P_{\mathrm{nad}} \, .
\end{equation}
Thus the time-dependence of $\R$ is directly determined by the non-adiabatic part of the scalar field's pressure perturbation on all scales. 
In the following we shall show that the non-adiabatic pressure perturbation does not vanish on all super-Hubble scales after a sudden transition from slow roll to ultra-slow roll.

\subsection{Separate-universe approach}

One might expect that the evolution of inhomogeneous scalar field perturbations on sufficiently long wavelengths, where spatial gradients become negligible, can be described locally by spatially-homogeneous perturbations of the background Klein--Gordon equation \eqref{eq:klein_gordon}. This is the separate-universe approach~\cite{Sasaki:1998ug,Wands:2000dp,Lyth:2004gb} commonly used to simplify the study of inhomogeneous perturbations on large scales. 

Considering a small {\em homogeneous} perturbation about the background scalar field, $\phi\to\phi(t)+\delta\phi_{\mathrm{h}}(t)$, and about the background spacetime (specifically perturbing the Hubble rate, $H\to H+\delta H$ given by \eqref{eq:Friedmann}, and the lapse function relating cosmic time to proper time, $\d t\to(1+A)\d t$), one obtains a linear equation for the homogeneous scalar field perturbation~\cite{Pattison:2019hef} 
\begin{equation}
        \Ddot{\delta\phi}_{\mathrm{h}} + 3H \dot{\delta\phi}_{\mathrm{h}} + \left[ \frac{\d^2 V}{\d \phi^2} - a^{-3} \frac{\d}{\d t} \left( \frac{a^3\dot\phi^2}{H} \right) \right] \delta\phi_{\mathrm{h}} = 0 \, .
\end{equation}
One can verify~\cite{Pattison:2019hef} that this is equivalent to setting $k\to0$ in the Sasaki--Mukhanov mode equation \eqref{eq:sasaki_mukhanov_equation} for $v_{\mathrm{h}}=a\delta\phi_{\mathrm{h}}$:
\begin{equation}
\label{eq:sasaki_mukhanov_homogeneous}
    v_{\mathrm{h}}'' + \mu^2 v_{\mathrm{h}} = 0 \, .
\end{equation}
From the form of the inhomogeneous mode equation \eqref{eq:sasaki_mukhanov_equation} we would expect this to be a good approximation on sufficiently large scales ($k^2\ll |\mu^2|$).

The general solution for corresponding homogeneous curvature perturbation, $\R_{\mathrm{h}}=v_{\mathrm{h}}/z$, is given by
\begin{equation}
\label{eq:R_super_horizon_solution_general_intregrated}
    \Rh =  C + D \int^{\eta_1}_{\eta}\frac{\d \tilde{\eta}}{z^2(\tilde{\eta})} \, ,
\end{equation}
where $C$ and $D$ are constants of integration. $C$ corresponds to the curvature perturbation at $\eta=\eta_1$, while $D$ determines the amplitude of the isocurvature perturbation at $\eta_1$. More generally $D$ can be identified with a non-adiabatic homogeneous perturbation since it controls $\RhDot$, see \Eq{eq:Rdot:deltaPnad}. 

The choice of $\eta_1$ in \eqref{eq:R_super_horizon_solution_general_intregrated} is arbitrary and is degenerate with $C$. 
If we set $\eta_1$ to be the end of inflation, which we take to be $\eta_1=0$, then the two terms appearing in \Eq{eq:R_super_horizon_solution_general_intregrated} 
can be identified as the homogeneous growing and decaying modes.
The growing mode is in fact constant, while the decaying mode decreases in slow roll where $1/z^2 \propto \eta^{2}$ and increases in ultra-slow roll where $1/z^2\propto \eta^{-4}$. It vanishes at the end of inflation by construction.

\subsubsection{\texorpdfstring{$k^2$}{k2}-expansion}
\label{sec:k2:exp}
One can include systematic corrections to the homogeneous solution by performing an expansion of the Sasaki--Mukhanov equation \eqref{eq:sasaki_mukhanov_equation} in $k^2$. To that end, it is convenient to first write down the equation of motion for $\Rk = v_k/z$ that follows from \Eq{eq:sasaki_mukhanov_equation}, namely 
\bea 
\label{eq:R:eom}
\Rk'' + 2\frac{z'}{z}\Rk' +k^2\Rk=0\, .
\eea

For any finite $k$ the decomposition into growing and decaying modes, $\R_k=\grow+\decay$, where $\grow$ and $\decay$ are themselves solutions of \Eq{eq:R:eom}, can be carried out order-by-order in $k^2$, by expanding
\bea
\label{eq:k_power_series_grow_and_decay}
    \grow(\eta) = \sum_n \grow_n(\eta)k^{2n} 
    \quad\text{and}\quad
    \decay(\eta) = \sum_n \decay_n(\eta)k^{2n}\, .  
\eea
At leading order in $k^2$ for a given wavemode
we recover the homogeneous behaviour~\eqref {eq:R_super_horizon_solution_general_intregrated}
\begin{equation}
\label{eq:R_super_horizon_solution_general_intregrated_grow_decay}
    \grow_0 \equiv  C_k \quad \text{and} \quad \decay_0(\eta) \equiv D_k \int^{0}_{\eta}\frac{\d \tilde{\eta}}{z^2(\tilde{\eta})} \, .
\end{equation}
By identifying terms of the same order in $k^2$ in \Eq{eq:R:eom}, one finds
\begin{equation}
        \grow_{n}'' + 2\frac{z'}{z}\grow_{n}'  = - \grow_{n-1}\, ,
\end{equation}
which leads to the recursive solution for the growing mode
\begin{equation}
    \label{eq_k_squared_correction}
    \grow_{n}(\eta)=\int_\eta^{\eta_3}\frac{\d \tilde{\eta}}{z^2(\tilde{\eta})}\int^{\tilde{\eta}}_{\eta_2} \d\dbtilde{\eta}\, z^2(\dbtilde{\eta}) \grow_{n-1}(\dbtilde{\eta}) \, ,
\end{equation}
with a similar expression for the decaying mode, $\decay_{n}(\eta)$. 

Two arbitrary integration limits have to be introduced, $\eta_2$ and $\eta_3$. We will require that the $k^2$-corrections vanish at the end of inflation at all orders in $k$, which sets $\eta_3=0$ in $\grow_{n}$, and we further impose that $\eta_3=0$ in $\decay_{n}$ as well, such that at the end of inflation the value of $\R_k$ corresponds to $C_k$ and the power spectrum of the curvature perturbation reads
\bea
\label{eq:PR:Ck}
\mathcal{P}_\R(k) = 
\frac{k^3}{2\pi^2} \left\vert C_k \right\vert^2\, .
\eea 

There is no unique prescription for $\eta_2$. In practice, one notices that changing $\eta_2$ in \Eq{eq_k_squared_correction} adds a contribution proportional to $\decay_0$ into $\grow_n$ and $\decay_n$. In particular, the $k^2$-correction to the growing mode, $\grow_{1}(\eta)$, is only defined up to terms proportional to the leading-order decaying mode, $\decay_0(\eta)$ \cite{Leach:2001zf}.

\subsubsection*{Solution for massless inflaton field in de Sitter}
\label{eq:de:Sitter}

In order to illustrate how the $k^2$-expansion works, let us consider 
the simple limit of a massless inflaton field in de Sitter, corresponding to $\epsilon_1\to0$, such that $H$ is a constant and $a=-1/(H\eta)$, and $\mu^2=
\to-2/\eta^2$ in \Eq{eq:z_prime_prime_by_z}.
This holds in both the slow-roll limit ($\epsilon_i\to0$ for all $i\geq1$) or the ultra-slow-roll limit ($\epsilon_1\to0$, $\epsilon_2\to-6$, and $\epsilon_i\to0$ for all $i\geq3$).
In this case \Eq{eq:R:eom} can be solved analytically
\begin{equation}
    \label{eq:R_slow_roll}
    \Rk = \frac{\mathrm{sign}({\dot{\phi}})}{2a \sqrt{k \epsilon_1}} \left[ \alpha_k \left( 1-\frac{i}{k \eta} \right) e^{-ik\eta} + \beta_k \left( 1+\frac{i}{k \eta} \right) e^{ik\eta} \right] \, .
\end{equation}
where the integration constants $\alpha_k$ and $\beta_k$ are Bogoliubov coefficients. In the Bunch--Davies vacuum \eqref{eq:bunch_davies}, $\alpha_k=1$ and $\beta_k=0$ but we will keep them general for now.

Let us perform the $k^2$-expansion in the slow-roll limit where $\epsilon_1$ can be considered to be a constant. First, from \Eq{eq:R_super_horizon_solution_general_intregrated_grow_decay}, one has $\grow_0=C_k$ and $\decay_0=-D_k H^2 \eta^3/(6\epsilon_1)$. Then, for the growing mode, one can show that the ansatz $\grow_n = C_k g_n \eta^{2n}$ satisfies \Eq{eq_k_squared_correction} if
\begin{equation}
    \label{eq:g_n}
    g_n = - \frac{g_{n-1}}{2n(2n-3)} \, .
\end{equation}
Note that in order to get this expression, we must set $\eta_2$ in such a way that the contribution coming from the lower bound in the inner integral of \Eq{eq_k_squared_correction} vanishes, which requires $\eta_2=-\infty$ (asymptotic past) for $n=1$ and $\eta_2=0$ (asymptotic future) for $n\geq 2$. Since $g_0=1$ (given that $\grow_0=C_k$ by definition), this leads to
\begin{equation}
    \label{eq:g_n_v2}
    g_n=\frac{(-1)^n (1-2n)}{(2n)!}\, .
\end{equation}
As a consequence, \Eq{eq:k_power_series_grow_and_decay} leads to
\bea
\label{eq:de_sitter_growing_mode}
\grow(\eta) = C_k \sum_{n=0}^\infty g_n(k\eta)^{2n}
= C_k \sum_{n=0}^\infty (-1)^n \frac{1-2n}{(2n)!} (k\eta)^{2n} = C_k \left[\cos(k\eta)+k\eta\sin(k\eta)\right]
\eea
where we have recognised the Taylor series of $\cos(k\eta)+k\eta\sin(k\eta)$.

For the decaying mode, likewise, the ansatz $\decay_n = -D_k H^2/(6\epsilon_1) d_n \eta^{3+2n}$ satisfies \Eq{eq_k_squared_correction} if $d_n=-d_{n-1}/[2n(2n+3)]$, where this time we have set $\eta_2=0$ for all $n$. Together with $d_0=1$, this leads to $d_n=6(-1)^n (n+1)/(2n+3)!$, and \Eq{eq:k_power_series_grow_and_decay} yields
\bea
\label{eq:de_sitter_decaying_mode}
\decay(\eta) = & -D_k \frac{H^2}{6\epsilon_1k^3} \sum_{n=0}^\infty d_n(k\eta)^{2n+3}
=  -D_k \frac{H^2}{2\epsilon_1k^3}\left[\sin(k\eta)-k\eta\cos(k\eta)\right]\, .
\eea
It is then straightforward to check that \Eq{eq:R_slow_roll} can indeed be rewritten as $\Rk=\grow+\decay$, provided one makes the identification
\begin{equation}
    \label{eq:C_D_slow_roll}
    C_k = \frac{iH\mathrm{sign}({\dot{\phi}})}{2 \sqrt{k^3 \epsilon_1}} (\alpha_k-\beta_k) \quad \text{and} \quad D_k = - \mathrm{sign}({\dot{\phi}})\frac{\sqrt{\epsilon_1 k^3}}{H}(\alpha_k+\beta_k ) \, .
\end{equation}

We can thus identify the homogeneous growing and decaying mode solutions \eqref{eq:R_super_horizon_solution_general_intregrated_grow_decay}
\begin{equation}
    \label{eq:R_de_sitter_homogeneous}
    \Rh = \frac{iH\mathrm{sign}({\dot{\phi}})}{2 \sqrt{k^3 \epsilon_1}} \left[ \left(\alpha_k-\beta_k \right) -i \left( \alpha_k + \beta_k \right) \frac{(k\eta)^3}{3}  \right] \, .
\end{equation}
In the slow-roll limit, $\epsilon_1$ is a constant\footnote{We note that \eqref{eq:R_de_sitter_homogeneous} follows from \eqref{eq:R_slow_roll} and thus holds more generally for a massless inflaton field in de Sitter, including both the ultra-slow-roll regime where $\epsilon_1$ is small, but time-dependent. However in the ultra-slow-roll limit, where $\sqrt{\epsilon_1} \propto \eta^3$, the identification of the constant and time-dependent terms in \eqref{eq:R_de_sitter_homogeneous} is swapped~\cite{Wands:1998yp}.}, and the first term in the square bracket is the constant growing mode. The second $\eta^3$ term is the homogeneous decaying mode. Note however that in slow roll at late times ($\eta\to0$) the leading $k^2$-correction to the growing mode, \ie the $n=1$ term in \eq{eq:de_sitter_growing_mode}, which decays as $\eta^2$, dominates over the homogeneous decaying mode, which decays as $\eta^3$.

The most common approximation for the description of perturbations on super-Hubble scales during inflation is thus to take the homogeneous growing mode for a massless field starting in the Bunch--Davies vacuum state ($\alpha_k=1$, $\beta_k=0$) in de Sitter 
\begin{equation}
\label{eq:R_de_sitter_homogeneous_growing}
    \Rh = \frac{iH\mathrm{sign}({\dot{\phi}})}{2 \sqrt{k^3 \epsilon_1}} \, .
\end{equation}
This is equivalent to setting $\delta\phi_{\textrm{h}}=iH/\sqrt{2k^3}$, leading to the power spectrum for long-wavelength field fluctuations evaluated at Hubble exit, $\mathcal{P}_{\delta\phi_h}=(H/2\pi)^2$.

\subsection{Super-Hubble matching}
\label{sec:super_horizon_matching}
\subsubsection{Homogeneous matching}
\label{sec:homog_matching}

The idea of the separate-universe approach \cite{Wands:2000dp} is to approximate the dynamics of perturbations on large scales by the homogeneous solutions, \ie to use only the $0^{\mathrm{th}}$-order terms in the $k^2$-expansion \eqref{eq:k_power_series_grow_and_decay} above a certain scale.
In practice, this can be implemented via a \textit{homogeneous-matching} procedure, where we use the solution to the full equation of motion~\eqref{eq:R:eom} below the matching scale, and match it to its homogeneous counterpart~\eqref{eq:R_super_horizon_solution_general_intregrated} above that scale. 
In what follows, we will express the matching scale in terms of the ratio, $\sigma$, of the wavenumber to the Hubble scale, \ie a wavemode $k$ crosses the matching scale when $k=\sigma a H$ and we choose $\sigma<1$ corresponding to matching on super-Hubble scales. 

At the matching scale, we require $\Rh=\R_{k*}$ and $\Rh'=\R_{k*}'$,
where the subscript ``$*$'' refers to quantities evaluated at the matching scale, hence they depend implicitly on $k$. This leads to the matching conditions
\bea
\label{eq:hom:matching}
\hat{C}_k = \R_{k*} +z^2_* \R_{k*}' \int_{\eta_*}^0\frac{\dd\tilde{\eta}}{z^2(\tilde{\eta})}
\quad\text{and}\quad \hat{D}_k = - z^2_*  \R_{k*}'\, ,
\eea
where hats denote quantities estimated by the homogeneous-matching procedure. 
In the above expressions, $\R_{k*}$ and $\R_{k*}'$ are obtained by solving the full linear dynamics, \Eq{eq:R:eom}, below the matching scale, taking initial conditions in the Bunch--Davies state at past infinity. 
At late times, $\R_k$ approaches $C_k$, and thus $\hat{C}_k$ in \Eq{eq:hom:matching} can be used to approximate the curvature perturbation at the end of inflation.

\subsubsection{Matching including \texorpdfstring{$k^2$}{k2}-corrections}
\label{sec:k_squared_corrections}

In practice, if we match soon after Hubble-crossing, \textsl{i.e.}, $\sigma\simeq1$, the values of $\R_{k*}$ and $\R_{k*}'$, and hence our estimates $\hat{C}_k$ and $\hat{D}_k$ in \Eq{eq:hom:matching}, are liable to be contaminated by gradient terms. One can thus improve our estimate of the solution on large scales by including the leading order $k^2$-corrections. While this can be done analytically in the slow-roll limit where $\epsilon_1$ is a constant, in general this would be computationally inefficient, as from \eq{eq_k_squared_correction}, a double integration is needed for the $k^2$ correction to the growing mode and a triple integration would be needed for the decaying-mode correction.

Instead, the freedom in the choice of the upper limit of the integral in \eq{eq:R_super_horizon_solution_general_intregrated} can be used to simplify the numerical computation so that the dominant $k^2$-correction will be the correction to the adiabatic solution. This suggests we should choose the adiabatic/isocurvature ($\adi / \iso$) split at the matching time, $\eta_1 = \eta_*$ in \eq{eq:R_super_horizon_solution_general_intregrated}. We then have 
\begin{equation}
\label{eq:R_super_horizon_solution_ad-iso}
    \Rh =  \adi_0 + \iso_0 \int^{\eta_*}_{\eta}\frac{\d \tilde{\eta}}{z^2(\tilde{\eta})} \, ,
\end{equation}
and hence we can estimate the leading-order terms in a $k^2$-expansion for the adiabatic and isocurvature modes at the matching scale, where
\begin{equation}
\label{eq:R_super_horizon_solution_general_intregrated_adi_iso}
    \hat{\adi}_0 = \R_{k*} \quad \text{and} \quad \hat{\iso}_0= - z^2_*  \R_{k*}' \, .
\end{equation}
If we further set $\eta_3 = \eta_2 = \eta_*$ in \eqref{eq_k_squared_correction} to obtain the $k^2$-correction to the adiabatic mode, 
\begin{equation}
    \label{eq_k_squared_correction_iso}
    \adi_{1}(\eta)= \adi_{0}\int_\eta^{\eta_*}\frac{\d \tilde{\eta}}{z^2(\tilde{\eta})}\int^{\tilde{\eta}}_{\eta_*} \d\dbtilde{\eta}\, z^2(\dbtilde{\eta}) \, ,
\end{equation}
then this correction makes no contribution to $\R_k$ and $\R_k '$ at the matching time, significantly simplifying the computation.

This choice gives a straightforward form for describing the behaviour of $\R_k$ for $\eta\geq\eta_*$ including the leading-order $k^2$-corrections
\bea
\label{eq:k_corrected:matching}
\hat{\R}_k (\eta) = \R_{k*} \left[1 - k^2\int_{\eta_*}^{\eta}\frac{\d \tilde{\eta}}{z^2(\tilde{\eta})}\int^{\tilde{\eta}}_{\eta_*}\d\dbtilde{\eta}\, z^2(\dbtilde{\eta}) \right] +z^2_* \R_{k*}' \int_{\eta_*}^\eta\frac{\dd\tilde{\eta}}{z^2(\tilde{\eta})}\, .
\eea
It is clear that if one sets $k^2=0$, then the homogeneous matching \eqref{eq:hom:matching} is recovered. This result was obtained in Ref.~\cite{Leach:2001zf}, where the super-Hubble enhancement of a mode between the matching time and the end of inflation due to $k^2$-corrections was calculated in a model with a transient ultra-slow-roll phase.

\subsubsection{The \texorpdfstring{$\delta N$}{δN} formalism}
\label{sec:delta:N}
An alternative method to implement the separate-universe approach is the $\delta N$ formalism~\cite{Starobinsky:1982ee, Starobinsky:1986fxa, Sasaki1996, Sasaki:1998ug, Lyth:2004gb}. This follows from the fact that, on large scales, the curvature perturbation can be identified with the fluctuation in the local expansion $N=\ln(a)$. In particular, the curvature perturbation on uniform-density hypersurfaces~\cite{Malik:2008im}, $\zeta$, can be identified with the perturbed expansion, $\delta N$, between an initial spatially-flat hypersurface during inflation and a final hypersurface of uniform energy density at the end of inflation. In the models that we consider, the perturbations become adiabatic on super-Hubble scales by the end of inflation, and the curvature perturbation on uniform-density hypersurfaces then coincides (up to a sign convention) with the comoving curvature perturbation~\cite{Wands:2000dp,Lyth:2004gb} 
\begin{equation}
\label{eq:R_derivative_non_adibatic}
    \zeta = -\R - \frac{3H}{2\dot{V}} \delta P_{\mathrm{nad}} \,.
\end{equation}

In practice, as in the homogeneous-matching procedure described in \Sec{sec:super_horizon_matching}, one solves for the full linear dynamics~\eqref{eq:R:eom} below the matching scale, to find $\R_{k*}$ and $\R_{k*}'$, and hence $\delta\phi_{k*}$ and $\dot{\delta \phi}_{k*}$ in the spatially-flat gauge (using $\delta\phi=v/a=\dot{\phi}\R/H$ in that gauge). Above the matching scale, the homogeneous dynamics is described by the function $N(\phi_\uin,\dot{\phi}_\uin)$, which returns the number of \efolds realised using the background equations of motion~\eqref{eq:klein_gordon} and~\eqref{eq:Friedmann} from the initial field values $(\phi_\uin,\dot{\phi}_\uin)$ up to a late-time hypersurface of uniform energy density (which in practice we take to be the end-of-inflation surface). One then evaluates the fluctuation in the expansion
\bea
    \label{eq:delta_N_formalism}
\delta N_k = N\left(\phi_*+\delta\phi_{k*},\dot\phi_*+\dot{\delta \phi}_{k*}\right)-N\left(\phi_*,\dot\phi_*\right)
\eea
and at late times, this gives the curvature perturbation $\zeta_k = \delta N_k$. 

The above formula can be expanded in the field perturbations $\delta\phi_k$ and $\dot{\delta \phi}_k$, leading to
\bea
\label{eq:deltaN:lin}
\delta N_k \simeq \frac{\partial N}{\partial\phi_\uin}\left(\phi_*,\dot\phi_*\right) \delta\phi_{k*} + \frac{\partial N}{\partial\dot\phi_\uin}\left(\phi_*,\dot\phi_*\right) \dot{\delta \phi}_{k*}\, .
\eea
At linear order, the $\delta N$ formalism is strictly equivalent to the homogeneous-matching procedure discussed in \Sec{sec:super_horizon_matching}. In \App{app:delta_N}, we prove this equivalence for the model discussed in \Sec{sec:starobinsky}. However, in principle one can use the fully non-linear background equations~\eqref{eq:klein_gordon} and~\eqref{eq:Friedmann}, which enables one to use \Eq{eq:delta_N_formalism} beyond linear order. In practice, we will not consider non-linear effects in what follows, so we will restrict our discussion to the homogeneous-matching procedure, given that it is equivalent to the linearised $\delta N$ formalism. 

\section{Piece-wise linear potential}
\label{sec:starobinsky}

Starobinsky's piece-wise linear  model~\cite{Starobinsky:1992ts} provides an illustrative example of a sudden transition, with the advantage of being fully tractable analytically. It has been extensively studied in the literature, see \eg \Refs{Leach:2001zf, Martin:2011sn, Martin:2014kja, Ahmadi:2022lsm, Pi:2022zxs}.

In this toy model, the potential function has the form
\begin{equation}
    \label{eq_starobinsky-potential}
    V (\phi) = \begin{cases}
    V_0 + A_+ (\phi - \phi_{\mathrm{T}}) & \text{for } \phi\geq \phi_{\mathrm{T}}\, ,\\ 
    V_0 + A_- (\phi - \phi_{\mathrm{T}})& \text{for } \phi < \phi_{\mathrm{T}} \, ,
\end{cases}
\end{equation}
with the constants $A_+> A_- > 0$. Initial conditions are set such that $\phi_\uin>\phi_{\mathrm{T}}$ and the system reaches the slow-roll attractor in the pre-transition phase of the dynamics. Immediately after the transition, the field velocity inherited from the $\phi>\phi_{\mathrm{T}}$ region is larger than the slow-roll velocity in the $\phi<\phi_{\mathrm{T}}$ region, so there is a transient phase of slow-roll violation, until a second slow-roll attractor is reached at late times. In the limit where $A_+\gg A_-$, this transient behaviour is an ultra-slow-roll phase.

\subsection{Homogeneous background}

We will consider the case where $V_0 \gg A_\pm \vert \phi-\phi_{\mathrm{T}}\vert$, which implies that $\epsilon_1 \ll 1$ throughout and that $H$ is effectively constant. In this de Sitter limit, the Klein--Gordon equation~\eqref{eq:klein_gordon} can be solved analytically to give
\begin{equation}
    \label{eq:klein_gordon_starobinsky_solution_with_eta}
    \phi (\eta) = \begin{cases}
    \frac{A_+}{3H^2}\ln{( -k_{\mathrm{T}}\eta )} + \phi_{\mathrm{T}} & \text{for } \eta \leq \eta_{\mathrm{T}} \, ,\\
    \frac{A_-}{3H^2}\ln{( -k_{\mathrm{T}}\eta )} +\frac{\Delta A}{9H^2}\left[1+( k_{\mathrm{T}}\eta )^{3} \right]  + \phi_{\mathrm{T}} & \text{for } \eta > \eta_{\mathrm{T}} \, ,
\end{cases}
\end{equation}
where $\Delta A = A_- - A_+<0$ and $\eta_{\mathrm{T}}$ is the comoving transition time. Here we have introduced the comoving scale $k_{\mathrm{T}}= -1/\eta_{\mathrm{T}}$ that crosses the Hubble radius at the transition time. We have assumed that inflation starts sufficiently far from $\phi_{\mathrm{T}}$ such that the slow-roll attractor has been reached well before the transition, and we have used the continuity of $\phi$ and $\dot\phi$ at the transition to set the integration constants for $\phi<\phi_{\mathrm{T}}$ (corresponding to $\eta > \eta_{\mathrm{T}}$).

The first and second Hubble-flow parameters (\ref{eq:hubble_flow_parameters}) are then
\begin{equation}
    \label{eq:epsilon_1_starobinsky}
    \epsilon_1 (\eta) = \begin{cases}
    \frac{A_+^2}{18H^4} & \text{for } \eta \leq \eta_{\mathrm{T}} \, ,\\
    \frac{\left(\Delta A k_{\mathrm{T}}^3\eta^3+A_-\right)^2}{18H^4} & \text{for } \eta > \eta_{\mathrm{T}} \, ,
\end{cases}
\end{equation}
and
\begin{equation}
    \label{eq:epsilon_2_starobinsky}
    \epsilon_2 (\eta) = \begin{cases}
   0 & \text{for } \eta \leq \eta_{\mathrm{T}} \, ,\\
    -\frac{6\Delta A k_{\mathrm{T}}^3\eta^3}{\Delta A k_{\mathrm{T}}^3\eta^3 + A_-} & \text{for } \eta > \eta_{\mathrm{T}} \, .
\end{cases}
\end{equation}
While $\epsilon_1$ is continuous, $\epsilon_2$ is discontinuous at the transition, due the discontinuity in $\dd V/\dd\phi$. In what follows, the pre- and post-transition behaviours will be labelled by the subscripts ``$+$'' and ``$-$'', respectively.

\subsection{Perturbations}
\label{eq:Staro:pert}

To find solutions of the mode equation~\eqref{eq:sasaki_mukhanov_equation} we first need to find the behaviour of $z=a\dot\phi/H$. From \Eq{eq:epsilon_1_starobinsky} we have
\begin{equation}
\label{eq:z_starobinsky}
    z(\eta) = \begin{cases}
    \frac{A_+}{3H^3} \eta^{-1} & \text{for } \eta \leq \eta_{\mathrm{T}} \, ,\\
    \frac{A_-+\Delta A k_{\mathrm{T}}^3\eta^3}{3H^3} \eta^{-1} & \text{for } \eta > \eta_{\mathrm{T}} \, .
\end{cases}
\end{equation}
One can check that $z$ is continuous at $\eta=\eta_{\mathrm{T}}$ but that its derivative is discontinuous.

The effective mass-squared $\mu^2=-z''/z$
appearing in the Sasaki--Mukhanov equation~\eqref{eq:sasaki_mukhanov_equation} can also be obtained from \Eq{eq:z_prime_prime_by_z}. 
Neglecting the sub-dominant contributions from $\epsilon_1\ll1$, one finds
\begin{equation}
    \label{eq:z_prime_prime_by_z_starobinsky}
    \mu^2 = \frac{1}{\eta^2} \left[ -2 + \frac{3\Delta A \eta_{\mathrm{T}}}{A_+} \delta_{\mathrm{D}}(\eta - \eta_{\mathrm{T}}) \right] \, ,
\end{equation}
where $\delta_{\mathrm{D}}$ is the Dirac $\delta$ distribution. Apart from at the transition, $\mu^2=-2/\eta^2$, but it experiences a sudden impulse at $\eta=\eta_{\mathrm{T}}$ due to the discontinuity in the second slow-roll parameter, $\epsilon_2$. Therefore the general solution for $\R_k$ is given by \eq{eq:R_slow_roll} both before and after the transition, but with different constants of integration.

\subsubsection*{Pre-transition}
Before the transition we set initial conditions in the Bunch--Davies vacuum~\eqref{eq:bunch_davies} at early times, corresponding to $\alpha_{k+}=1$ and $\beta_{k+}=0$ in \eq{eq:R_slow_roll}, leading to
\begin{equation}
    \label{eq:R_starobinsky_pre_transition}
    \Rk = -\frac{iH}{2 \sqrt{k^3 \epsilon_{1+}}} ( 1+ik \eta ) e^{-ik\eta} \, .
\end{equation}
In the absence of a transition (\ie if $\Delta A=0$), the amplitude of the growing and decaying modes \eqref{eq:R_super_horizon_solution_general_intregrated_grow_decay} can be identified from \Eq{eq:C_D_slow_roll}. One finds 
\begin{equation}
    \label{eq:starobinsiky_C_D_pre_transition}
    \widetilde{C}_{k} = -\frac{3iH^3}{A_+\sqrt{2k^3}} \quad \text{and} \quad \widetilde{D}_{k} = \frac{A_+}{3H^3 \sqrt{2k^3}}k^3 \, .
\end{equation}
Here, tildes are used to denote the \emph{apparent} growing and decaying modes in the pre-transition phase.
\subsubsection*{Post-transition}
\label{subsec:staro_post}
After the transition, the behaviour of the perturbations is still given by the general solution, \Eq{eq:R_slow_roll}, but with different values for the Bogoliubov coefficients compared to the pre-transition phase.

Both $\Rk$ and $\Rk'$ are continuous at the transition, since $z'/z$ remains finite in \Eq{eq:R:eom}.
However, although the Sasaki--Mukhanov variable, $v_k=\Rk/z$ is continuous at the transition, its time derivative $v_k'$ is discontinuous due to the Dirac $\delta$-function in the effective mass-squared, $\mu^2$ in \Eq{eq:z_prime_prime_by_z_starobinsky}, which appears in the mode equation~\Eq{eq:sasaki_mukhanov_equation}.
The jump in $v_k'$ is given by
\bea
    \label{eq:sasaki_mukhano_derivative_transition}
   v_{k-}'(\eta_{\mathrm{T}} + h) - v_{k+}'(\eta_{\mathrm{T}} - h)  =  &\int^{\eta_{\mathrm{T}} + h}_{\eta_{\mathrm{T}} - h}\d \eta\, v_k''(\eta) \\ 
   = &  \int^{\eta_{\mathrm{T}} + h}_{\eta_{\mathrm{T}} - h}\d\eta\, \left[
   \frac{2}{\eta^2} - \frac{3\Delta A}{A_+ \eta_{\mathrm{T}}} \delta_{\mathrm{D}}(\eta - \eta_{\mathrm{T}})- k^2 \right]v_k(\eta) \\ &
  \underset{h\to 0}{\longrightarrow}  - \frac{3\Delta A}{A_+ \eta_{\mathrm{T}}} v_k\left(\eta_{\mathrm{T}}\right)
    \, .
\eea
This leads to the new Bogoliubov coefficients after the transition~\cite{Starobinsky:1992ts,Martin:2014kja}
\begin{equation}
    \label{eq:starobinsky_R_post_transition_alpha}
    \alpha_{k-} = 1 + \frac{3i}{2}\frac{\Delta A k_{\mathrm{T}}}{A_+ k} \left( 1 + \frac{k_{\mathrm{T}}^2}{k^2} \right) \, ,
\end{equation}
\begin{equation}
    \label{eq:starobinsky_R_post_transition_beta}
    \beta_{k-} = -\frac{3i}{2}\frac{\Delta A k_{\mathrm{T}}}{A_+ k} \left( 1 + i\frac{k_{\mathrm{T}}}{k} \right)^2 e^{2ik/k_{\mathrm{T}}} \, .
\end{equation}
We see that the sudden transition is non-adiabatic ($\beta_{k-}\neq0$), leading to particle production on sub-Hubble scales and time-dependence of the curvature perturbation, even on super-Hubble scales.

The homogeneous growing and decaying modes can be obtained by rewriting the homogeneous part of the general solution, \Eq{eq:R_slow_roll}, in the form~\eqref{eq:R_super_horizon_solution_general_intregrated}, which gives
\bea 
\label{eq:starobinsiky_C_D_post_transition}
C_k =& -\frac{3iH^3}{A_-\sqrt{2k^3} }\left(\alpha_{k-}-\beta_{k-}\right) ,\\
D_k = & \frac{ A_- k^3}{3 H^3\sqrt{2 k^3}}\left[\alpha_{k-}+\beta_{k-}-3i \frac{\Delta A}{A_-} \left(\frac{k_{\mathrm{T}}}{k}\right)^3 \left(\alpha_{k-}-\beta_{k-}\right)\right] .
\eea 
In the absence of a transition, $\Delta A=0$, one recovers $\beta_{k-}=0$ and the apparent pre-transition growing and decaying modes given in \Eq{eq:starobinsiky_C_D_pre_transition}. To better understand the effect of the transition, it is helpful to expand \Eq{eq:starobinsiky_C_D_post_transition}, using \Eq{eq:starobinsky_R_post_transition_alpha} and \Eq{eq:starobinsky_R_post_transition_beta}, in various limits.

\subsubsection*{Scales $k\ll k_{\mathrm{T}}$}

When $k\ll k_{\mathrm{T}}$, \ie for scales that exit the Hubble radius long before the transition, by Taylor expanding \Eq{eq:starobinsiky_C_D_post_transition}, one finds
\bea
\label{eq:Ck:kllkt}
C_k =& -\frac{3 i H^3}{\sqrt{2 k^3} A_+} \left\lbrace 1 + \frac{2}{5} \frac{\Delta A }{A_- } \left(\frac{k}{k_{\mathrm{T}}}\right)^2\left[1+\frac{5i}{6}\frac{k}{k_{\mathrm{T}}}+\mathcal{O} \left(\frac{k^2}{k^2_{\mathrm{T}}}\right)\right]\right\rbrace\, ,\\
D_k =& \frac{A_- k^3}{3H^3\sqrt{2k^3}}\left[3i \frac{\Delta A}{A_-} \left(1+\frac{3}{5}\frac{\Delta A}{A_+}\right)\frac{k_{\mathrm{T}}}{k}+\frac{A_+}{A_-}+ \mathcal{O}\left(\frac{k}{k_{\mathrm{T}}}\right) \right]\, .
\eea
Note that the term $\Delta A/A_-$ may be large (this is the case if $A_- \ll A_+$), which leads us to distinguish three sub-cases. 

If $k/k_{\mathrm{T}} \ll \sqrt{A_-/\vert \Delta A\vert}$, then the non-adiabatic, decaying mode $D_k$ is suppressed in \Eq{eq:Ck:kllkt} and $C_k$ reduces to the apparent growing mode in the pre-transition phase, \Eq{eq:starobinsiky_C_D_pre_transition}. These scales are sufficiently far outside the Hubble scale at the time of the transition that they are not affected by the sudden jump and $\R_k$ remains constant after the transition. This corresponds to the left-hand scale-invariant plateau shown in the left-hand panel of \Fig{fig:starobinsky_homogenoues_matching}.

For $\sqrt{A_- /\vert \Delta A\vert} \sim k/k_{\mathrm{T}}$, the two leading terms in \Eq{eq:Ck:kllkt} for $C_k$ are of the same order and can cancel, since $\Delta A/A_-<0$. The higher-order terms are strongly suppressed for $k/k_{\mathrm{T}} \ll 1$. This explains the well-known dip in the power spectrum \eqref{eq:PR:Ck}, as shown in \fig{fig:starobinsky_homogenoues_matching}.

If $\sqrt{A_- /\vert \Delta A\vert} \ll k/k_{\mathrm{T}}$, then the term proportional to $(k/k_{\mathrm{T}})^2$ dominates in \Eq{eq:Ck:kllkt}, which leads to the intermediate region in \Fig{fig:starobinsky_homogenoues_matching} where the power spectrum scales as $\mathcal{P}_\R\propto k^4$~\cite{Byrnes:2018txb, Carrilho:2019oqg}. These scales are super-Hubble at the time of the transition, but they have spent too little time outside the Hubble radius to suppress gradients terms to the level that would be necessary to make the non-adiabatic term irrelevant after the transition. Therefore, as first observed in \Refa{Leach:2001zf}, the presence of gradient terms at the sudden transition can leave a strong imprint on super-Hubble scales. More precisely, the scale above which $\R$ is approximately constant is not the Hubble radius but is somewhat larger, given by $\sqrt{\vert\Delta A \vert /A_- }$ times the Hubble radius\footnote{The reason why the ratio $\Delta A/A_-$ appears can be understood from \Eq{eq:epsilon_1_starobinsky}, which can be rewritten as $\epsilon_{1-} = A_-^2/(18 H^4)[1-\Delta A/A_- \ee^{-3(N-N_{\mathrm{T}})}]^2$. This shows that the ultra-slow-roll phase after the transition lasts for of order $\ln(\Delta A/A_-)/3$ \efolds before it returns to slow roll.\label{eq:NUSR}}.

Since this behaviour is key to understanding the main result of this paper, let us further comment on it. As already explained, $\R$ and $\R'$ are continuous at the transition. Let us check whether or not this is the case for their homogeneous counterparts, \ie let us compare the apparent homogeneous solution before the transition \eqref{eq:starobinsiky_C_D_pre_transition} to the homogeneous solution after the transition \eqref{eq:Ck:kllkt}
\bea
    \label{eq:discontinuity_of_R_homogeneous}
    \frac{\Delta \Rh}{\widetilde{\R}_{\mathrm{h}}(\eta_{\mathrm{T}})} \equiv
    \frac{\Rh(\eta_{\mathrm{T}}) - \widetilde{\R}_{\mathrm{h}}(\eta_{\mathrm{T}})}{\widetilde{\R}_{\mathrm{h}}(\eta_{\mathrm{T}})} &\simeq -\frac{3}{5}\frac{\Delta A}{A_+}\left(\frac{k}{k_{\mathrm{T}}}\right)^2\, , \\
     \frac{\Delta \Rh'}{\widetilde{\R}_{\mathrm{h}}'(\eta_{\mathrm{T}})} \equiv
    \frac{\R_{\mathrm{h}}'(\eta_{\mathrm{T}}) - \widetilde{\R}_{\mathrm{h}}'(\eta_{\mathrm{T}})}{\widetilde{\R}_{\mathrm{h}}'(\eta_{\mathrm{T}})}&\simeq 3i \frac{\Delta A}{A_+}\left( 1+\frac{3}{5}\frac{\Delta A}{A_+} \right) \frac{k_{\mathrm{T}}}{k}\, .
\eea
Since $\vert \Delta A/A_+\vert <1$, the relative jump in $\R_\mathrm{h}$ at the transition is suppressed by $(k/k_{\mathrm{T}})^2$, which is indeed small for scales that are super-Hubble at the time the transition. However the relative jump in $\R_{\mathrm{h}}'$ is enhanced by $k_{\mathrm{T}}/k$, even if the absolute jump is suppressed by $k/k_{\mathrm{T}}$. 

\begin{figure}
\begin{center}
        \includegraphics[width=\figurewidth\textwidth]{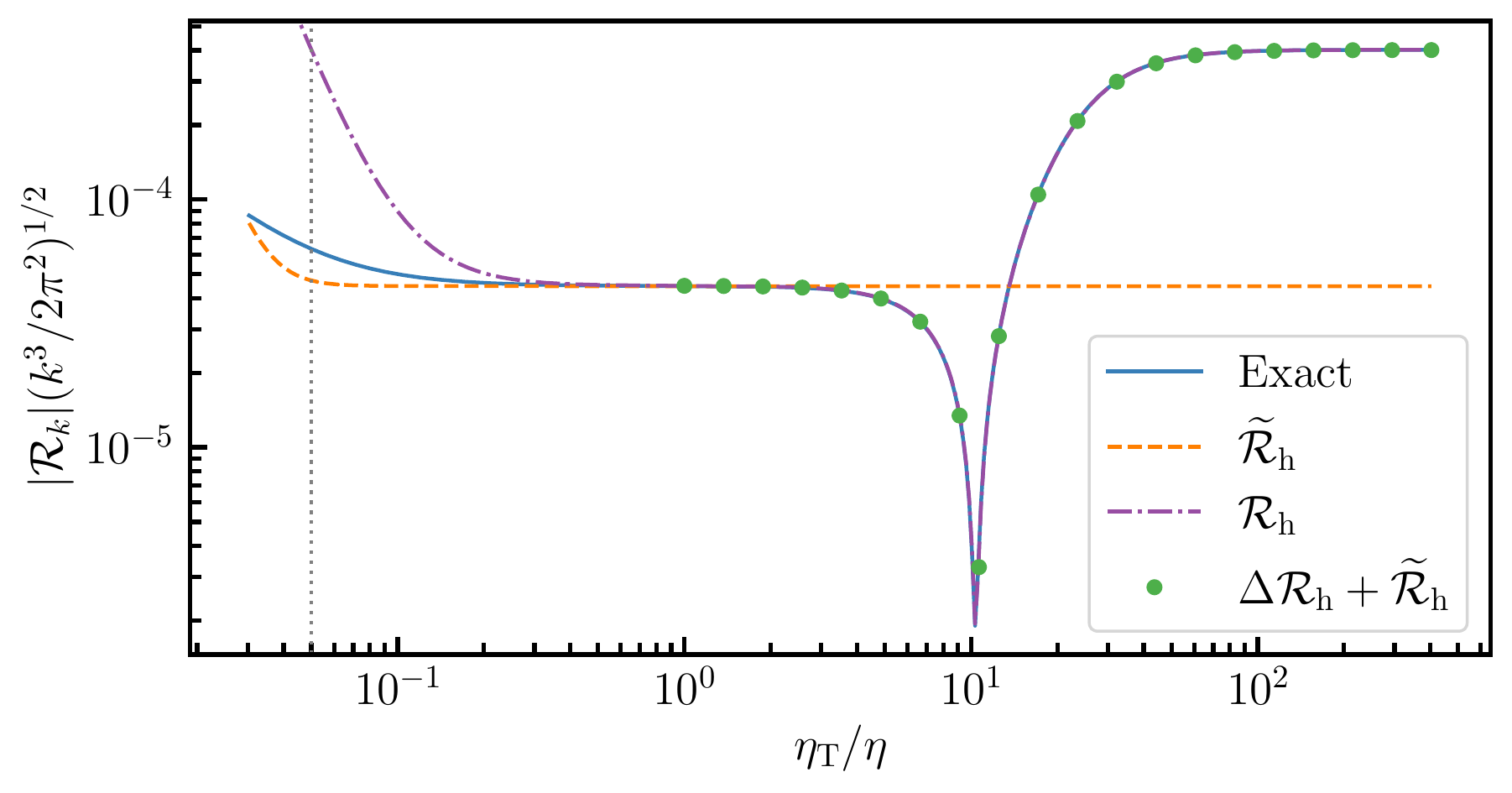}
        \caption{The evolution of the curvature perturbation $\Rk$ with comoving time $\eta$ for a wavemode with $k=0.05\,k_{\mathrm{T}}$ in the piece-wise linear potential (\ref{eq_starobinsky-potential}) with $A_-/A_+ = 10^{-4}$.
        The sudden transition from slow roll to ultra-slow roll occurs at $\eta_{\mathrm{T}}=-1/k_{\mathrm{T}}$. The solid blue curve is found by numerically solving the full Sasaki--Mukhanov equation \eqref{eq:sasaki_mukhanov_equation}. The dashed orange curve corresponds to the homogeneous solution \eqref{eq:R_super_horizon_solution_general_intregrated}
        with the apparent growing and decaying mode amplitudes before the transition, \Eq{eq:starobinsiky_C_D_pre_transition}, and the dot-dashed purple curve corresponds to the homogeneous solution \eqref{eq:R_super_horizon_solution_general_intregrated} with the growing and decaying mode amplitudes after the transition \eq{eq:starobinsiky_C_D_post_transition}. The green data points are found by adding the discontinuous jump at the transition (\ref{eq:discontinuity_of_R_homogeneous}) to the homogeneous solution \eqref{eq:R_super_horizon_solution_general_intregrated}
        with the apparent growing and decaying mode amplitudes before the transition, \Eq{eq:starobinsiky_C_D_pre_transition}. The dotted vertical line shows the Hubble-exit time, when $aH=0.05\, k_{\mathrm{T}}$.}
\label{fig:homogeneous_mode}
\end{center}
\end{figure}

This explains what is observed in \fig{fig:homogeneous_mode}, where the apparent homogeneous solution found before the transition, $\widetilde{\R}_{\mathrm{h}}$, remains constant post-transition, while the actual homogeneous solution after the transition, $\Rh$, shows rapid time evolution. Also note that the actual homogeneous solution after the transition, $\Rh$, propagated back to Hubble-exit shows an order of magnitude difference compared to the apparent pre-transition homogeneous solution, $\widetilde{\R}_{\mathrm{h}}$, at that time. 
But if we track the change to the homogeneous behaviour, $\Delta \Rh$ and $\Delta \Rh'$ induced at the transition, and propagate that forward using the homogeneous equation of motion, the correct post-transition behaviour on super-Hubble scales is recovered. This suggests we can keep track of the homogeneous behaviour, even if there is a sudden transition, by including the correct non-adiabatic jump coming from gradient terms at the transition.

\subsubsection*{Scales $k\gg k_{\mathrm{T}}$}
Finally we note that for $k\gg k_{\mathrm{T}}$, \ie scales that exit the Hubble radius some time after the transition, one has 
\bea
\label{eq:Ck:kggkt}
C_k = -\frac{3 i H^3}{\sqrt{2 k^3} A_-} \left[1+ \frac{3i}{2} \frac{\Delta A}{A_+} \frac{k_{\mathrm{T}}}{k}\left(1+\ee^{2i \frac{k}{k_{\mathrm{T}}}} \right)+\mathcal{O}\left(\frac{k_{\mathrm{T}}^2}{k^2}\right)\right]\, .
\eea
Here, $\vert \Delta A/A_+ \vert $ is always smaller than one, hence the first term in the square bracket dominates and one recovers the slow-roll result in the post-transition region (\ie \Eq{eq:Ck:kggkt} reduces to \Eq{eq:starobinsiky_C_D_pre_transition} where $A_+$ is replaced with $A_-$).
This corresponds to the right-hand scale-invariant plateau in the left panel of \Fig{fig:starobinsky_homogenoues_matching}, with damped sinusoidal modulation of the power spectrum for $k\gtrsim  k_{\mathrm{T}}$. 
\subsection{Homogeneous matching}
Let us now employ the homogeneous-matching procedure described in \Sec{sec:super_horizon_matching} to estimate the homogeneous perturbations in the Starobinsky piece-wise linear model. Using \Eq{eq:z_starobinsky}
to evaluate the terms involving $z(\eta)$, \Eq{eq:hom:matching} gives rise to
\bea
\label{eq:HM:gen:Staro}
\hat{C}_k = 
\begin{cases}
\R_{k*} + \left[1-\frac{\Delta A}{A_-}\left(\frac{k}{\sigma k_{\mathrm{T}}}\right)^3\right] \frac{\sigma}{3k} \R'_{k*}
& \text{for } k\leq \sigma k_{\mathrm{T}}
 \vspace{0.3em}\\
\R_{k*} + \left[1-\frac{\Delta A}{A_-}\left(\frac{k}{\sigma k_{\mathrm{T}}}\right)^{-3}\right]  \frac{\sigma}{3k} \R'_{k*}
& \text{for } k> \sigma k_{\mathrm{T}}
\end{cases} \, ,
\eea
where we recall that $\R_{k*}$ and $\R_{k*}'$ correspond to the curvature perturbation and its time derivative at the time where $k$ crosses the matching scale (\ie when $k=\sigma a H$), as given by the full perturbation theory established in \Sec{eq:Staro:pert}. In \fig{fig:starobinsky_homogenoues_matching}, we show the power spectrum of $\R$ and the evolution of a single mode with $k = 0.05\, k_\mathrm{T}$ calculated using the homogeneous matching procedure for different values of $\sigma$.

At this stage, it is worth highlighting the calculation performed in \App{app:delta_N}, where the $\delta N$ program described in \Sec{sec:delta:N} has been implemented for Starobinsky's piece-wise linear model. In the linear version of the calculation, one obtains \Eq{eq:deltaN:linear:Staro}, which exactly coincides with \Eq{eq:HM:gen:Staro}. This illustrates the statement made above that, for linear perturbations, the homogeneous-matching procedure and the $\delta N$ approach are equivalent.
\subsubsection*{Matching before the transition}

For $k\leq \sigma k_{\mathrm{T}}$, the matching scale is crossed before the transition, and inserting \Eq{eq:R_starobinsky_pre_transition} into the first entry of \Eq{eq:HM:gen:Staro} leads to
\bea
\label{eq:Ck:HM:ksmall}
\hat{C}_k = -\frac{3iH^3\ee^{i\sigma}}{\sqrt{2k^3}A_+}\left[1-i\sigma-\frac{\sigma^2}{3}+\frac{\sigma^2}{3}\frac{\Delta A}{A_-} \left(\frac{k}{\sigma k_{\mathrm{T}}}\right)^3\right] .
\eea
This has to be compared with the full homogeneous solution after the transition given by \Eq{eq:starobinsiky_C_D_post_transition}, which reduces to \Eq{eq:Ck:kllkt} in the limit where $k\ll k_{\mathrm{T}}$. If matching is done in the super-Hubble regime, $\sigma\ll 1$, one observes different behaviours depending on how $k/k_{\mathrm{T}}$ compares with $\sqrt{A_-/|\Delta A|}$.

If $k/k_{\mathrm{T}}\ll \sqrt{A_-/|\Delta A|}$, the late-time curvature perturbation obtained from homogeneous matching, \Eq{eq:Ck:HM:ksmall}, reduces to the full result~\eqref{eq:Ck:kllkt}, up to ${\cal O}(\sigma^2)$ corrections. The usual separate-universe approach is therefore able to reproduce the left-hand scale-invariant plateau in \Fig{fig:starobinsky_homogenoues_matching}, since the curvature perturbation remains a constant on super-Hubble scales even during the ultra-slow-roll phase.

If $\sqrt{A_-/|\Delta A|} \ll k/k_{\mathrm{T}}\ll 1$, the $k^2$ term becomes dominant in \Eq{eq:Ck:kllkt}, and cannot be reproduced by the terms in \Eq{eq:Ck:HM:ksmall}. The usual separate-universe approach thus fails in this case. The reason is that, in the exact result, \Eq{eq:Ck:kllkt}, the $k^2$ term comes from gradient corrections to the growing mode, see \Sec{sec:k2:exp}. However, this contribution is not accounted for in the homogeneous matching, \Eqs{eq:HM:gen:Staro} and \eqref{eq:Ck:HM:ksmall},  since it does not appear in the homogeneous solution \eqref{eq:R_super_horizon_solution_general_intregrated}. Homogeneous matching only captures the leading order of the decaying mode, which scales as $k^3$. This is demonstrated in \fig{fig:starobinsky_homogenoues_matching} for both the power spectrum and the time evolution of a particular mode.

\subsubsection*{Matching after the transition}

\begin{figure}
\begin{center}
        \includegraphics[width=\halffigurewidth\textwidth]{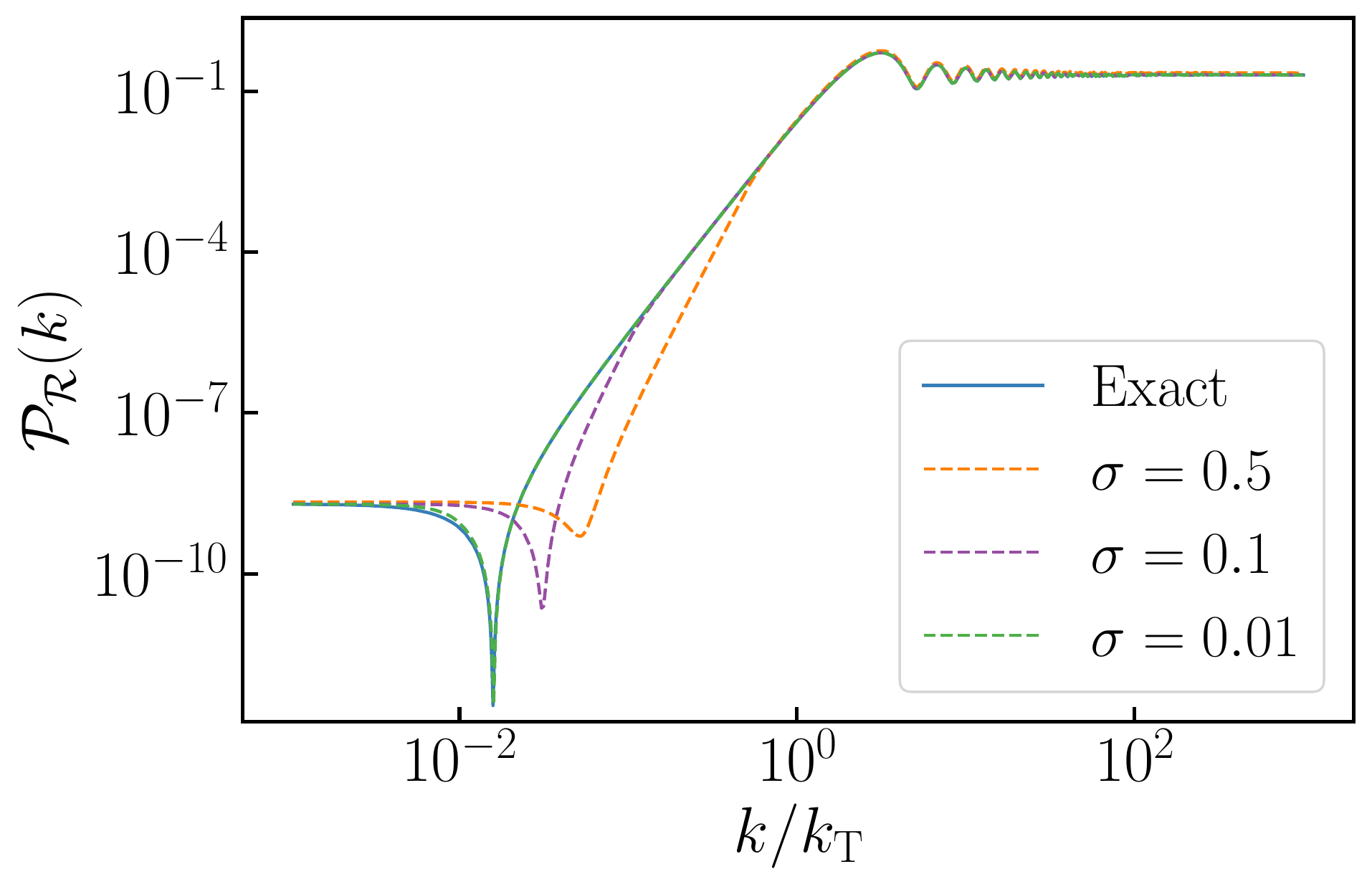}
        \includegraphics[width=\halffigurewidth\textwidth]{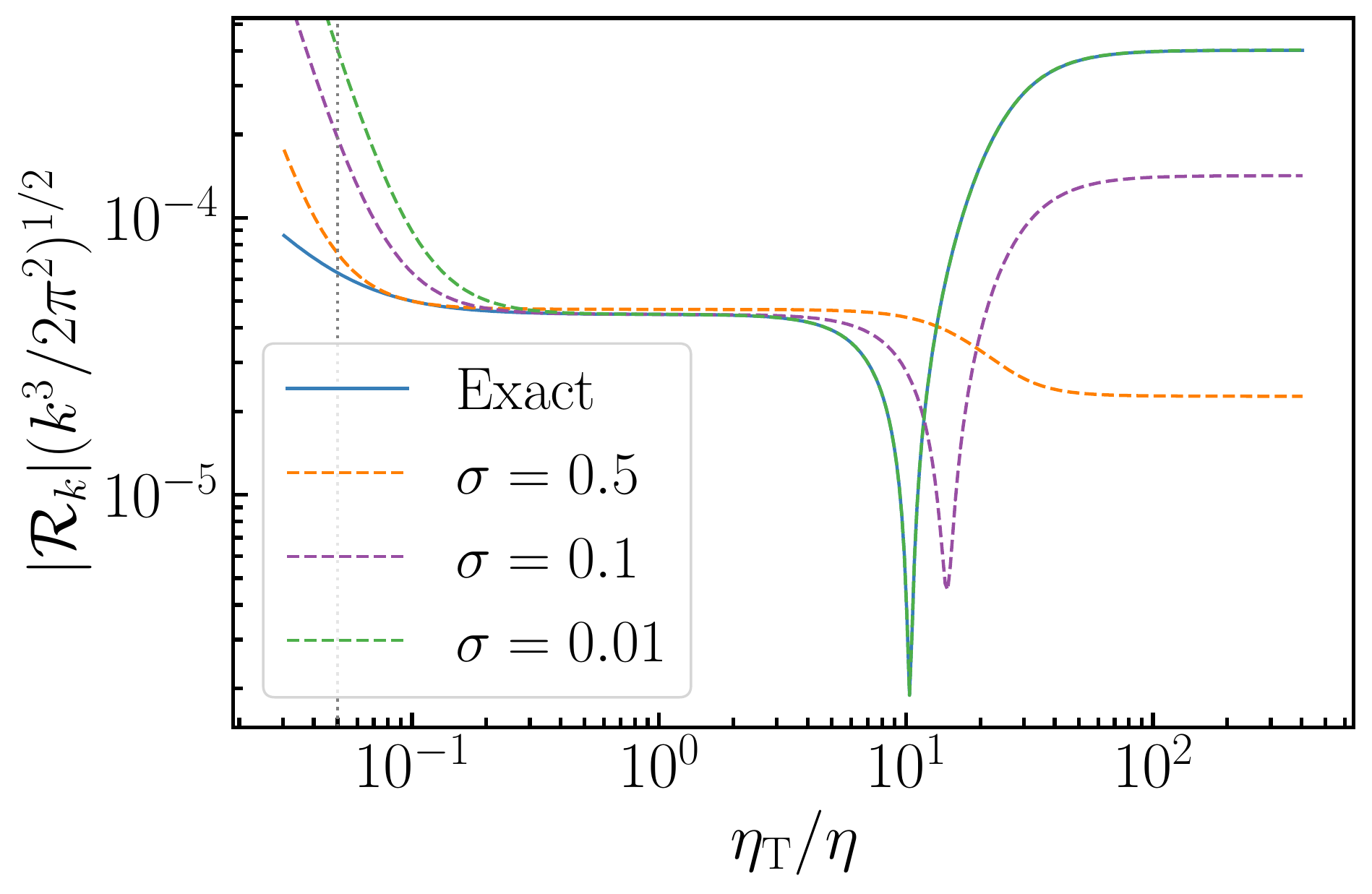}
        \caption{Left: the power spectrum \eqref{eq:PR:Ck} of the curvature perturbation, $\R$, against wavenumber, $k/k_{\mathrm{T}}$ for the piece-wise linear potential (\ref{eq_starobinsky-potential}) with $A_-/A_+ = 10^{-4}$. Right: the evolution of $\R_k$ for a mode with $k = 0.05\, k_{\mathrm{T}}$, along with the homogeneous solution resulting from matching at different values of $\sigma=k/aH$. In both panels, the solid blue curve is found by numerically solving the full mode equation \eq{eq:sasaki_mukhanov_equation} and the dashed curves are obtained numerically using the matching procedure detailed in \Sec{sec:super_horizon_matching}. The dotted vertical line is Hubble-exit time. The ultra-slow-roll period starts at $\eta_{\mathrm{T}}=-1/k_{\mathrm{T}}$.}
\label{fig:starobinsky_homogenoues_matching}
\end{center}
\end{figure}

For $k> \sigma k_{\mathrm{T}}$, the matching scale is crossed after the transition and one may use \Eq{eq:R_slow_roll}, together with \Eq{eq:epsilon_1_starobinsky} for $\epsilon_1(\eta)$ for $\eta \geq \eta_{\mathrm{T}}$, to evaluate $\R_{k*}$ and $\R_{k*}'$ in the second entry of \Eq{eq:HM:gen:Staro}. This leads to
\begin{equation}
    \label{eq:Ck:HM:largek}
    \hat{C}_k = -\frac{3iH^3}{A_-\sqrt{2k^3}}\left[\alpha_{k-}\ee^{i\sigma}\left(1-i\sigma-\frac{\sigma^2}{3}\right)-\beta_{k-}\ee^{-i\sigma}\left(1+i\sigma-\frac{\sigma^2}{3}\right)\right] ,
\end{equation}
where $\alpha_{k-}$ and $\beta_{k-}$ after the transition are given in \Eqs{eq:starobinsky_R_post_transition_alpha} and~\eqref{eq:starobinsky_R_post_transition_beta}. At leading order in $\sigma$, this reduces to $\hat{C}_k\simeq -3iH^3(\alpha_{k-}-\beta_{k-})/(A_-\sqrt{2k^3})$, which coincides with the exact expression, \Eq{eq:C_D_slow_roll}. As a consequence, up to ${\cal O}(\sigma^2)$ corrections, the homogeneous-matching procedure is able to correctly reproduce the late-time behaviour of the curvature perturbation if the matching is performed after the transition, as shown in \fig{fig:starobinsky_homogenoues_matching}.\vspace{\baselineskip}

Let us summarise our findings. The above results show that the homogeneous-matching procedure fails for the scales within the range $\sqrt{A_- /\vert \Delta A\vert} < k/k_{\mathrm{T}} < 1$ \emph{if} they are matched before the transition, and succeeds otherwise as long as $\sigma\ll 1$. Therefore, in order to match the dangerous set of scales after the transition, one requires
\bea
\label{eq:cond:sigma}
\sigma\ll \mathrm{min}\left(1,\sqrt{\frac{A_-}{\vert \Delta A \vert}}\right)\, .
\eea 
As mentioned in Footnote \ref{eq:NUSR}, $A_-/\Delta A$ can be related to the number of \efolds during the ultra-slow-roll phase, $N_{\mathrm{USR}}$, so we expect the above condition to generalise to 
\bea 
\label{eq:cond:sigma:gen}
\sigma\ll \ee^{-\frac{3}{2}N_{\mathrm{USR}}} \, ,
\eea
in more general models.
This condition is more stringent than the one usually encountered in the absence of sudden transitions, $\sigma\ll 1$, but it ensures the validity of the separate-universe approach after homogeneous matching. 

\subsection{Including \texorpdfstring{$k^2$}{k2}-corrections}

\begin{figure}
\begin{center}
        \includegraphics[width=\halffigurewidth\textwidth]{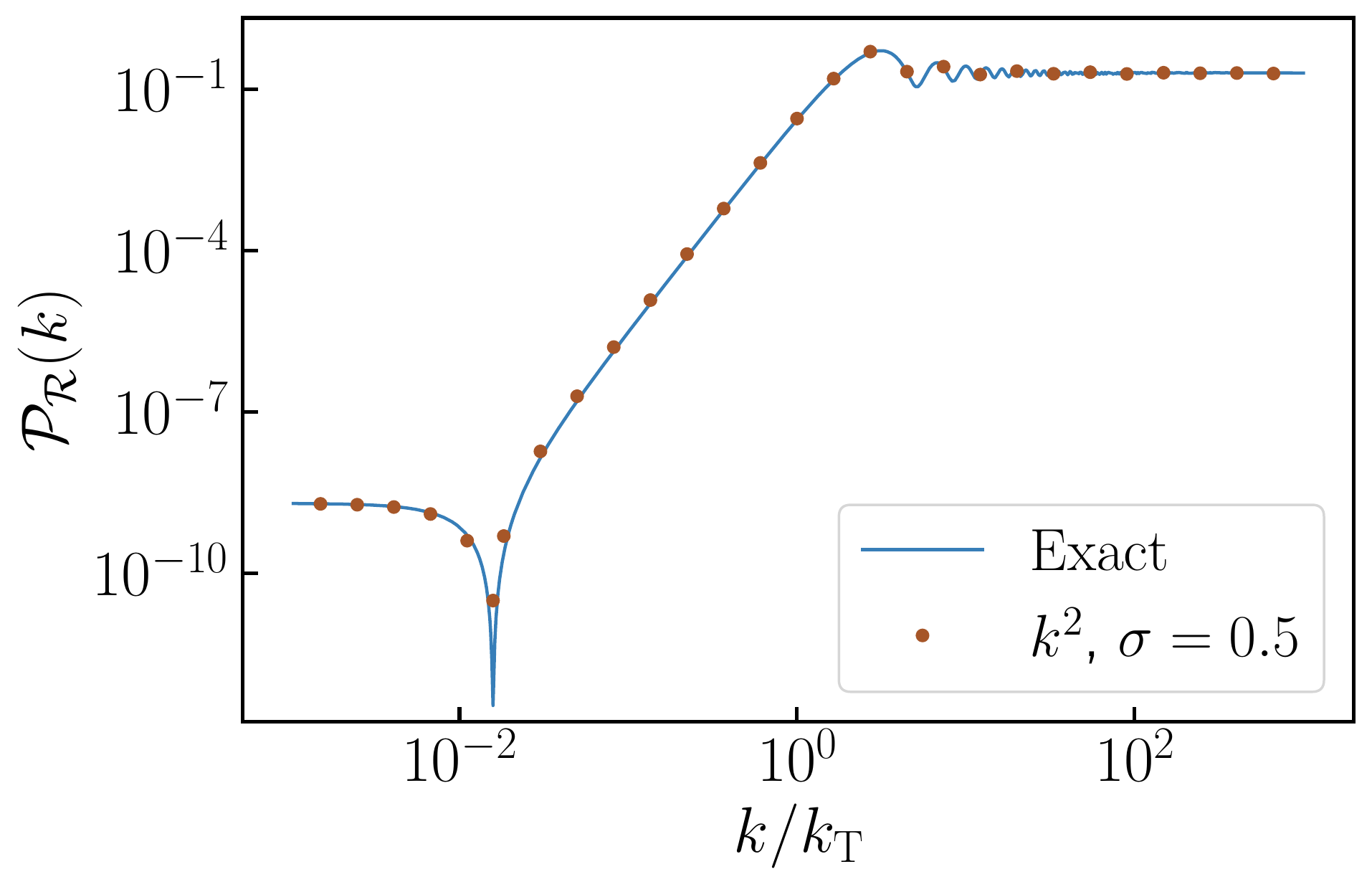}
        \includegraphics[width=\halffigurewidth\textwidth]{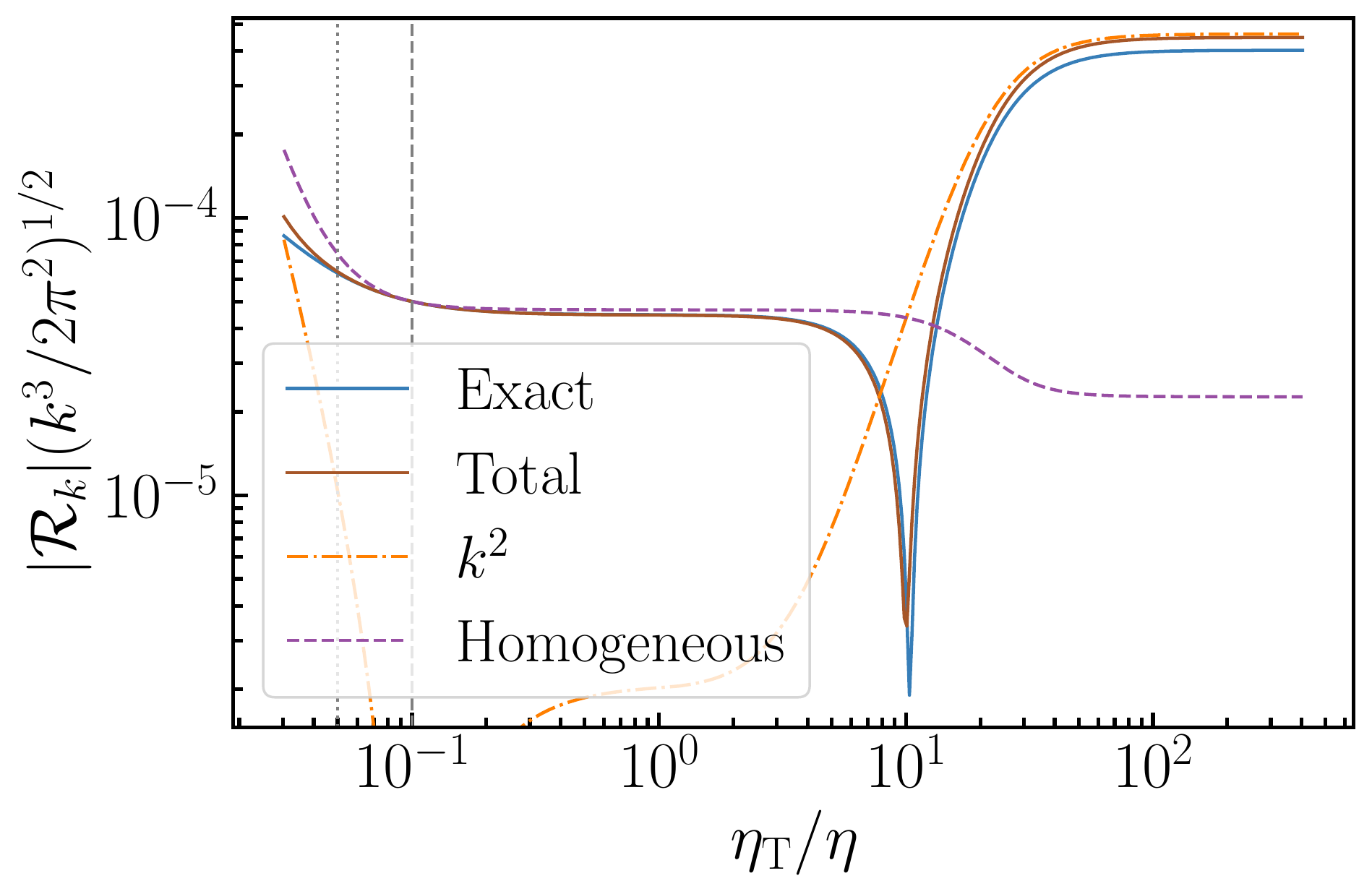}
        \caption{Left: the power spectrum \eqref{eq:PR:Ck} of the curvature perturbation $\R$ for the piece-wise linear potential (\ref{eq_starobinsky-potential}) with $A_-/A_+ = 10^{-4}$. The solid blue curve is found by numerically solving \eq{eq:sasaki_mukhanov_equation} and the brown points are found by including the $k^2$ term in the adiabatic mode at matching, as detailed in \Sec{sec:k_squared_corrections}. Right: evolution of the components of \eq{eq:k_corrected:matching} plotted along with the full solution of the mode equation for $k=0.05\, k_{\mathrm{T}}$. The $k^2$ curve is the $k^2$-term, ``Homogeneous'' is the $k^2$-independent component, and ``Total'' is the sum of the two. The dotted and dashed vertical lines correspond to Hubble-exit and matching times, respectively. The ultra-slow-roll period starts at $\eta_{\mathrm{T}}=-1/k_{\mathrm{T}}$.}
\label{fig:starobinsky_ps_k_squared}
\end{center}
\end{figure}

As discussed below \Eq{eq:C_D_slow_roll}, the $k^2$-correction to the growing mode can dominate over the leading-order decaying mode at late times. As a consequence, the late-time non-adiabatic behaviour may be dominated by contributions coming from $k^2$-corrections in the gradient expansion, hence the separate-universe approach fails on some super-Hubble scales in the presence of a sudden transition. One can further check the validity of this statement by including the leading $k^2$-corrections in the matching procedure, as was discussed in \Sec{sec:k_squared_corrections}. This has been shown in \fig{fig:starobinsky_ps_k_squared} matching very close to Hubble-exit, $\sigma=0.5$. This gives an excellent approximation to the exact power spectrum, confirming the $k^2$-corrections are indeed needed to fully explain the super-Hubble evolution in the presence of a sudden transition after the matching.

By comparing the $k^2$-corrected reconstruction shown in \fig{fig:starobinsky_ps_k_squared} with the homogeneous matching used in \fig{fig:starobinsky_homogenoues_matching}, it is clear that the post-transition behaviour is recovered well (within $\sim 10\%$) even for $\sigma=0.5$, when the $k^2$-corrections are included. This is in contrast to the homogeneous matching case, where a $\sigma$-value corresponding to matching after the transition ($\sigma< 0.05$) is required. This clearly shows that the $k^2$-corrections at the transition are required to describe the non-adiabatic homogeneous behaviour on super-Hubble scales after the transition. If $\sigma=0.1$ is used for the $k^2$-corrected \eq{eq:k_corrected:matching}, the late-time behaviour is reconstructed within $\sim 1\%$.

\section{Smooth potential with a Gaussian bump}
\label{sec:swagat}

Let us consider a model with a smooth slow-roll to ultra-slow-roll transition. One way this can be achieved is by the addition of a Gaussian bump to a working slow-roll potential, giving a model which is both in agreement with CMB observations and can produce a significant abundance of PBHs~\cite{Mishra:2019pzq}. One such potential is given by
\begin{equation}
\label{eq:swagat_potential}
    V(\phi) = V_0 \frac{\phi^2}{m^2 + \phi^2}\left\lbrace1 + K\exp \left[ -\frac{1}{2}\frac{(\phi - \phi_0)^2}{\Sigma^2} \right] \right\rbrace \, ,
\end{equation}
with $m=0.5 \mpl$, $K = 1.876 \times 10^{-3}$, $\phi_0 = 2.005\mpl$ and $\Sigma = 1.993\times 10^{-2} \mpl$~\cite{Mishra:2019pzq}. Note that all of the above precision is needed to produce an asteroid-mass PBH abundance accounting for all of the dark matter, due to fine-tuning~\cite{Cole:2023wyx}. 
This model has a period of ultra-slow roll of $\sim 3$ \efolds, before smoothly transitioning to a phase of constant roll. The transition from slow roll to ultra-slow roll takes $\sim 1$ \efold. As such, one cannot easily propagate modes through the transition analytically, and a numerical approach is required.

\begin{figure}
\begin{center}
        \includegraphics[width=\halffigurewidth\textwidth]{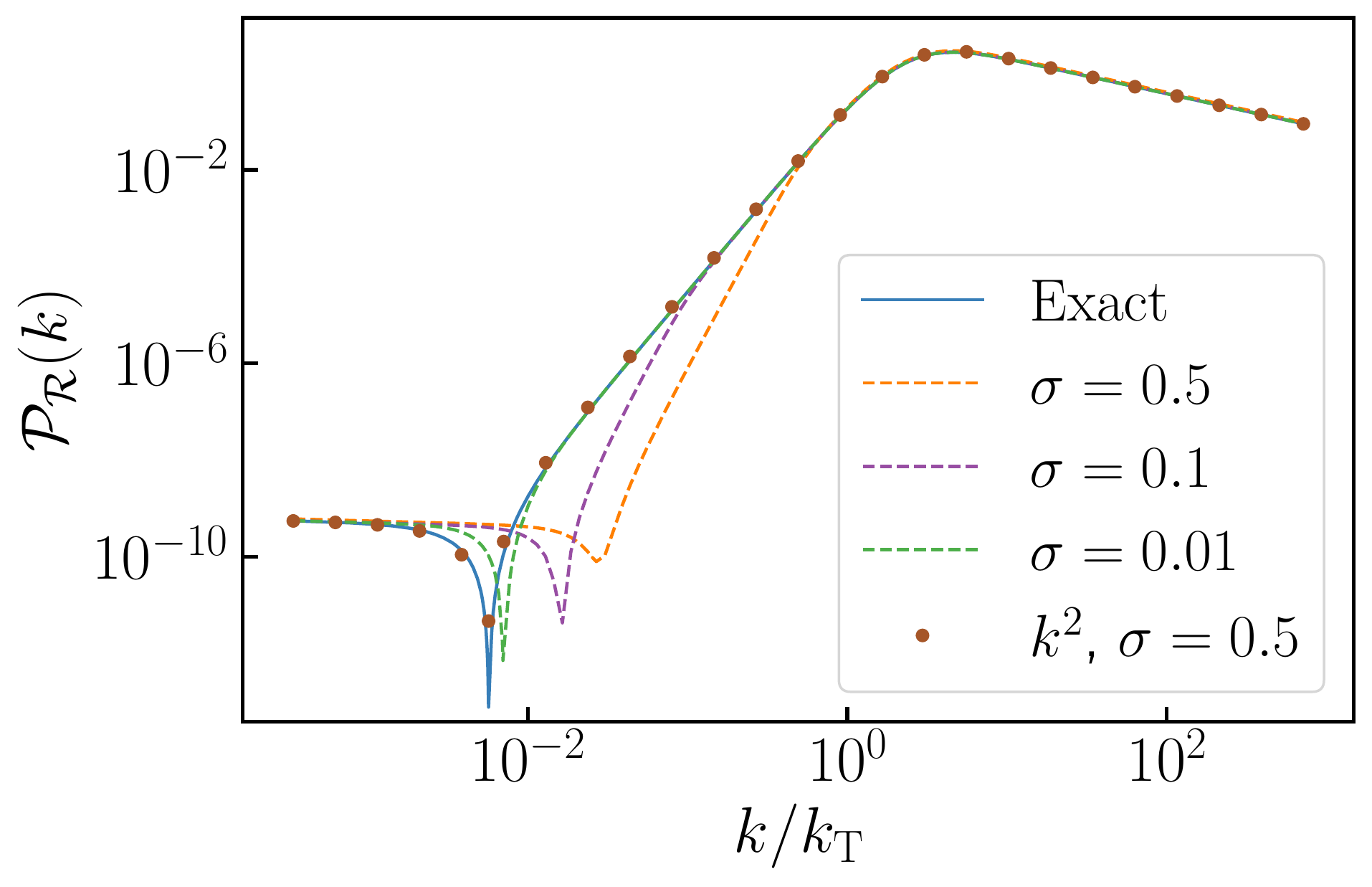}
        \includegraphics[width=\halffigurewidth\textwidth]{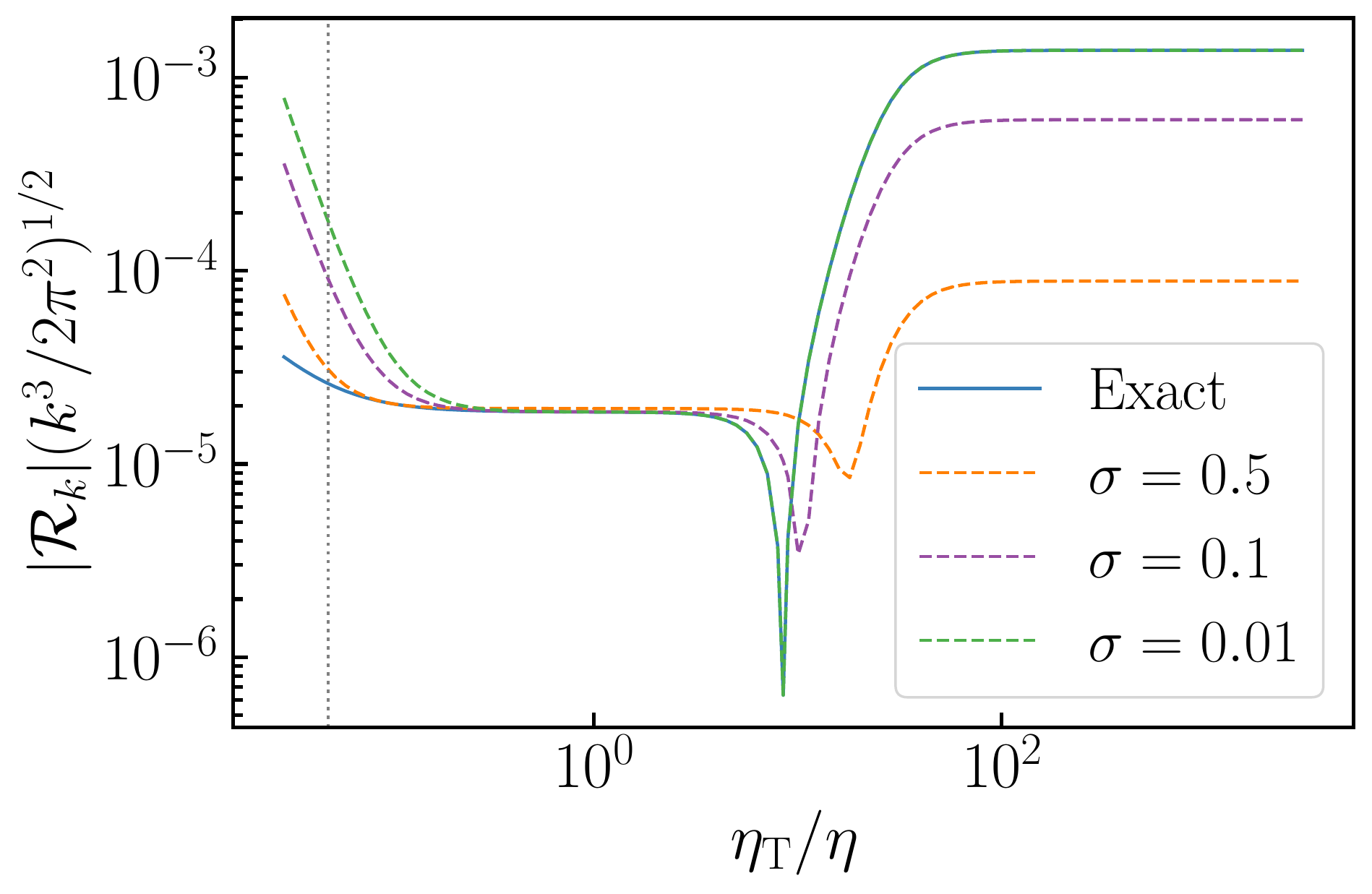}
        \caption{Left: the power spectrum \eqref{eq:PR:Ck} of the curvature perturbation $\R$ for a potential with a Gaussian bump \eqref{eq:swagat_potential}. Right: the evolution of $\R$ for a mode with $k=0.05\, k_{\mathrm{T}}$. The solid blue curve is found by numerically solving \eq{eq:sasaki_mukhanov_equation}, the dashed curves are found using the numerical matching procedure detailed in \Sec{sec:super_horizon_matching}, and the brown data points are found by including the $k^2$ correction \eqref{eq:k_corrected:matching}. The dotted vertical line is Hubble-exit time. The ultra-slow-roll period ($\epsilon_2 < -3$) starts at $\eta_{\mathrm{T}}=-1/k_{\mathrm{T}}$.}
\label{fig:swagat_general}
\end{center}
\end{figure}

Determining the homogeneous growing and decaying modes can be done numerically using the homogeneous-matching procedure, detailed in \eq{eq:hom:matching}. This is shown in \fig{fig:swagat_general}. Phenomenologically we see similar behaviour to the previously discussed piece-wise linear model. The power spectrum is most accurately reconstructed when matching sufficiently long after Hubble-exit, $\sigma=0.01$, although it is still not able to fully reproduce the dip behaviour. This is because $\sigma=0.01$ is not sufficient to satisfy the condition given in \eq{eq:cond:sigma:gen} for this model, as it has a longer duration of ultra-slow-roll inflation compared to the piece-wise linear potential shown above. The $k^2$-corrected data points, found using \eq{eq:k_corrected:matching} to do the matching, are accurate even when matching soon after Hubble-exit for $\sigma=0.5$. This shows the effect of gradient terms on the homogeneous behaviour on super-Hubble scales at the sudden transition from slow roll to ultra-slow roll, in smooth models, not just in the idealised, instant transition case.

The interpretation of the evolution of the mode seen in the right-hand panel in \fig{fig:swagat_general} is the same as for the piece-wise linear model if we define the transition to ultra-slow roll to happen when $\epsilon_2<-3$ first occurs. If the homogeneous matching is done before the transition time, the early-time behaviour is recovered but the behaviour after the transition is not. Equally, matching after the transition reproduces the late-time behaviour (more accurately for smaller $\sigma$) but does not reproduce the early-time behaviour. This shows numerically that the sudden transition to ultra-slow roll has resulted in the behaviour of the homogeneous growing/decaying modes changing due to gradient terms at the transition.

In Appendix~\ref{app:eemeli} we show another example of a model with a smooth potential that leads to a sudden transition from slow roll to ultra-slow roll. In that case we see that a potential with an inflection point shows the same qualitative behaviour for modes before and after a sudden transition from slow roll to ultra-slow roll, suggesting that this is a common feature of such models.

\section{Implications for stochastic inflation}
\label{sec:discussion}

In the stochastic approach to studying inflationary dynamics~\cite{Starobinsky:1986fx,Starobinsky:1994bd} quantum field perturbations originating on small, sub-Hubble scales are incorporated into the locally-homogeneous background classical field in a local patch after crossing the coarse-graining scale, giving the local background field a stochastic kick.
In effect the coarse-graining introduces a matching scale, $k=\sigma aH$, between the solution to the full mode equation on small scales and a homogeneous solution used on larger scales.
In the standard case the stochastic kick is included soon after Hubble exit, corresponding to $\sigma\sim1$. But we have shown that a sudden transition leads to a change in the state of the field on a finite range of super-Hubble scales. 
If there is a discontinuity in the homogeneous field perturbation after coarse-graining, then the standard stochastic approach breaks down. 

One approach to address this issue is to coarse-grain only after the transition or on sufficiently large super-Hubble scales such that any discontinuity in the homogeneous solution at the transition is minimal. If the condition given in \eq{eq:cond:sigma:gen} is used, then the correct post-transition solution for $\delta \phi_{\textrm{h}}$ is captured for all modes. For example in Refs.~\cite{De:2020hdo, Figueroa:2020jkf, Figueroa:2021zah, Ahmadi:2022lsm, Mishra:2023lhe} $\sigma=0.01$ is chosen, in which case coarse-graining captures the transition-induced change to $\delta \phi_{\textrm{h}}$ for modes with $k \gtrsim 0.01\,  k_{\mathrm{T}}$, which is sufficient to avoid any significant discontinuity for a sudden transition leading to an ultra-slow-roll period of $\lesssim3$ \efolds. 

However, incorporating the stochastic kick into the long-wavelength evolution a long time after Hubble exit fails to account for nonlinear effects due to field fluctuations close to the Hubble scale, so is likely to under-estimate the full stochastic effects. 
This may be particularly important when modelling the transition to ultra-slow roll as this is a transient regime. For example, this could reduce the probability of large, rare fluctuations in $\delta N$ which may be associated with PBH production.
We could attempt to evolve the post-transition homogeneous solution back to the Hubble-exit time, and apply the corresponding stochastic kick at that time in order to include the non-adiabatic behaviour induced after the transition. However, this gives an erroneously large non-adiabatic kick before the transition, at a time when the super-Hubble evolution was in fact adiabatic.
This can be seen by comparing $\widetilde{\R}_{\textrm{h}}$ and $\R_{\textrm{h}}$ in \fig{fig:homogeneous_mode} near Hubble exit. The enhanced non-adiabatic decaying mode of $\R_{\textrm{h}}$ is induced by the transition and was not present when the mode crossed the Hubble scale. 

A possible solution is to apply {\em two} kicks for each mode with $k<k_{\rm T}$: the first applied in the usual stochastic way at or soon after Hubble exit before the transition, and the second, correlated kick applied at or just after the transition to reproduce the correct non-adiabatic behaviour post-transition. As we already saw in \fig{fig:homogeneous_mode}, for linear perturbations, the correct behaviour before and after the transition can be recovered if we include the discontinuity in the homogeneous solution at the transition.
The pre-transition kick can be found in a number of ways. The simplest is to use the slow-roll result, $\delta \phi_{\textrm{h}} \simeq iH/\sqrt{2k^3}$ at Hubble exit. Alternatively, this can be done numerically, either directly matching $\delta \phi_{\textrm{h}}$ to $\delta \phi_k$, or using a Bessel function approximation at Hubble exit as described in Appendix~\ref{app:bessel}. The latter allows $\delta \phi_{\textrm{h}}$ to be found close to Hubble-scale crossing and it explicitly removes spurious gradient terms from the homogeneous solutions. 
The post-transition non-adiabatic kick would need to be found from the solution to the full mode equations at the transition. This could again either be done directly from a numerical solution, or with the Bessel function approximation. 
Since the discontinuity in $\delta \phi_{\textrm{h}}$ occurs at the transition for super-Hubble modes, $k<k_{\rm T}$, the second kicks for all these modes would be applied at the same time. Thus a single non-adiabatic kick at the transition, found by summing the contributions from all of the $k<k_{\rm T}$ modes, could incorporate the non-adiabatic change to the homogeneous behaviour induced by the transition.

Finally we note that in the stochastic-inflation formalism it is assumed that the quantum field fluctuations can be described by a classical noise. This is consistent with the initial adiabatic vacuum state becoming a highly squeezed state on super-Hubble scales~\cite{Polarski1996, Lesgourgues:1996jc, Kiefer:2008ku}. However, the discontinuity in the homogeneous solution at the transition may have implications for the quantum-to-classical transition~\cite{Martin:2015qta, Martin:2017zxs, Pattison:2019hef, Martin:2021znx}. A full study of classicalisation at such transitions is beyond the scope of this paper, and we leave further investigation of this interesting question for future work.

\section{Conclusion}
\label{sec:conclusions}

Models of inflation incorporating a transition from slow roll to ultra-slow roll have been intensively studied recently as a mechanism to produce primordial black holes (PBHs). As well as using linear perturbation theory to study the power spectrum produced in such models, non-perturbative approaches, such as the $\delta N$ or stochastic $\delta N$ formalisms, have been used to determine the full probability distribution function for the curvature perturbation. These typically rely on the separate-universe approach, using solutions to the spatially-homogeneous equations of motion to model the evolution of inhomogeneities beyond the Hubble scale.

In this paper we have shown that the separate-universe approach breaks down on a finite range of super-Hubble scales at a sudden transition during inflation from slow roll to ultra-slow roll. The sudden transition leads to particle production in the inflaton field on sub-Hubble scales and non-adiabatic pressure perturbations on super-Hubble scales.

We emphasise that the separate-universe approach works in a piece-wise fashion, before the transition and in the ultra-slow-roll phase after\footnote{It was previously shown in Ref.~\cite{Pattison:2019hef} that the separate universe approach applies beyond slow roll, in contrast to the conclusions of Ref.~\cite{Cruces:2018cvq}. In general one must allow for perturbations in the lapse function on large scales, which vanish only in the slow-roll limit where $\epsilon_1\to0$.}, but 
a solution to the spatially-homogeneous equations is unable to reproduce the discontinuous change in the super-Hubble behaviour at the transition, which is due to small, but finite spatial gradients still present on super-Hubble scales. The role of spatial gradients even on super-Hubble scales in a model of ultra-slow-roll inflation was first noted in Ref.~\cite{Leach:2001zf}.

\begin{figure}
\begin{center}
        \includegraphics[width=\figurewidth\textwidth]{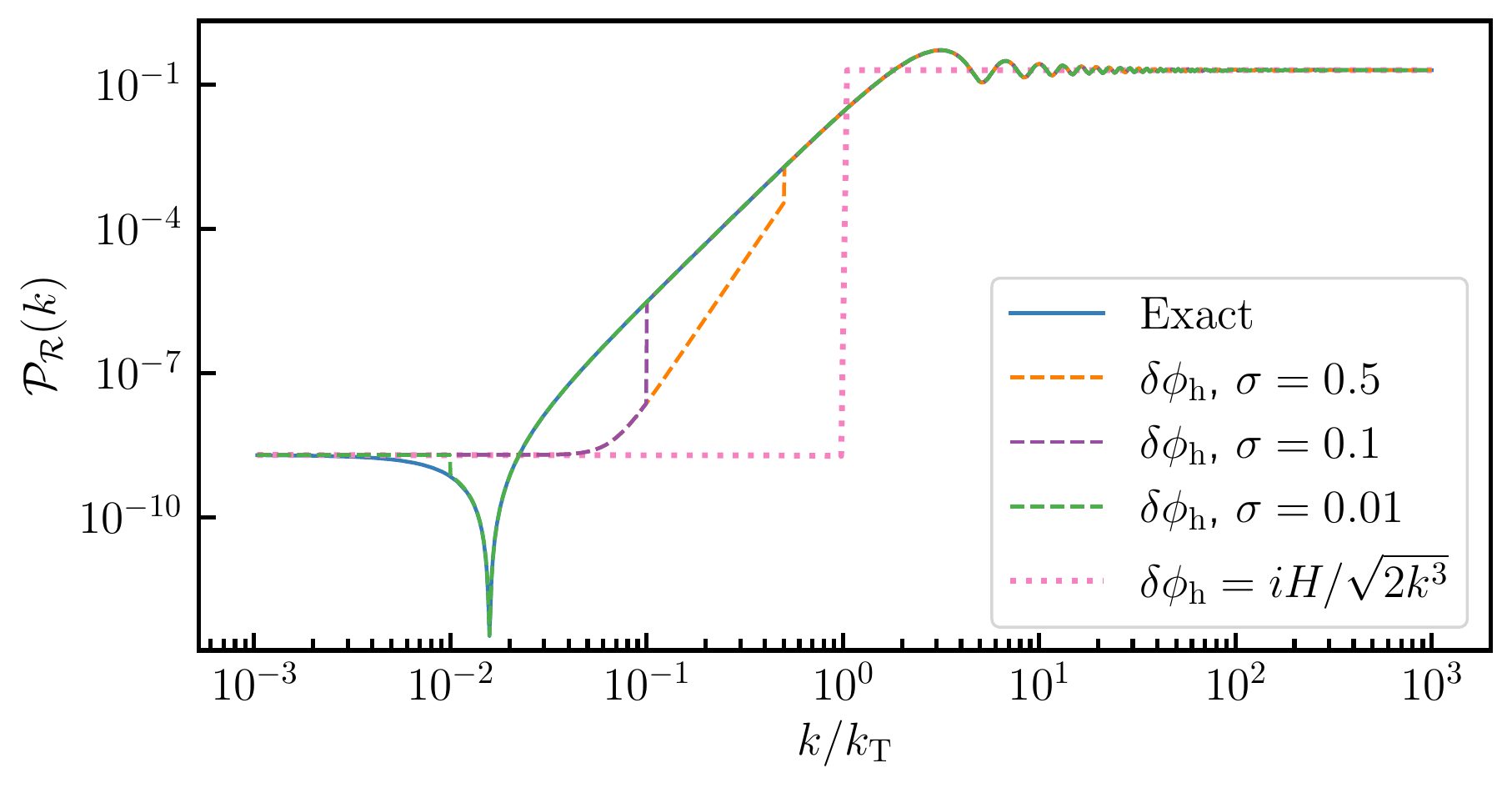}
        \caption{The power spectrum \eqref{eq:PR:Ck} of the curvature perturbation $\R$ for the piece-wise linear potential (\ref{eq_starobinsky-potential}) with $A_-/A_+ = 10^{-4}$. The solid blue curve is found by numerically solving the full mode equation \eqref{eq:sasaki_mukhanov_equation}. The dashed curves use the linear $\delta N$ formalism \eqref{eq:deltaN:lin} with $\delta \phi_{\mathrm{h}}=\dot{\phi}\Rh/H$ given in a piece-wise manner, where \eq{eq:starobinsiky_C_D_pre_transition} is used before the transition and \eq{eq:starobinsiky_C_D_post_transition} after. The $\delta N$ formalism is applied at $k=\sigma aH$. The pink dotted curve uses the result \eqref{eq:R_de_sitter_homogeneous_growing} for a massless field at Hubble-exit in de Sitter, $\delta \phi_{\mathrm{h}} = iH/\sqrt{2k^3}$ with $\sigma =1$. The ultra-slow-roll period starts at comoving time $\eta_{\mathrm{T}}=-1/k_{\mathrm{T}}$.}
\label{fig:delta_N_power_spectrum}
\end{center}
\end{figure}

The discontinuity in the homogeneous behaviour was shown analytically in \eqref{eq:discontinuity_of_R_homogeneous} for an idealised test case corresponding to the piece-wise linear potential \eqref{eq_starobinsky-potential}. In this case the general solution to the full mode equation before and after the transition is given in \eqref{eq:R_slow_roll} allowing us to identify the homogeneous solution \eqref{eq:R_de_sitter_homogeneous} from a formal gradient expansion \eqref{eq:k_power_series_grow_and_decay}.
We have also demonstrated numerically that a discontinuity in the homogeneous solution occurs at a sudden transition in models with a smooth feature in the potential (either a Gaussian bump, or an inflection point, shown in Figs.~\ref{fig:swagat_general} and \ref{fig:eemeli_general} respectively) using numerical matching to estimate the homogeneous solution after Hubble exit. 

This discontinuous behaviour at the transition time, $\eta_{\mathrm{T}}=-1/k_{\mathrm{T}}$, has implications for the $\delta N$ formalism, where one uses the homogeneous solution to follow the evolution of the perturbed universe on super-Hubble scales. The naïve approach, using the Bunch--Davies vacuum state for a massless field in de Sitter at Hubble-exit to set the amplitude of fluctuations in the spatially-homogeneous field,
$\mathcal{P}^{1/2}_{\delta\phi_h}=H/2\pi$,
fails completely to describe the growth of the power spectrum for $k<k_{\mathrm{T}}$ or accurately give the peak value at $k\sim k_{\mathrm{T}}$. See \fig{fig:delta_N_power_spectrum}, where the error in the power spectrum peak is $\sim 60\%$ if $\mathcal{P}^{1/2}_{\delta\phi_h}=H/2\pi$ is used, but less than 0.1\% for the correct post-transition homogeneous $\delta \phi_{\mathrm{h}}$. This shows the importance of non-adiabatic evolution even on sub-horizon scales at the transition.
%


We can only reconstruct the full behaviour by matching fluctuations in the homogeneous field to the mode functions at (or soon after) Hubble-exit if $k^2$-corrections \eqref{eq:k_corrected:matching} are included at the matching, showing the essential role of spatial gradients on super-Hubble scales at the transition. If the corrected homogeneous solution is used post-transition in the classical $\delta N$ formalism \eqref{eq:delta_N_formalism}, including a non-adiabatic decaying mode (\ie including both $\delta\phi_{k*}$ and $\dot{\delta \phi}_{k*}$), the correct linear power spectrum can be recovered at the end of inflation. This suggests that a modified separate-universe approach might still be used, and non-linear perturbations could be found using the classical $\delta N$ formalism in the presence of a sudden transition, if applied in a piece-wise manner. We leave investigation of higher-order effects in the classical $\delta N$ formalism to future work.

In the stochastic-inflation formalism, applying the separate universe approach in a piece-wise manner would correspond to coarse-graining only on scales very much larger than the Hubble radius, such that all the affected modes are coarse-grained after the transition. However this may under estimate stochastic effects by failing to include all super-Hubble fluctuations in the coarse-grained field before the transition. An alternative method would be to coarse-grain close to the Hubble scale before the transition, but then to include the discontinuity in the homogeneous behaviour by summing all the affected modes into a single non-adiabatic kick to the coarse-grained field at the transition. We leave the implementation of a modified stochastic-inflation approach in such models for future work.

\section*{Note Added} 

While this work was in preparation Ref.~\cite{Domenech:2023dxx} appeared, where it was shown that the usual $\delta N$ formalism could only be applied sufficiently long after a sudden transition, such that $\delta \phi_{\mathrm{h}}$ is dominated by the growing mode, which is consistent with our results.

\acknowledgments

The authors would like to thank Laura Iacconi, Eemeli Tomberg, Diego Cruces, David Mulryne and the participants of Cosmo'23 for insightful discussions. This work was supported by the Science and Technology Facilities Council (grant numbers ST/T506345/1 and ST/W001225/1). For the purpose of open access, the authors have applied a Creative Commons Attribution (CC-BY) licence to any Author Accepted Manuscript version arising from this work. Supporting research data are available on reasonable request from the corresponding author, Joseph Jackson.

\appendix

\section{\texorpdfstring{$\delta N$}{δN} for piece-wise linear model}
\label{app:delta_N}
In this section, we implement the linearised version of the $\delta N$ approach presented in \Sec{sec:delta:N}, to Starobinsky's piece-wise linear model discussed in \Sec{sec:starobinsky}. The first step is to derive the function $ N(\phi_\uin, \dot{\phi}_\uin)$, which returns the number of \efolds realised from the spatially-homogeneous initial field configuration given by $\phi_\uin$ and $\dot{\phi}_\uin$ to the end of inflation. This can be done using the general solution of the Klein--Gordon equation~\eqref{eq:klein_gordon} with $N$ as the time variable, namely
\begin{equation}
    \label{eq:starobinsky_phi_solution_N_general}
    \phi (N) = -\frac{A_{\pm}}{3H^2}N + B_{\pm}^{(1)}+ B_{\pm}^{(2)}e^{-3N} \, .
\end{equation}
The integration constants $B_{\pm}^{(1)}$ and $B_{\pm}^{(2)}$ depend on the initial phase-space coordinates
$(\phi_\uin, \dot{\phi}_\uin)$. 

Let us first consider the case where $\phi_\uin<\phi_{\mathrm{T}}$. By differentiating \Eq{eq:starobinsky_phi_solution_N_general} with respect to time, 
\begin{equation}
    \label{eq:starobinsky_phi_dot_solution_N_general}
    \dot{\phi} = -\frac{A_{-}}{3H} -3 H B_{-}^{(2)}e^{-3N} \, ,
\end{equation}
and combining \Eqs{eq:starobinsky_phi_solution_N_general} and \eqref{eq:starobinsky_phi_dot_solution_N_general}, one finds
\begin{equation}
    \label{eq:starobinsky_N_function_initial}
    N = -\frac{3H^2}{A_{-}} \left(\phi + \frac{\dot{\phi}}{3H} - B_{-}^{(1)} + \frac{A_{-}}{9H^2} \right) .
\end{equation}
The constant $B_{-}^{(1)}$ can be found by requiring that $\phi(0)=\phi_\uin$ in \Eq{eq:starobinsky_phi_solution_N_general} and $\dot{\phi}(0) = \dot{\phi}_\uin$ in \Eq{eq:starobinsky_phi_dot_solution_N_general}, leading to
\begin{equation}
    \label{eq:starobinsky_N_function_B_1_minus}
    B_-^{(1)} = \phi_\uin + \frac{\dot{\phi}_\uin}{3H} +  \frac{A_-}{9H^2} \, .
\end{equation}
The number of \efolds from $(\phi_\uin, \dot{\phi}_\uin)$ to the end of inflation at $(\phi_\uend, \dot{\phi}_\uend)$, is then
\begin{equation}
\label{eq:starobinsky_solution_N_function_minus}
    N_-(\phi_\uin, \dot{\phi}_\uin) = \frac{3H^2}{A_{-}} \left(\phi_\uin - \phi_{\mathrm{end}} + \frac{\dot{\phi}_\uin-\dot{\phi}_{\mathrm{end}}}{3H} \right) \, .
\end{equation}

For $\phi_\uin>\phi_{\mathrm{T}}$, the total number of \efolds~can be decomposed into $N_{\mathrm{T}}(\phi_\uin, \dot{\phi}_\uin)$, the number of \efolds~until the transition, and $N_{-}[\phi_{\mathrm{T}}, \dot{\phi}_{\mathrm{T}}(\phi_\uin, \dot{\phi}_\uin)]$, the number of \efolds~from the transition to the end of inflation. Note that the field velocity at the transition, $\dot{\phi}_{\mathrm{T}}$, is a function of the initial phase-space coordinate $(\phi_\uin, \dot{\phi}_\uin)$. 
For $N_{\mathrm{T}}(\phi_\uin, \dot{\phi}_\uin)$, one can solve \Eq{eq:starobinsky_phi_solution_N_general} for $N$ with $+$ as a subscript and find
\begin{equation}
    \label{eq:starobinsky_N_function_pre_transition}
    N_{\mathrm{T}}\left(\phi_\uin, \dot{\phi}_\uin\right) =  \frac{3H^2}{A_{+}} \left(B_+^{(1)}  - \phi_{\mathrm{T}} \right)  + \frac{1}{3}W_0 \left[ \frac{9H^2B_+^{(2)}}{A_+}e^{-\frac{9H^2}{A_+}\left(B_+^{(1)}  - \phi_{\mathrm{T}} \right)} \right] \, ,
\end{equation}
where $W_0(x)$ is the $0^{\text{th}}$ branch of the Lambert function. In this expression, $B_+^{(1)}$ and $B_{+}^{(2)}$ are set as before by requiring that $\phi(0)=\phi_\uin$ and $\dot\phi(0)=\dot\phi_\uin$ in \Eqs{eq:starobinsky_phi_solution_N_general} and \eqref{eq:starobinsky_phi_dot_solution_N_general}, leading to 
\begin{equation}
    \label{eq:starobinsky_N_function_B_1_and_2_plus}
    B_+^{(1)} = \phi_\uin + \frac{\dot{\phi}_\uin}{3H} + \frac{A_+}{9H^2} \quad \text{and} \quad B_+^{(2)} = -\frac{\dot{\phi}_\uin}{3H} - \frac{A_+}{9H^2} \, .
\end{equation}
For $N_{-}[\phi_{\mathrm{T}}, \dot{\phi}_{\mathrm{T}}(\phi_\uin, \dot{\phi}_\uin)]$, by differentiating \Eq{eq:starobinsky_phi_solution_N_general} with respect to time and evaluating the result at the time $N_{\mathrm{T}}(\phi_\uin, \dot{\phi}_\uin)$ given in \Eq{eq:starobinsky_N_function_pre_transition}, one first finds
\begin{equation}
    \label{eq:starobinsky_phi_dot_transition}
    \dot{\phi}_{\mathrm{T}}(\phi_\uin, \dot{\phi}_\uin) =  -\frac{A_{+}}{3H}\left\{1+ W_0 \left[ \frac{9H^2B_+^{(2)}}{A_+}e^{-\frac{9H^2}{A_+}\left(B_+^{(1)}  - \phi_{\mathrm{T}} \right)} \right] \right\}  \, .
\end{equation}
Replacing $\phi_\uin$ by $\phi_{\mathrm{T}}$ and $\dot{\phi}_\uin$ by the above expression in \Eq{eq:starobinsky_solution_N_function_minus} then gives
\bea
\label{eq:starobinsky_solution_N_function_minus2}
    N_{-}[\phi_{\mathrm{T}}, \dot{\phi}_{\mathrm{T}}(\phi_\uin, \dot{\phi}_\uin)] =&  \frac{3H^2}{A_{-}} \left( \phi_{\mathrm{T}} - \phi_{\mathrm{end}} -\frac{\dot{\phi}_{\mathrm{end}}}{3H} \right) 
    \\ & -\frac{A_{+}}{3A_-}\left\{1+ W_0 \left[ \frac{9H^2B_+^{(2)}}{A_+}e^{-\frac{9H^2}{A_+}\left(B_+^{(1)}  - \phi_{\mathrm{T}} \right)} \right] \right\} \, .
\eea
Note that the dependence on $\phi_\uin$ and $\dot{\phi}_\uin$ is implicit through $B_+^{(1)}$ and $B_+^{(2)}$. One can summarise the above results as
\begin{equation}
    \label{eq:starobinsky_solution_N_function_final}
    N(\phi, \dot{\phi}) = \begin{cases}
    N_{\mathrm{T}}(\phi, \dot{\phi}) + N_{-}[\phi_{\mathrm{T}}, \dot{\phi}_{\mathrm{T}}(\phi, \dot{\phi})]& \text{if } \phi \geq \phi_{\mathrm{T}} \, ,\\
     N_-(\phi, \dot{\phi}) & \text{if } \phi < \phi_{\mathrm{T}}  \, ,
\end{cases}
\end{equation}
where $N_{\mathrm{T}}(\phi, \dot{\phi})$, $N_{-}[\phi_{\mathrm{T}}, \dot{\phi}_{\mathrm{T}}(\phi, \dot{\phi})]$ and $N_-(\phi, \dot{\phi})$ are given above, and where hereafter the subscripts ``$\uin$'' are dropped for convenience. 

The next step is to differentiate the number of realised \efolds with respect to the initial field value and its velocity. One obtains
\begin{equation}
    \label{eq:starobinsky_linear_delta_N_dN_by_dphi}
    \frac{\partial N}{\partial \phi} = \begin{cases}
    \frac{3H^2}{A_+}\left[1-\frac{\Delta A}{A_-} \frac{W_0(x)}{1+W_0(x)}\right] & \text{if } \phi \geq \phi_{\mathrm{T}} \, ,\\
     \frac{3H^2}{A_-}, & \text{if } \phi < \phi_{\mathrm{T}}  \, ,
\end{cases}
\end{equation}
and
\begin{equation}
    \label{eq:starobinsky_linear_delta_N_dN_by_dotdphi}
    \frac{\partial N}{\partial \dot{\phi}} = \begin{cases}
    \frac{H}{A_+}\left[1-\frac{\Delta A}{A_-}\left(1+\frac{A_+}{9H^2 B_+^{(2)}}\right) \frac{W_0(x)}{1+W_0(x)}\right] & \text{if } \phi \geq \phi_{\mathrm{T}} \, ,\\
     \frac{H}{A_-}, & \text{if } \phi < \phi_{\mathrm{T}}  \, ,
\end{cases}
\end{equation}
where
\begin{equation}
    \label{eq:Lambert_func_argument}
    x \equiv  \frac{9H^2B_+^{(2)}}{A_+}e^{-\frac{9H^2}{A_+}\left(B_+^{(1)}  - \phi_{\mathrm{T}} \right)} \, .
\end{equation}

Then, for a given scale $k$, one has to evaluate the above derivatives at the time when that scale crosses out the $\sigma$-Hubble radius, \ie when $k=\sigma a H$. In the de Sitter limit, this occurs at a time (denoted by a subscript ``$*$'' hereafter) $\eta_*(k)=\sigma/k$, or equivalently, $N_*(k)= N_{\mathrm{T}}+  \ln [k/(\sigma k_{\mathrm{T}})]$. From \Eq{eq:klein_gordon_starobinsky_solution_with_eta}, one has
\begin{equation}
    \label{eq:klein_gordon_starobinsky_solution_with_k}
    \phi_* (k) = \begin{cases}
    -\frac{A_+}{3H^2}\ln{\Big( \frac{k}{\sigma k_{\mathrm{T}}}\Big)} + \phi_{\mathrm{T}} & \text{for } k \leq \sigma k_{\mathrm{T}} \, ,\\
    -\frac{A_-}{3H^2}\ln{\Big( \frac{k}{\sigma k_{\mathrm{T}}}\Big)} +\frac{\Delta A}{9H^2}\Bigg[1-\Big( \frac{k}{\sigma k_{\mathrm{T}}} \Big)^{-3} \Bigg]  + \phi_{\mathrm{T}} & \text{for } k > \sigma k_{\mathrm{T}} \, ,
\end{cases}
\end{equation}
and 
\begin{equation}
    \label{eq:klein_gordon_starobinsky_solution_with_k_time_derivative}
    \dot{\phi}_* (k) = \begin{cases}
    -\frac{A_+}{3H} & \text{for } k \leq \sigma k_{\mathrm{T}} \, ,\\
    -\frac{A_-}{3H} +\frac{\Delta A}{3H}\Big( \frac{k}{\sigma k_{\mathrm{T}}} \Big)^{-3} & \text{for } k > \sigma k_{\mathrm{T}} \, .
\end{cases}
\end{equation}
For $k\leq \sigma k_{\mathrm{T}}$, one has $\phi_* \geq \phi_{\mathrm{T}}$ and inserting \Eqs{eq:klein_gordon_starobinsky_solution_with_k} and~\eqref{eq:klein_gordon_starobinsky_solution_with_k_time_derivative} into \Eq{eq:starobinsky_N_function_B_1_and_2_plus} leads to $B_{+*}^{(1)}=\phi_{\mathrm{T}}+A_+/(3H^2)\ln(\sigma k_{\mathrm{T}}/k)$ and $B_{+*}^{(2)}=0$. Therefore, $x_*=0$ for those scales, and using that $W_0(x)=x+\mathcal{O}(x^2)$, one obtains
\bea
\left.\frac{\partial N}{\partial\phi}\right\vert_*=
\begin{cases}
\frac{3H^2}{A_+} & \text{for } k\leq \sigma k_{\mathrm{T}}\\
\frac{3H^2}{A_-} & \text{for } k > \sigma k_{\mathrm{T}}\\
\end{cases}
\eea
and
\bea
\left.\frac{\partial N}{\partial\dot\phi}\right\vert_*=
\begin{cases}
\frac{H}{A_+}\left[1-\frac{\Delta A}{A_-}\left(\frac{k}{\sigma k_{\mathrm{T}}}\right)^3\right] & \text{for } k\leq \sigma k_{\mathrm{T}}\\
\frac{H}{A_-} & \text{for } k > \sigma k_{\mathrm{T}}\\
\end{cases}\, .
\eea

The next step is to evaluate $\delta\phi_k$ and $\delta\dot{\phi}_k$ at the $\sigma$-Hubble crossing time, within the full linear cosmological perturbation theory. Recalling that $v=a\delta\phi$ in the spatially-flat gauge, and given that $\R=v/z$ where $z=\mathrm{sign}(\dot{\phi})\sqrt{2a^2\epsilon_1}$, one has 
\bea
\delta\phi_k=-\sqrt{2\epsilon_1}\R_k 
\quad\text{and}\quad
\delta\dot\phi_k = -\sqrt{2\epsilon_1} \left(\dot{\R}_k+\frac{\epsilon_2}{2} H \R_k \right) .
\eea
For $k\leq \sigma k_{\mathrm{T}}$ (\ie $\phi_*\geq \phi_{\mathrm{T}}$), from \Eqs{eq:epsilon_1_starobinsky} and~\eqref{eq:epsilon_2_starobinsky} one has $\epsilon_{1*}=A_+^2/(18 H^4)$ while $\epsilon_{2*}$ can be discarded, leading to 
\bea
\delta\phi_{k*}=\frac{-A_+}{3H^2} \R_{k*}
\quad\text{and}\quad
\delta\dot\phi_{k*}=\frac{-A_+}{3H^2} \dot\R_{k*}\, .
\eea
For $k> \sigma k_{\mathrm{T}}$ (\ie $\phi_*< \phi_{\mathrm{T}}$), one obtains 
\bea
\epsilon_{1*}=\frac{A_-^2}{18H^4}\left[1-\frac{\Delta A}{A_-}\left(\frac{\sigma k_{\mathrm{T}}}{k}\right)^3\right]^2
\quad\text{and}\quad
\epsilon_{2*}= 6 \frac{\frac{\Delta A}{A_-}\left(\frac{\sigma k_{\mathrm{T}}}{k}\right)^3}{1-\frac{\Delta A}{A_-}\left(\frac{\sigma k_{\mathrm{T}}}{k}\right)^3}\, ,
\eea
leading to
\bea
\delta\phi_{k*}= \frac{-A_-}{3H^2}\left[1-\frac{\Delta A}{A_-}\left(\frac{\sigma k_{\mathrm{T}}}{k}\right)^3\right] \R_{k*}
\eea
and
\bea
\delta\dot\phi_{k*}= -\frac{\Delta A}{H}\left(\frac{\sigma k_{\mathrm{T}}}{k}\right)^3 \R_{k*}-\frac{A_-}{3H^2}\left[1-\frac{\Delta A}{A_-}\left(\frac{\sigma k_{\mathrm{T}}}{k}\right)^3\right] \dot\R_{k*}\, .
\eea
Combining the above results into \Eq{eq:deltaN:lin}, one finally obtains
\bea
\delta N_k = 
\begin{cases}
-\R_{k*} - \left[1-\frac{\Delta A}{A_-}\left(\frac{k}{\sigma k_{\mathrm{T}}}\right)^3\right] \frac{\dot\R_{k*}}{3H}
& \text{for } k\leq \sigma k_{\mathrm{T}}\\
-\R_{k*} - \left[1-\frac{\Delta A}{A_-}\left(\frac{k}{\sigma k_{\mathrm{T}}}\right)^{-3}\right] \frac{\dot\R_{k*}}{3H}
& \text{for } k> \sigma k_{\mathrm{T}}
\end{cases}\, .
\label{eq:deltaN:linear:Staro}
\eea
This expression agrees with the one obtained for $\hat{C}_k$ (\ie the late-time value of $\R_k = -\zeta_k$) in the homogeneous-matching procedure, see \Eq{eq:HM:gen:Staro}. This proves that the homogeneous-matching procedure and the linearised $\delta N$ formalism are strictly equivalent. This should be clear from the way these two approaches are formulated (namely, they both rely on using linear cosmological perturbation theory below the matching scale and the homogeneous dynamics above that scale), but here the correspondence is shown explicitly in the case of the Starobinsky piece-wise linear potential. 
\section{Example: inflection-point model}
\label{app:eemeli}
\begin{figure}
\begin{center}
        \includegraphics[width=\figurewidth\textwidth]{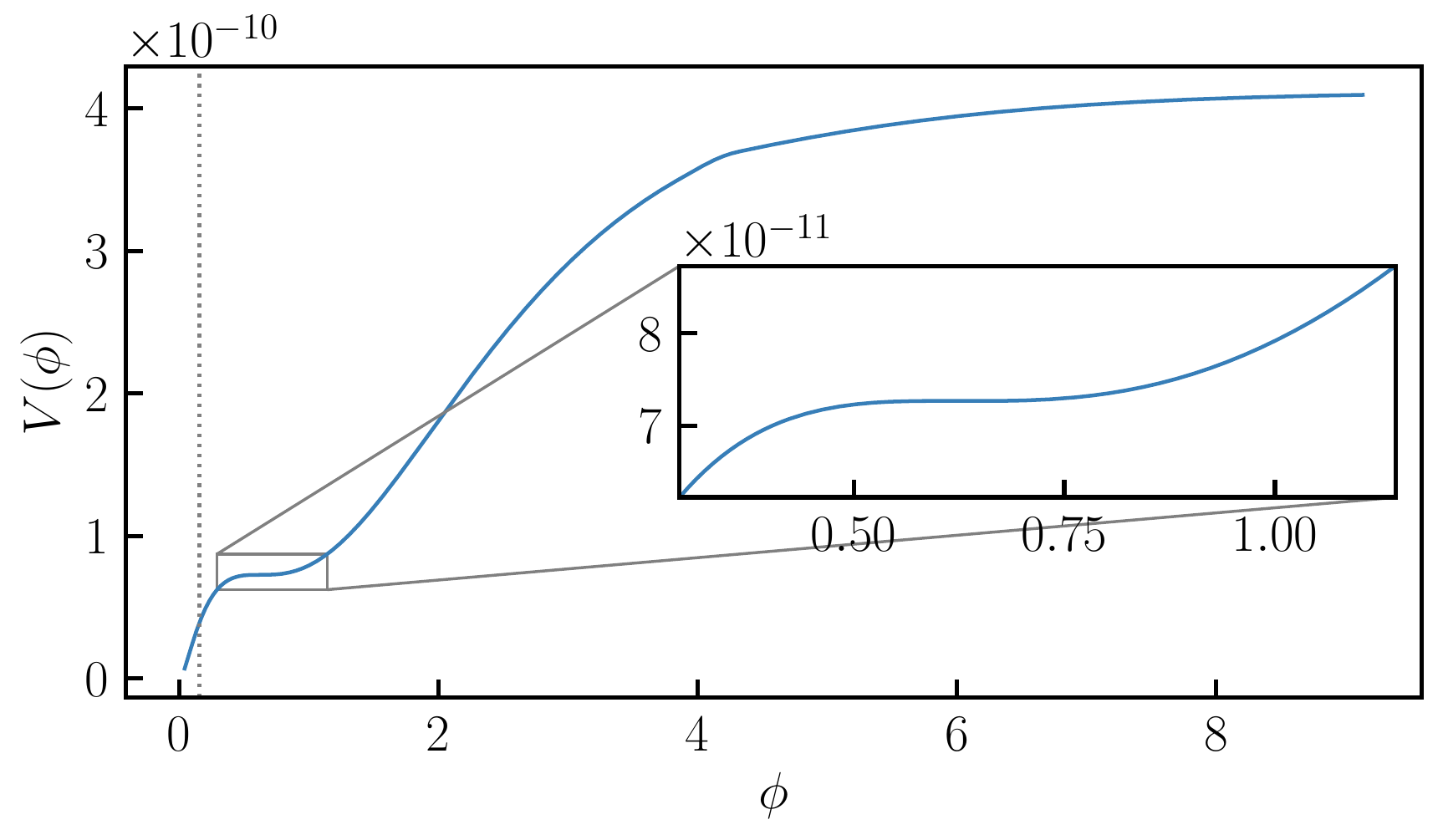}
        \caption{The potential for Higgs inflation modified by quantum corrections~\cite{Rasanen:2018fom}. The parameters are chosen such that sufficient PBHs of asteroid mass are produced to provide all of the dark matter, giving the highlighted inflection point. The dotted vertical line is where inflation ends.}
\label{fig:eemeli_potential}
\end{center}
\end{figure}

A potential with an inflection point can also lead to a phase of ultra-slow-roll inflation and significant PBH production, with many different models proposed, see \eg \cite{Garcia-Bellido:2017mdw, Di:2017ndc, Ballesteros:2017fsr, Geller:2022nkr}. Here we will investigate a model motivated by Higgs inflation with quantum corrections~\cite{Rasanen:2018fom}, which has been used for simulations of stochastic inflation~\cite{Figueroa:2020jkf, Figueroa:2021zah}. The potential in Ref.~\cite{Rasanen:2018fom} does not have a closed analytical form, but the authors kindly provided numerical data for $V(\phi)$ to reproduce the asteroid mass PBH case, which is plotted in \fig{fig:eemeli_potential}. There is a bump near $\phi=4$ due to an interpolation, but this has little effect on the dynamics at the inflection point. This model has a phase of ultra-slow roll of $\sim 3$ \efolds followed by a smooth transition to a constant-roll phase.

\begin{figure}
\begin{center}
        \includegraphics[width=\halffigurewidth\textwidth]{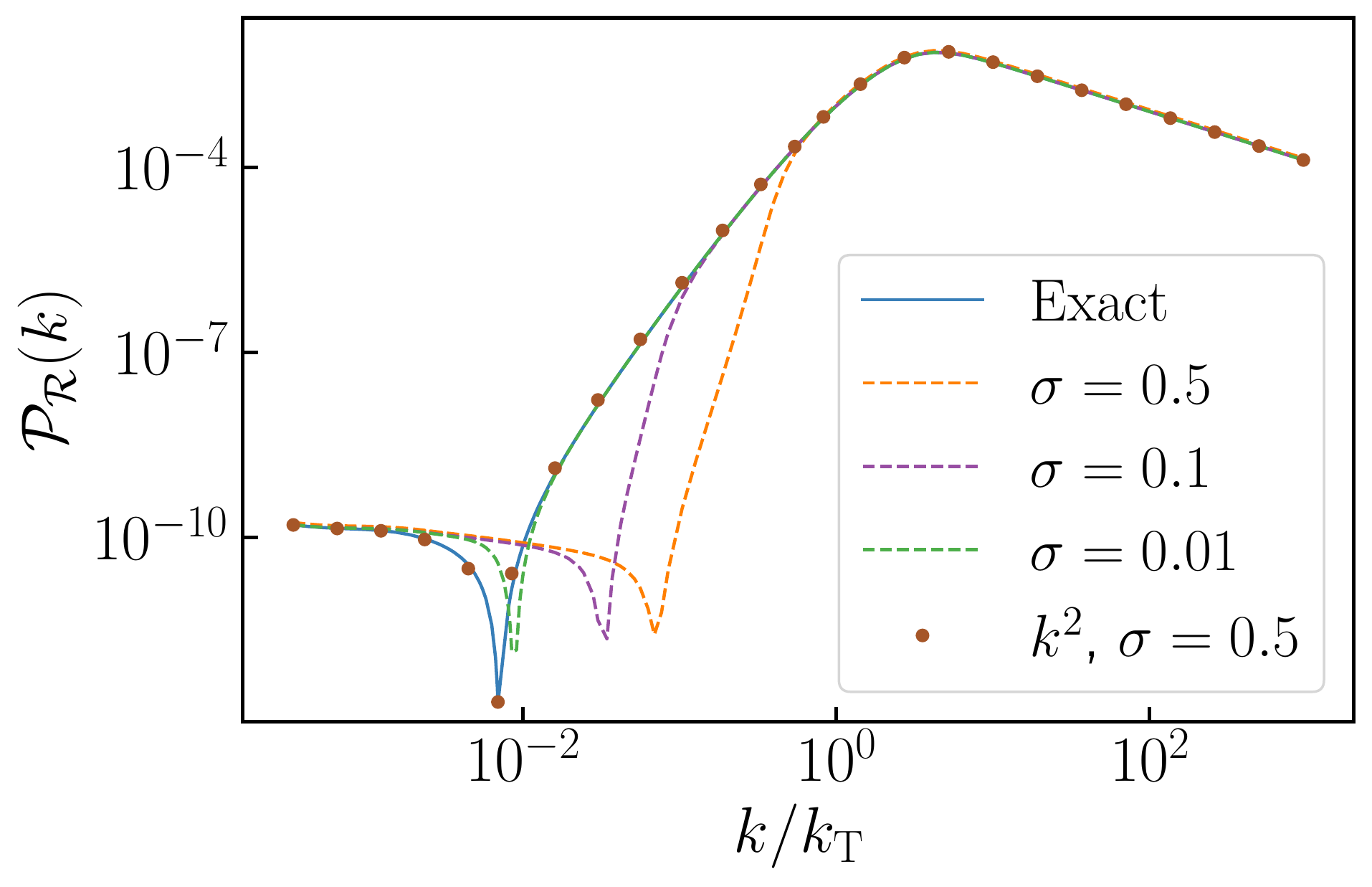}
        \includegraphics[width=\halffigurewidth\textwidth]{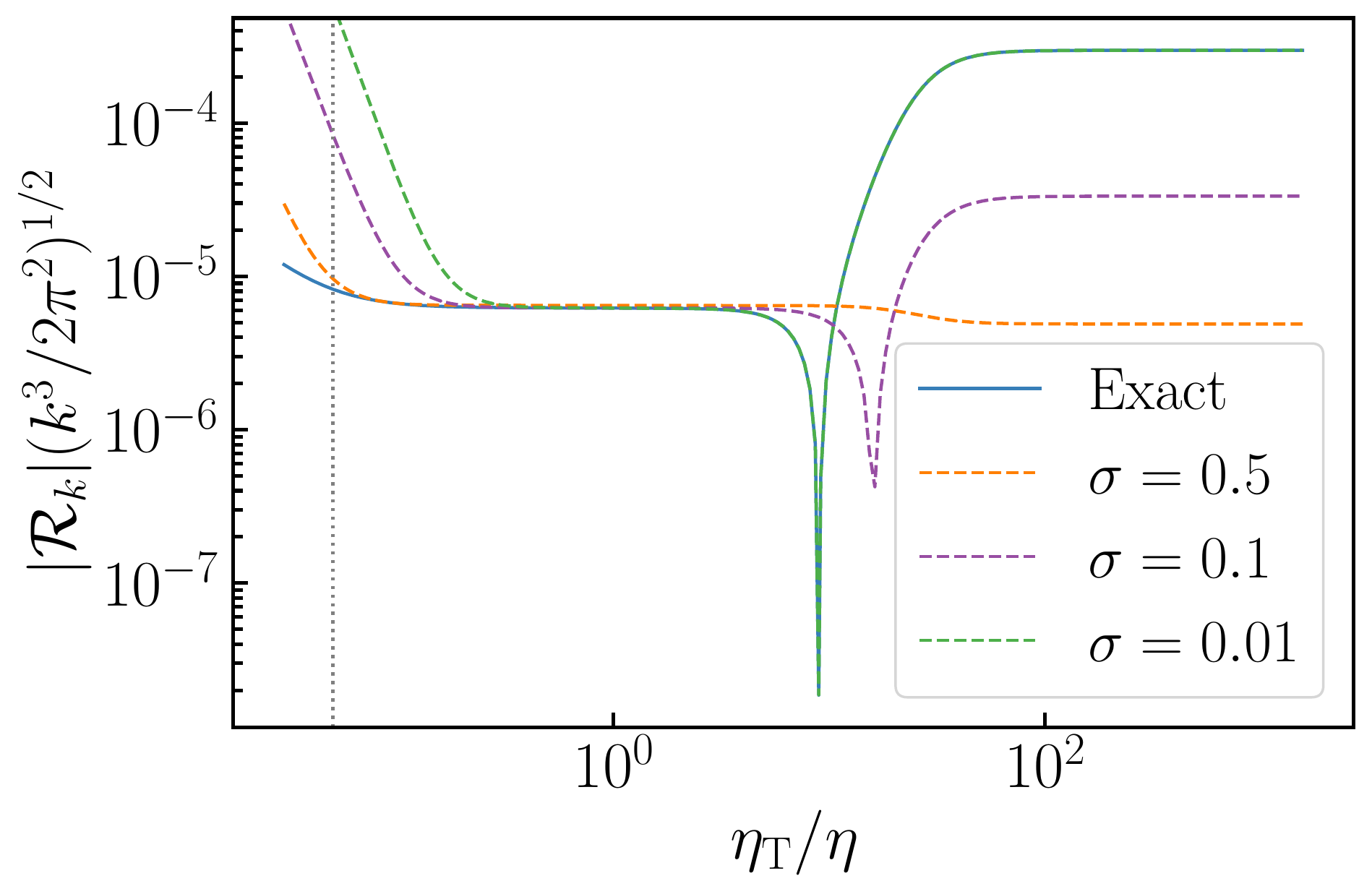}
        \caption{Left: the curvature perturbation power spectrum at the end of inflation for the inflection-point model motivated by Higgs inflation~\cite{Rasanen:2018fom}. Right: the evolution of $\R$ for the mode with $k=0.05k_{\mathrm{T}}$. The solid blue curve is found by numerically solving \eq{eq:sasaki_mukhanov_equation}, the dashed curves are found using the numerical matching detailed in \Sec{sec:super_horizon_matching}, and the brown data points are found by including the $k^2$ correction (\ref{eq:k_corrected:matching}). The dotted vertical line is Hubble-exit time. The ultra-slow-roll period ($\epsilon_2 < -3$) starts at $\eta_{\mathrm{T}}=-1/k_{\mathrm{T}}$.}
\label{fig:eemeli_general}
\end{center}
\end{figure}

Determining the growing and decaying modes can again be done numerically using the homogeneous matching procedure detailed in \eq{eq:hom:matching}, the result of which is shown in \fig{fig:eemeli_general}. Both the power spectrum and mode evolution plots show similar behaviour to the Gaussian bump discussed in \Sec{sec:swagat}. Therefore the same conclusions drawn from the Gaussian bump model also apply here. This demonstrates that the dynamics discussed in this paper are a general feature of single-field models with a sudden transition from slow roll to ultra-slow roll, followed by a slow-roll or constant-roll phase to the end of inflation.

\section{Bessel functions}
\label{app:bessel}

Let us introduce the parameter
\begin{equation}
\label{eq:nu_squared}
    \nu^2 = -\frac{\mu^2}{(aH)^2} + \frac{1}{4} = \frac{9}{4} - \epsilon_1 + \frac{3}{2} \epsilon_2 -  \frac{1}{2} \epsilon_1\epsilon_2 + \frac{1}{4} \epsilon_2^2 + \frac{1}{2} \epsilon_2\epsilon_3 \, ,
\end{equation}
where $\mu^2$ is given in \eq{eq:z_prime_prime_by_z}. If $\nu^2$ is a constant, then \eq{eq:sasaki_mukhanov_equation} has an exact solution~\cite{Stewart1993}
\begin{equation}
\label{eq:sasai_mukhanov_bessel}
    v_k = \sqrt{-\eta} [A_k J_{\nu}(-k\eta) + B_k Y_{\nu}(-k\eta)] \, ,
\end{equation}
where
\begin{equation}
    \label{eq:bessel_functions}
    J_{\nu}(x) = \bigg( \frac{x}{2} \bigg)^{\nu} \sum_{n=0}^{\infty} \bigg( \frac{x}{2} \bigg)^{2n} \frac{(-1)^n}{n! \Gamma (\nu + n + 1)} \hspace{0.8em} \text{and} \hspace{0.8em} Y_{\nu}(x) = \frac{J_{\nu}(x) \cos{(\nu \pi)} - J_{-\nu}(x)}{\sin{(\nu \pi)} } \, ,
\end{equation}
and $\Gamma$ is the gamma function. In the standard case the integration constants $A_k$ and $B_k$ are set by using Bunch--Davies initial conditions \eqref{eq:bunch_davies}. This is the more general form of \eq{eq:R_slow_roll}, and reproduces it when $\nu=3/2$. As both $J_{\nu}$ and $Y_{\nu}$ have power series expansions, the growing and decaying mode interpretation of \eq{eq:k_power_series_grow_and_decay} applies. One can also propagate \eq{eq:sasai_mukhanov_bessel} through the sharp transition of \Sec{sec:starobinsky}, using \eq{eq:sasaki_mukhano_derivative_transition}, to find the post-transition $A_k$ and $B_k$ in an equivalent manner to finding $\alpha_k$ and $\beta_k$. While the computation is more complex, one again finds $A_k$ and $B_k$ are different to the pre-transition values. This shows the conclusions of \Sec{sec:starobinsky} are valid beyond the de Sitter approximation.

\begin{figure}
\begin{center}
        \includegraphics[width=\halffigurewidth\textwidth]{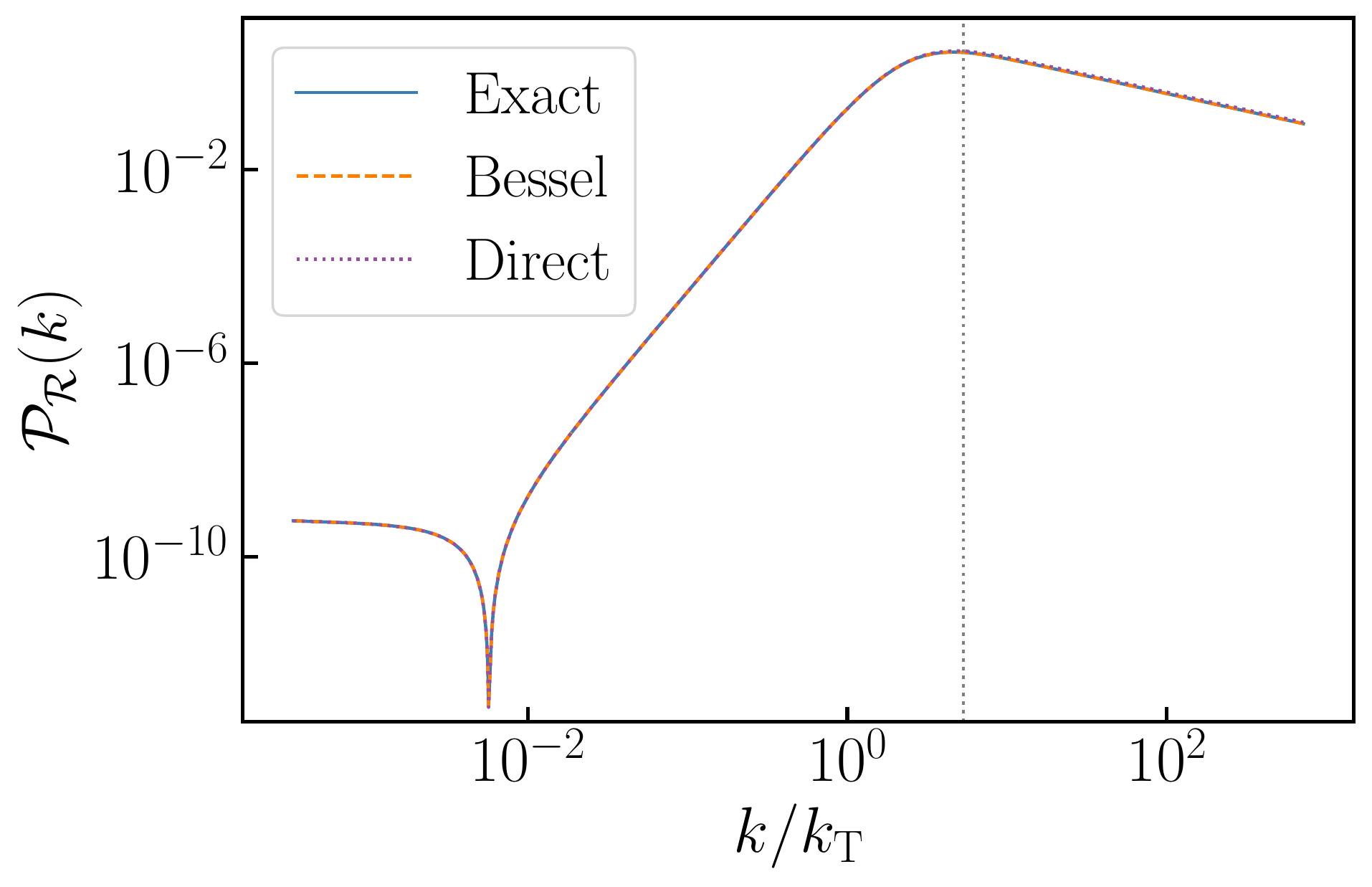}
        \includegraphics[width=\halffigurewidth\textwidth]{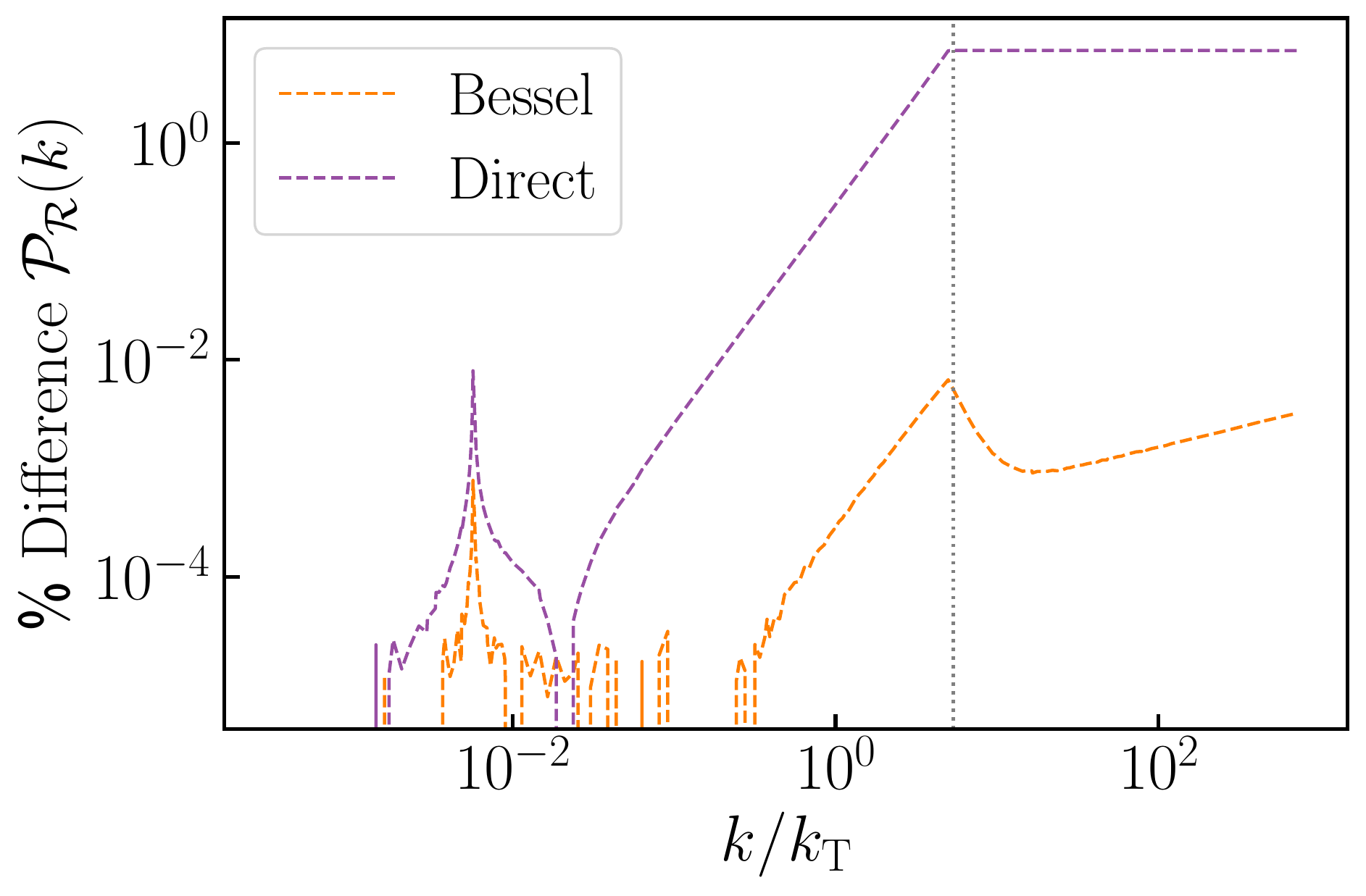}
        \caption{Left panel: the curvature perturbation power spectrum at the end of inflation for a potential with a Gaussian bump (\ref{eq:swagat_potential}). The solid blue curve is found by numerically solving the full mode equation \eq{eq:sasaki_mukhanov_equation}. The orange dashed line uses the intermediate step of finding the homogeneous behaviour using the Bessel functions, see Eqs. (\ref{eq:bessel_homogeneous_C}) and (\ref{eq:bessel_homogeneous_D}), while the purple dotted line uses the homogeneous matching \eqref{eq:hom:matching} directly. 
        The matching time, $\eta_*$, is when $\nu^2$ \eqref{eq:nu_squared} first becomes constant post-transition (for modes to the left of the dotted vertical line), or when $k=0.5aH$ for modes to the right of the dotted vertical line.
        Right: the percent difference between the two matching prescriptions and the full numerical result. The ultra-slow-roll period ($\epsilon_2 < -3$) starts at $\eta_{\mathrm{T}}=-1/k_{\mathrm{T}}$. }
\label{fig:swagat_bessel}
\end{center}
\end{figure}

The solution given in \eq{eq:sasai_mukhanov_bessel} is valid during any period of constant $\nu^2$, as is the case during both slow roll and ultra-slow roll. This is true even if following a period when $\nu^2$ is varying, in a piece-wise manner. Then, $A_k$ and $B_k$ can be found by matching onto the full solution for $v_k$ at a time $\eta_{*}$, using
\begin{equation}
\label{eq:bessel_matching}
\begin{split}
    A_k &= \frac{1}{2(-\eta_*)^{3/2}}\frac{2 \eta_* Y_{\nu}(-k\eta_*) v_{k*}' -  Y_{\nu}(-k\eta_*)v_k -2\eta_* [Y_{\nu}(-k\eta_*)]'v_{k*}}{J_{\nu}[Y_{\nu}(-k\eta_*)]' - Y_{\nu}(-k\eta_*)[J_{\nu}(-k\eta_*)]'}\, , \\
    B_k &= -\frac{1}{2(-\eta_*)^{3/2}}\frac{2 \eta_* J_{\nu}(-k\eta_*) v_{k*}' -  J_{\nu}(-k\eta_*)v_k -2\eta_* [J_{\nu}(-k\eta_*)]'v_{k*}}{J_{\nu}[Y_{\nu}(-k\eta_*)]' - Y_{\nu}(-k\eta_*)[J_{\nu}(-k\eta_*)]'} \, .
\end{split}
\end{equation}
This allows the homogeneous matching of \Sec{sec:super_horizon_matching} to be done more accurately with the following steps. Assuming no further sudden transitions occur, then the homogeneous growing and decaying mode coefficients of \eq{eq:R_super_horizon_solution_general_intregrated} will be constant until the end of inflation. This is true even if there is a slow variation in $\nu^2$ later. The leading order expansion of \eq{eq:sasai_mukhanov_bessel} can then be used, along with $\R = v_k/z$ and the homogeneous matching \eqref{eq:hom:matching}, to give
\begin{multline}
\label{eq:bessel_homogeneous_C}
    \hat{C}_k =   \frac{\sqrt{\eta_*}}{z_*}\left(  a_k \eta_*^{-\nu} + b_k \eta_*^{\nu} \right) + \\
    z_*\sqrt{\eta_*} \left\{ a_k \left[ \left( \frac{1}{2} - \nu \right) \eta_*^{-1} - \frac{z'}{z} \right] \eta_*^{-\nu} + b_k \left[ \left( \frac{1}{2} + \nu \right) \eta_*^{-1} - \frac{z'}{z} \right] \eta_*^{\nu}\right\} \int_{\eta_*}^0\frac{\dd\tilde{\eta}}{z^2(\tilde{\eta})} \, ,
\end{multline}
\begin{equation}
\label{eq:bessel_homogeneous_D}
    \hat{D}_k =  - z_*\sqrt{\eta_*} \left\{ a_k \left[ \left( \frac{1}{2} - \nu \right) \eta_*^{-1} - \frac{z'}{z} \right] \eta_*^{-\nu} + b_k \left[ \left( \frac{1}{2} + \nu \right) \eta_*^{-1} - \frac{z'}{z} \right] \eta_*^{\nu}\right\} \, ,
\end{equation}
where
\begin{equation}
\label{eq:bessel_homogeneous_coefficents}
\begin{split}
    a_k & =  - \bigg[ B_k\frac{\Gamma(\nu)}{\pi} \bigg]\bigg(\frac{-k}{2} \bigg)^{-\nu}  , \\
    b_k & = \bigg[ \frac{A_k}{\Gamma(\nu+1)} - B_k\frac{\Gamma(-\nu)\cos{(\nu \pi )}}{\pi} \bigg]\bigg(\frac{-k}{2} \bigg)^{\nu}\, .
\end{split}
\end{equation}
With this approach, any contamination from gradient terms can be effectively removed. This improves the accuracy of the method, and allows the post-transition homogeneous behaviour to be found closer to Hubble exit (corresponding to a matching scale, $k=\sigma aH$, with larger $\sigma$). 

The method detailed above has been applied numerically in \fig{fig:swagat_bessel} for the Gaussian bump model, see \Sec{sec:swagat}. The power spectrum at the end of inflation has been reconstructed to test the accuracy of this method against applying \eq{eq:hom:matching} directly. For modes which would have left the matching scale ($k=0.5aH$) before $\nu^2$ becomes constant, the matching has been done at $\eta_*=\eta_{\mathrm{T}}/10$, when $\nu^2$ becomes constant post-transition. The later modes are simply matched when $k=0.5aH$ (\ie $\sigma=0.5$). The dotted line shows when the matching time changes. Using Eqs.~\eqref{eq:bessel_homogeneous_C} and \eqref{eq:bessel_homogeneous_D}, which effectively removes any gradient terms, produces excellent agreement with the full numerical result. The error is less than $0.01\%$ throughout, while directly applying \eq{eq:hom:matching} gives a percent level error. Both have vanishing errors for modes which left the Hubble scale long before the transition.

Periods of constant $\nu^2$ after the slow-roll to ultra-slow-roll transition are present in many models. This is the case for all potentials investigated in this paper, as can be seen, using \eq{eq:nu_squared}, in \fig{fig:z_double_prime}. Therefore it is expected that the approach will be applicable in a wide variety of models.

\bibliographystyle{JHEP-edit}
\bibliography{main.bib}

\providecommand{\href}[2]{#2}\begingroup\raggedright\begin{thebibliography}{100}

\bibitem{Starobinsky:1980te}
A.~A. Starobinsky, \emph{A new type of isotropic cosmological models without
  singularity},
  \href{https://doi.org/10.1016/0370-2693(80)90670-X}{\emph{Physics Letters B}
  {\bfseries 91}{\bfseries (1)} (1980) 99}.

\bibitem{Sato:1980yn}
K.~Sato, \emph{{First-order phase transition of a vacuum and the expansion of
  the Universe}}, \href{https://doi.org/10.1093/mnras/195.3.467}{\emph{Monthly
  Notices of the Royal Astronomical Society} {\bfseries 195}{\bfseries (3)}
  (1981) 467}.

\bibitem{Guth:1980zm}
A.~H. Guth, \emph{{Inflationary universe: A possible solution to the horizon
  and flatness problems}},
  \href{https://doi.org/10.1103/PhysRevD.23.347}{\emph{Physical Review D}
  {\bfseries 23}{\bfseries (2)} (1981) 347}.

\bibitem{Linde:1981mu}
A.~D. Linde, \emph{{A new inflationary universe scenario: A possible solution
  of the horizon, flatness, homogeneity, isotropy and primordial monopole
  problems}}, \href{https://doi.org/10.1016/0370-2693(82)91219-9}{\emph{Physics
  Letters B} {\bfseries 108}{\bfseries (6)} (1982) 389}.

\bibitem{Albrecht:1982wi}
A.~Albrecht and P.~J. Steinhardt, \emph{{Cosmology for Grand Unified Theories
  with Radiatively Induced Symmetry Breaking}},
  \href{https://doi.org/10.1103/PhysRevLett.48.1220}{\emph{Physical Review
  Letters} {\bfseries 48}{\bfseries (17)} (1982) 1220}.

\bibitem{Linde:1983gd}
A.~D. Linde, \emph{Chaotic inflation},
  \href{https://doi.org/10.1016/0370-2693(83)90837-7}{\emph{Physics Letters B}
  {\bfseries 129}{\bfseries (3--4)} (1983) 177}.

\bibitem{Mukhanov:1981xt}
V.~F. Mukhanov and G.~V. Chibisov, \emph{Quantum fluctuations and a nonsingular
  universe},
  \href{http://www.jetpletters.ru/ps/1510/article_23079.shtml}{\emph{Journal of
  Experimental and Theoretical Physics Letters} {\bfseries 33}{\bfseries (10)}
  (1981) 532}.

\bibitem{Mukhanov:1982nu}
V.~F. Mukhanov and G.~V. Chibisov, \emph{Vacuum energy and large-scale
  structure of the universe},
  \href{http://jetp.ras.ru/cgi-bin/e/index/e/56/2/p258?a=list}{\emph{Journal of
  Experimental and Theoretical Physics} {\bfseries 56}{\bfseries (2)} (1982)
  258}.

\bibitem{Starobinsky:1982ee}
A.~A. Starobinsky, \emph{Dynamics of phase transition in the new inflationary
  universe scenario and generation of perturbations},
  \href{https://doi.org/10.1016/0370-2693(82)90541-X}{\emph{Physics Letters B}
  {\bfseries 117}{\bfseries (3--4)} (1982) 175}.

\bibitem{Guth:1982ec}
A.~H. Guth and S.-Y. Pi, \emph{{Fluctuations in the New Inflationary
  Universe}}, \href{https://doi.org/10.1103/PhysRevLett.49.1110}{\emph{Physical
  Review Letters} {\bfseries 49}{\bfseries (15)} (1982) 1110}.

\bibitem{Hawking:1982cz}
S.~Hawking, \emph{The development of irregularities in a single bubble
  inflationary universe},
  \href{https://doi.org/10.1016/0370-2693(82)90373-2}{\emph{Physics Letters B}
  {\bfseries 115}{\bfseries (4)} (1982) 295}.

\bibitem{Bardeen:1983qw}
J.~M. Bardeen, P.~J. Steinhardt and M.~S. Turner, \emph{Spontaneous creation of
  almost scale-free density perturbations in an inflationary universe},
  \href{https://doi.org/10.1103/PhysRevD.28.679}{\emph{Physical Review D}
  {\bfseries 28}{\bfseries (4)} (1983) 679}.

\bibitem{Lyth:2009zz}
D.~H. Lyth and A.~R. Liddle,
  \href{https://doi.org/10.1017/CBO9780511819209}{\emph{{The Primordial Density
  Perturbation: Cosmology, Inflation and the Origin of Structure}}}, Cambridge
  University Press (2009).

\bibitem{Baumann:2022mni}
D.~Baumann, \href{https://doi.org/10.1017/9781108937092}{\emph{Cosmology}},
  Cambridge University Press (2022).

\bibitem{Ade:2015xua}
{\scshape Planck collaboration}, \emph{{Planck 2015 results. XIII. Cosmological
  parameters}},
  \href{https://doi.org/10.1051/0004-6361/201525830}{\emph{Astronomy \&
  Astrophysics} {\bfseries 594} (2016) A13}
  [\href{https://arxiv.org/abs/1502.01589}{{\ttfamily 1502.01589}}].

\bibitem{Ade:2015lrj}
{\scshape Planck collaboration}, \emph{{Planck 2015 results. XX. Constraints on
  inflation}},
  \href{https://doi.org/10.1051/0004-6361/201525898}{\emph{Astronomy \&
  Astrophysics} {\bfseries 594} (2016) A20}
  [\href{https://arxiv.org/abs/1502.02114}{{\ttfamily 1502.02114}}].

\bibitem{Planck2018}
{\scshape Planck collaboration}, \emph{{Planck 2018 results. VI. Cosmological
  parameters}},
  \href{https://doi.org/10.1051/0004-6361/201833910}{\emph{Astronomy \&
  Astrophysics} {\bfseries 641} (2020) A6}
  [\href{https://arxiv.org/abs/1807.06209}{{\ttfamily 1807.06209}}].

\bibitem{Chluba:2015bqa}
J.~Chluba, J.~Hamann and S.~P. Patil, \emph{{Features and new physical scales
  in primordial observables: Theory and observation}},
  \href{https://doi.org/10.1142/S0218271815300232}{\emph{International Journal
  of Modern Physics D} {\bfseries 24}{\bfseries (10)} (2015) 1530023}
  [\href{https://arxiv.org/abs/1505.01834}{{\ttfamily 1505.01834}}].

\bibitem{Caprini:2018mtu}
C.~Caprini and D.~G. Figueroa, \emph{Cosmological backgrounds of gravitational
  waves}, \href{https://doi.org/10.1088/1361-6382/aac608}{\emph{Classical and
  Quantum Gravity} {\bfseries 35}{\bfseries (16)} (2018) 163001}
  [\href{https://arxiv.org/abs/1801.04268}{{\ttfamily 1801.04268}}].

\bibitem{Domenech:2021ztg}
G.~Domenech, \emph{{Scalar Induced Gravitational Waves Review}},
  \href{https://doi.org/10.3390/universe7110398}{\emph{Universe} {\bfseries
  7}{\bfseries (11)} (2021) 398}
  [\href{https://arxiv.org/abs/2109.01398}{{\ttfamily 2109.01398}}].

\bibitem{Ozsoy:2023ryl}
O.~Özsoy and G.~Tasinato, \emph{{Inflation and Primordial Black Holes}},
  \href{https://doi.org/10.3390/universe9050203}{\emph{Universe} {\bfseries
  9}{\bfseries (5)} (2023) 203}
  [\href{https://arxiv.org/abs/2301.03600}{{\ttfamily 2301.03600}}].

\bibitem{NANOGrav:2023gor}
{\scshape NANOGrav collaboration}, \emph{{The NANOGrav 15 yr Data Set: Evidence
  for a Gravitational-wave Background}},
  \href{https://doi.org/10.3847/2041-8213/acdac6}{\emph{The Astrophysical
  Journal Letters} {\bfseries 951}{\bfseries (1)} (2023) L8}
  [\href{https://arxiv.org/abs/2306.16213}{{\ttfamily 2306.16213}}].

\bibitem{Antoniadis:2023rey}
{\scshape EPTA collaboration and InPTA collaboration}, \emph{{The second data
  release from the European Pulsar Timing Array III. Search for gravitational
  wave signals}},
  \href{https://doi.org/10.1051/0004-6361/202346844}{\emph{Astronomy \&
  Astrophysics} {\bfseries 678} (2023) A50}
  [\href{https://arxiv.org/abs/2306.16214}{{\ttfamily 2306.16214}}].

\bibitem{Xu:2023wog}
H.~Xu et~al., \emph{{Searching for the Nano-Hertz Stochastic Gravitational Wave
  Background with the Chinese Pulsar Timing Array Data Release I}},
  \href{https://doi.org/10.1088/1674-4527/acdfa5}{\emph{Research in Astronomy
  and Astrophysics} {\bfseries 23}{\bfseries (7)} (2023) 075024}
  [\href{https://arxiv.org/abs/2306.16216}{{\ttfamily 2306.16216}}].

\bibitem{NANOGrav:2023hvm}
{\scshape NANOGrav collaboration}, \emph{{The NANOGrav 15 yr Data Set: Search
  for Signals from New Physics}},
  \href{https://doi.org/10.3847/2041-8213/acdc91}{\emph{The Astrophysical
  Journal Letters} {\bfseries 951}{\bfseries (1)} (2023) L11}
  [\href{https://arxiv.org/abs/2306.16219}{{\ttfamily 2306.16219}}].

\bibitem{Carr:2016drx}
B.~Carr, F.~Kühnel and M.~Sandstad, \emph{Primordial black holes as dark
  matter}, \href{https://doi.org/10.1103/PhysRevD.94.083504}{\emph{Physical
  Review D} {\bfseries 94}{\bfseries (8)} (2016) 083504}
  [\href{https://arxiv.org/abs/1607.06077}{{\ttfamily 1607.06077}}].

\bibitem{Carr:2020gox}
B.~Carr, K.~Kohri, Y.~Sendouda and J.~Yokoyama, \emph{Constraints on primordial
  black holes}, \href{https://doi.org/10.1088/1361-6633/ac1e31}{\emph{Reports
  on Progress in Physics} {\bfseries 84}{\bfseries (11)} (2021) 116902}
  [\href{https://arxiv.org/abs/2002.12778}{{\ttfamily 2002.12778}}].

\bibitem{Carr:2020xqk}
B.~Carr and F.~Kühnel, \emph{{Primordial Black Holes as Dark Matter: Recent
  Developments}},
  \href{https://doi.org/10.1146/annurev-nucl-050520-125911}{\emph{Annual Review
  of Nuclear and Particle Science} {\bfseries 70} (2020) 355}
  [\href{https://arxiv.org/abs/2006.02838}{{\ttfamily 2006.02838}}].

\bibitem{Green:2020jor}
A.~M. Green and B.~J. Kavanagh, \emph{Primordial black holes as a dark matter
  candidate}, \href{https://doi.org/10.1088/1361-6471/abc534}{\emph{Journal of
  Physics G} {\bfseries 48}{\bfseries (4)} (2021) 043001}
  [\href{https://arxiv.org/abs/2007.10722}{{\ttfamily 2007.10722}}].

\bibitem{Byrnes:2018txb}
C.~T. Byrnes, P.~S. Cole and S.~P. Patil, \emph{Steepest growth of the power
  spectrum and primordial black holes},
  \href{https://doi.org/10.1088/1475-7516/2019/06/028}{\emph{Journal of
  Cosmology and Astroparticle Physics} {\bfseries 2019}{\bfseries (06)} (2019)
  028} [\href{https://arxiv.org/abs/1811.11158}{{\ttfamily 1811.11158}}].

\bibitem{Cole:2022xqc}
P.~S. Cole, A.~D. Gow, C.~T. Byrnes and S.~P. Patil, \emph{Steepest growth
  re-examined: repercussions for primordial black hole formation},  (2022)
  [\href{https://arxiv.org/abs/2204.07573}{{\ttfamily 2204.07573}}].

\bibitem{Karam:2022nym}
A.~Karam \textit{et~al.}, \emph{Anatomy of single-field inflationary models for
  primordial black holes},
  \href{https://doi.org/10.1088/1475-7516/2023/03/013}{\emph{Journal of
  Cosmology and Astroparticle Physics} {\bfseries 2023}{\bfseries (03)} (2023)
  013} [\href{https://arxiv.org/abs/2205.13540}{{\ttfamily 2205.13540}}].

\bibitem{Sasaki:1986hm}
M.~Sasaki, \emph{{Large Scale Quantum Fluctuations in the Inflationary
  Universe}}, \href{https://doi.org/10.1143/PTP.76.1036}{\emph{Progress of
  Theoretical Physics} {\bfseries 76}{\bfseries (5)} (1986) 1036}.

\bibitem{Mukhanov:1988jd}
V.~F. Mukhanov, \emph{Quantum theory of gauge-invariant cosmological
  perturbations},
  \href{http://jetp.ras.ru/cgi-bin/e/index/e/67/7/p1297?a=list}{\emph{Journal
  of Experimental and Theoretical Physics} {\bfseries 67}{\bfseries (7)} (1988)
  1297}.

\bibitem{Kristiano:2022maq}
J.~Kristiano and J.~Yokoyama, \emph{{Ruling Out Primordial Black Hole Formation
  From Single-Field Inflation}},  (2022)
  [\href{https://arxiv.org/abs/2211.03395}{{\ttfamily 2211.03395}}].

\bibitem{Ota:2022hvh}
A.~Ota, M.~Sasaki and Y.~Wang, \emph{{Scale-invariant enhancement of
  gravitational waves during inflation}},
  \href{https://doi.org/10.1142/S0217732323500633}{\emph{Modern Physical
  Letters A} {\bfseries 38}{\bfseries (12n13)} (2023) 2350063}
  [\href{https://arxiv.org/abs/2209.02272}{{\ttfamily 2209.02272}}].

\bibitem{Choudhury:2023vuj}
S.~Choudhury, M.~R. Gangopadhyay and M.~Sami, \emph{{No-go for the formation of
  heavy mass Primordial Black Holes in Single Field Inflation}},  (2023)
  [\href{https://arxiv.org/abs/2301.10000}{{\ttfamily 2301.10000}}].

\bibitem{Motohashi:2023syh}
H.~Motohashi and Y.~Tada, \emph{Squeezed bispectrum and one-loop corrections in
  transient constant-roll inflation},
  \href{https://doi.org/10.1088/1475-7516/2023/08/069}{\emph{Journal of
  Cosmology and Astroparticle Physics} {\bfseries 2023}{\bfseries (08)} (2023)
  069} [\href{https://arxiv.org/abs/2303.16035}{{\ttfamily 2303.16035}}].

\bibitem{Firouzjahi:2023ahg}
H.~Firouzjahi and A.~Riotto, \emph{{Primordial Black Holes and Loops in
  Single-Field Inflation}},  (2023)
  [\href{https://arxiv.org/abs/2304.07801}{{\ttfamily 2304.07801}}].

\bibitem{Franciolini:2023lgy}
G.~Franciolini, A.~J. Iovino, M.~Taoso and A.~Urbano, \emph{{One loop to rule
  them all: Perturbativity in the presence of ultra slow-roll dynamics}},
  (2023) [\href{https://arxiv.org/abs/2305.03491}{{\ttfamily 2305.03491}}].

\bibitem{Fumagalli:2023loc}
J.~Fumagalli \textit{et~al.}, \emph{One-loop infrared rescattering by enhanced
  scalar fluctuations during inflation},  (2023)
  [\href{https://arxiv.org/abs/2307.08358}{{\ttfamily 2307.08358}}].

\bibitem{Fumagalli:2023hpa}
J.~Fumagalli, \emph{{Absence of one-loop effects on large scales from small
  scales in non-slow-roll dynamics}},  (2023)
  [\href{https://arxiv.org/abs/2305.19263}{{\ttfamily 2305.19263}}].

\bibitem{Tada:2023rgp}
Y.~Tada, T.~Terada and J.~Tokuda, \emph{Cancellation of quantum corrections on
  the soft curvature perturbations},  (2023)
  [\href{https://arxiv.org/abs/2308.04732}{{\ttfamily 2308.04732}}].

\bibitem{Starobinsky:1986fxa}
A.~A. Starobinskii, \emph{{Multicomponent de Sitter (inflationary) stages and
  the generation of perturbations}},
  \href{http://www.jetpletters.ru/ps/1419/article_21563.shtml}{\emph{Journal of
  Experimental and Theoretical Physics Letters} {\bfseries 42}{\bfseries (3)}
  (1985) 152}.

\bibitem{Sasaki1996}
M.~Sasaki and E.~D. Stewart, \emph{{A General Analytic Formula for the Spectral
  Index of the Density Perturbations Produced during Inflation}},
  \href{https://doi.org/10.1143/PTP.95.71}{\emph{Progress of Theoretical
  Physics} {\bfseries 95}{\bfseries (1)} (1996) 71}
  [\href{https://arxiv.org/abs/astro-ph/9507001}{{\ttfamily
  astro-ph/9507001}}].

\bibitem{Sasaki:1998ug}
M.~Sasaki and T.~Tanaka, \emph{{Super-Horizon Scale Dynamics of Multi-Scalar
  Inflation}}, \href{https://doi.org/10.1143/PTP.99.763}{\emph{Progress of
  Theoretical Physics} {\bfseries 99}{\bfseries (5)} (1998) 763}
  [\href{https://arxiv.org/abs/gr-qc/9801017}{{\ttfamily gr-qc/9801017}}].

\bibitem{Lyth:2004gb}
D.~H. Lyth, K.~A. Malik and M.~Sasaki, \emph{A general proof of the
  conservation of the curvature perturbation},
  \href{https://doi.org/10.1088/1475-7516/2005/05/004}{\emph{Journal of
  Cosmology and Astroparticle Physics} {\bfseries 2005}{\bfseries (05)} (2005)
  004} [\href{https://arxiv.org/abs/astro-ph/0411220}{{\ttfamily
  astro-ph/0411220}}].

\bibitem{Wands:2000dp}
D.~Wands, K.~A. Malik, D.~H. Lyth and A.~R. Liddle, \emph{New approach to the
  evolution of cosmological perturbations on large scales},
  \href{https://doi.org/10.1103/PhysRevD.62.043527}{\emph{Physical Review D}
  {\bfseries 62}{\bfseries (4)} (2000) 043527}
  [\href{https://arxiv.org/abs/astro-ph/0003278}{{\ttfamily
  astro-ph/0003278}}].

\bibitem{Lyth:2003im}
D.~H. Lyth and D.~Wands, \emph{Conserved cosmological perturbations},
  \href{https://doi.org/10.1103/PhysRevD.68.103515}{\emph{Physical Review D}
  {\bfseries 68}{\bfseries (10)} (2003) 103515}
  [\href{https://arxiv.org/abs/astro-ph/0306498}{{\ttfamily
  astro-ph/0306498}}].

\bibitem{Lyth:2005fi}
D.~H. Lyth and Y.~Rodriguez, \emph{{Inflationary Prediction for Primordial
  Non-Gaussianity}},
  \href{https://doi.org/10.1103/PhysRevLett.95.121302}{\emph{Physical Review
  Letters} {\bfseries 95}{\bfseries (12)} (2005) 121302}
  [\href{https://arxiv.org/abs/astro-ph/0504045}{{\ttfamily
  astro-ph/0504045}}].

\bibitem{Artigas:2021zdk}
D.~Artigas, J.~Grain and V.~Vennin, \emph{Hamiltonian formalism for
  cosmological perturbations: the~separate-universe approach},
  \href{https://doi.org/10.1088/1475-7516/2022/02/001}{\emph{Journal of
  Cosmology and Astroparticle Physics} {\bfseries 2022}{\bfseries (02)} (2022)
  001} [\href{https://arxiv.org/abs/2110.11720}{{\ttfamily 2110.11720}}].

\bibitem{Salopek:1990jq}
D.~S. Salopek and J.~R. Bond, \emph{Nonlinear evolution of long-wavelength
  metric fluctuations in inflationary models},
  \href{https://doi.org/10.1103/PhysRevD.42.3936}{\emph{Physical Review D}
  {\bfseries 42}{\bfseries (12)} (1990) 3936}.

\bibitem{Rigopoulos:2003ak}
G.~I. Rigopoulos and E.~P.~S. Shellard, \emph{Separate universe approach and
  the evolution of nonlinear superhorizon cosmological perturbations},
  \href{https://doi.org/10.1103/PhysRevD.68.123518}{\emph{Physical Review D}
  {\bfseries 68}{\bfseries (12)} (2003) 123518}
  [\href{https://arxiv.org/abs/astro-ph/0306620}{{\ttfamily
  astro-ph/0306620}}].

\bibitem{Biagetti:2018pjj}
M.~Biagetti, G.~Franciolini, A.~Kehagias and A.~Riotto, \emph{Primordial black
  holes from inflation and quantum diffusion},
  \href{https://doi.org/10.1088/1475-7516/2018/07/032}{\emph{Journal of
  Cosmology and Astroparticle Physics} {\bfseries 2018}{\bfseries (07)} (2018)
  032} [\href{https://arxiv.org/abs/1804.07124}{{\ttfamily 1804.07124}}].

\bibitem{Firouzjahi:2020jrj}
H.~Firouzjahi, A.~Nassiri-Rad and M.~Noorbala, \emph{Stochastic nonattractor
  inflation}, \href{https://doi.org/10.1103/PhysRevD.102.123504}{\emph{Physical
  Review D} {\bfseries 102}{\bfseries (12)} (2020) 123504}
  [\href{https://arxiv.org/abs/2009.04680}{{\ttfamily 2009.04680}}].

\bibitem{Cai:2022erk}
Y.-F. Cai \textit{et~al.}, \emph{{Highly non-Gaussian tails and primordial
  black holes from single-field inflation}},
  \href{https://doi.org/10.1088/1475-7516/2022/12/034}{\emph{Journal of
  Cosmology and Astroparticle Physics} {\bfseries 2022}{\bfseries (12)} (2022)
  034} [\href{https://arxiv.org/abs/2207.11910}{{\ttfamily 2207.11910}}].

\bibitem{Pi:2022ysn}
S.~Pi and M.~Sasaki, \emph{{Logarithmic Duality of the Curvature
  Perturbation}},
  \href{https://doi.org/10.1103/PhysRevLett.131.011002}{\emph{Physical Review
  Letters} {\bfseries 131}{\bfseries (1)} (2023) 011002}
  [\href{https://arxiv.org/abs/2211.13932}{{\ttfamily 2211.13932}}].

\bibitem{Hooshangi:2023kss}
S.~Hooshangi, M.~H. Namjoo and M.~Noorbala, \emph{Tail diversity from
  inflation},  (2023) [\href{https://arxiv.org/abs/2305.19257}{{\ttfamily
  2305.19257}}].

\bibitem{Enqvist:2008kt}
K.~Enqvist, S.~Nurmi, D.~Podolsky and G.~I. Rigopoulos, \emph{On the
  divergences of inflationary superhorizon perturbations},
  \href{https://doi.org/10.1088/1475-7516/2008/04/025}{\emph{Journal of
  Cosmology and Astroparticle Physics} {\bfseries 2008}{\bfseries (04)} (2008)
  025} [\href{https://arxiv.org/abs/0802.0395}{{\ttfamily 0802.0395}}].

\bibitem{Fujita:2013cna}
T.~Fujita, M.~Kawasaki, Y.~Tada and T.~Takesako, \emph{A new algorithm for
  calculating the curvature perturbations in stochastic inflation},
  \href{https://doi.org/10.1088/1475-7516/2013/12/036}{\emph{Journal of
  Cosmology and Astroparticle Physics} {\bfseries 2013}{\bfseries (12)} (2013)
  036} [\href{https://arxiv.org/abs/1308.4754}{{\ttfamily 1308.4754}}].

\bibitem{Vennin:2015hra}
V.~Vennin and A.~A. Starobinsky, \emph{Correlation functions in stochastic
  inflation}, \href{https://doi.org/10.1140/epjc/s10052-015-3643-y}{\emph{The
  European Physical Journal C} {\bfseries 75} (2015) 413}
  [\href{https://arxiv.org/abs/1506.04732}{{\ttfamily 1506.04732}}].

\bibitem{Leach:2001zf}
S.~M. Leach, M.~Sasaki, D.~Wands and A.~R. Liddle, \emph{Enhancement of
  superhorizon scale inflationary curvature perturbations},
  \href{https://doi.org/10.1103/PhysRevD.64.023512}{\emph{Physical Review D}
  {\bfseries 64}{\bfseries (2)} (2001) 023512}
  [\href{https://arxiv.org/abs/astro-ph/0101406}{{\ttfamily
  astro-ph/0101406}}].

\bibitem{Starobinsky:1992ts}
A.~A. Starobinskii, \emph{Spectrum of adiabatic perturbations in the universe
  when there are singularities in the inflation potential},
  \href{http://www.jetpletters.ru/ps/1276/article_19291.shtml}{\emph{Journal of
  Experimental and Theoretical Physics Letters} {\bfseries 55}{\bfseries (9)}
  (1992) 489}.

\bibitem{Schwarz:2001vv}
D.~J. Schwarz, C.~A. Terrero-Escalante and A.~A. García, \emph{Higher order
  corrections to primordial spectra from cosmological inflation},
  \href{https://doi.org/10.1016/S0370-2693(01)01036-X}{\emph{Physics Letters B}
  {\bfseries 517}{\bfseries (3--4)} (2001) 243}
  [\href{https://arxiv.org/abs/astro-ph/0106020}{{\ttfamily
  astro-ph/0106020}}].

\bibitem{Leach:2002ar}
S.~M. Leach, A.~R. Liddle, J.~Martin and D.~J. Schwarz, \emph{Cosmological
  parameter estimation and the inflationary cosmology},
  \href{https://doi.org/10.1103/PhysRevD.66.023515}{\emph{Physical Review D}
  {\bfseries 66}{\bfseries (2)} (2002) 023515}
  [\href{https://arxiv.org/abs/astro-ph/0202094}{{\ttfamily
  astro-ph/0202094}}].

\bibitem{Dimopoulos:2017ged}
K.~Dimopoulos, \emph{Ultra slow-roll inflation demystified},
  \href{https://doi.org/10.1016/j.physletb.2017.10.066}{\emph{Physics Letters
  B} {\bfseries 775} (2017) 262}
  [\href{https://arxiv.org/abs/1707.05644}{{\ttfamily 1707.05644}}].

\bibitem{Pattison:2018bct}
C.~Pattison, V.~Vennin, H.~Assadullahi and D.~Wands, \emph{The attractive
  behaviour of ultra-slow-roll inflation},
  \href{https://doi.org/10.1088/1475-7516/2018/08/048}{\emph{Journal of
  Cosmology and Astroparticle Physics} {\bfseries 2018}{\bfseries (08)} (2018)
  048} [\href{https://arxiv.org/abs/1806.09553}{{\ttfamily 1806.09553}}].

\bibitem{Malik:2008im}
K.~A. Malik and D.~Wands, \emph{Cosmological perturbations},
  \href{https://doi.org/10.1016/j.physrep.2009.03.001}{\emph{Physics Reports}
  {\bfseries 475}{\bfseries (1--4)} (2009) 1}
  [\href{https://arxiv.org/abs/0809.4944}{{\ttfamily 0809.4944}}].

\bibitem{Weinberg:2003sw}
S.~Weinberg, \emph{Adiabatic modes in cosmology},
  \href{https://doi.org/10.1103/PhysRevD.67.123504}{\emph{Physical Review D}
  {\bfseries 67}{\bfseries (12)} (2003) 123504}
  [\href{https://arxiv.org/abs/astro-ph/0302326}{{\ttfamily
  astro-ph/0302326}}].

\bibitem{Kofman:1994rk}
L.~Kofman, A.~Linde and A.~A. Starobinsky, \emph{Reheating after inflation},
  \href{https://doi.org/10.1103/PhysRevLett.73.3195}{\emph{Physical Review
  Letters} {\bfseries 73}{\bfseries (24)} (1994) 3195}
  [\href{https://arxiv.org/abs/hep-th/9405187}{{\ttfamily hep-th/9405187}}].

\bibitem{Bassett:2005xm}
B.~A. Bassett, S.~Tsujikawa and D.~Wands, \emph{Inflation dynamics and
  reheating}, \href{https://doi.org/10.1103/RevModPhys.78.537}{\emph{Reviews of
  Modern Physics} {\bfseries 78}{\bfseries (2)} (2006) 537}
  [\href{https://arxiv.org/abs/astro-ph/0507632}{{\ttfamily
  astro-ph/0507632}}].

\bibitem{Gordon:2000hv}
C.~Gordon, D.~Wands, B.~A. Bassett and R.~Maartens, \emph{Adiabatic and entropy
  perturbations from inflation},
  \href{https://doi.org/10.1103/PhysRevD.63.023506}{\emph{Physical Review D}
  {\bfseries 63}{\bfseries (2)} (2001) 023506}
  [\href{https://arxiv.org/abs/astro-ph/0009131}{{\ttfamily
  astro-ph/0009131}}].

\bibitem{Romano:2015vxz}
A.~E. Romano, S.~Mooij and M.~Sasaki, \emph{{Adiabaticity and gravity theory
  independent conservation laws for cosmological perturbations}},
  \href{https://doi.org/10.1016/j.physletb.2016.02.054}{\emph{Physics Letters
  B} {\bfseries 755} (2016) 464}
  [\href{https://arxiv.org/abs/1512.05757}{{\ttfamily 1512.05757}}].

\bibitem{Pattison:2019hef}
C.~Pattison, V.~Vennin, H.~Assadullahi and D.~Wands, \emph{Stochastic inflation
  beyond slow roll},
  \href{https://doi.org/10.1088/1475-7516/2019/07/031}{\emph{Journal of
  Cosmology and Astroparticle Physics} {\bfseries 2019}{\bfseries (07)} (2019)
  031} [\href{https://arxiv.org/abs/1905.06300}{{\ttfamily 1905.06300}}].

\bibitem{Wands:1998yp}
D.~Wands, \emph{Duality invariance of cosmological perturbation spectra},
  \href{https://doi.org/10.1103/PhysRevD.60.023507}{\emph{Physical Review D}
  {\bfseries 60}{\bfseries (2)} (1999) 023507}
  [\href{https://arxiv.org/abs/gr-qc/9809062}{{\ttfamily gr-qc/9809062}}].

\bibitem{Martin:2011sn}
J.~Martin and L.~Sriramkumar, \emph{{The scalar bi-spectrum in the Starobinsky
  model: The equilateral case}},
  \href{https://doi.org/10.1088/1475-7516/2012/01/008}{\emph{Journal of
  Cosmology and Astroparticle Physics} {\bfseries 2012}{\bfseries (01)} (2012)
  008} [\href{https://arxiv.org/abs/1109.5838}{{\ttfamily 1109.5838}}].

\bibitem{Martin:2014kja}
J.~Martin, L.~Sriramkumar and D.~K. Hazra, \emph{Sharp inflaton potentials and
  bi-spectra: effects of smoothening the discontinuity},
  \href{https://doi.org/10.1088/1475-7516/2014/09/039}{\emph{Journal of
  Cosmology and Astroparticle Physics} {\bfseries 2014}{\bfseries (09)} (2014)
  039} [\href{https://arxiv.org/abs/1404.6093}{{\ttfamily 1404.6093}}].

\bibitem{Ahmadi:2022lsm}
N.~Ahmadi \textit{et~al.}, \emph{Quantum diffusion in sharp transition to
  non-slow-roll phase},
  \href{https://doi.org/10.1088/1475-7516/2022/08/078}{\emph{Journal of
  Cosmology and Astroparticle Physics} {\bfseries 2022}{\bfseries (08)} (2022)
  078} [\href{https://arxiv.org/abs/2207.10578}{{\ttfamily 2207.10578}}].

\bibitem{Pi:2022zxs}
S.~Pi and J.~Wang, \emph{{Primordial black hole formation in Starobinsky's
  linear potential model}},
  \href{https://doi.org/10.1088/1475-7516/2023/06/018}{\emph{Journal of
  Cosmology and Astroparticle Physics} {\bfseries 2023}{\bfseries (06)} (2023)
  018} [\href{https://arxiv.org/abs/2209.14183}{{\ttfamily 2209.14183}}].

\bibitem{Carrilho:2019oqg}
P.~Carrilho, K.~A. Malik and D.~J. Mulryne, \emph{Dissecting the growth of the
  power spectrum for primordial black holes},
  \href{https://doi.org/10.1103/PhysRevD.100.103529}{\emph{Physical Review D}
  {\bfseries 100}{\bfseries (10)} (2019) 103529}
  [\href{https://arxiv.org/abs/1907.05237}{{\ttfamily 1907.05237}}].

\bibitem{Mishra:2019pzq}
S.~S. Mishra and V.~Sahni, \emph{Primordial black holes from a tiny bump/dip in
  the inflaton potential},
  \href{https://doi.org/10.1088/1475-7516/2020/04/007}{\emph{Journal of
  Cosmology and Astroparticle Physics} {\bfseries 2020}{\bfseries (04)} (2020)
  007} [\href{https://arxiv.org/abs/1911.00057}{{\ttfamily 1911.00057}}].

\bibitem{Cole:2023wyx}
P.~S. Cole, A.~D. Gow, C.~T. Byrnes and S.~P. Patil, \emph{Primordial black
  holes from single-field inflation: a fine-tuning audit},
  \href{https://doi.org/10.1088/1475-7516/2023/08/031}{\emph{Journal of
  Cosmology and Astroparticle Physics} {\bfseries 2023}{\bfseries (08)} (2023)
  031} [\href{https://arxiv.org/abs/2304.01997}{{\ttfamily 2304.01997}}].

\bibitem{Starobinsky:1986fx}
A.~A. Starobinsky,
  \href{https://doi.org/10.1007/3-540-16452-9\_6}{\emph{{Stochastic De Sitter
  (Inflationary) Stage in the Early Universe}}, } in \emph{Lecture Notes in
  Physics (Field Theory, Quantum Gravity and Strings)}, vol.~246, p.~107, 1988.

\bibitem{Starobinsky:1994bd}
A.~A. Starobinsky and J.~Yokoyama, \emph{{Equilibrium state of a
  self-interacting scalar field in the de Sitter background}},
  \href{https://doi.org/10.1103/PhysRevD.50.6357}{\emph{Physical Review D}
  {\bfseries 50}{\bfseries (10)} (1994) 6357}
  [\href{https://arxiv.org/abs/astro-ph/9407016}{{\ttfamily
  astro-ph/9407016}}].

\bibitem{De:2020hdo}
A.~De and R.~Mahbub, \emph{Numerically modeling stochastic inflation in
  slow-roll and beyond},
  \href{https://doi.org/10.1103/PhysRevD.102.123509}{\emph{Physical Review D}
  {\bfseries 102}{\bfseries (12)} (2020) 123509}
  [\href{https://arxiv.org/abs/2010.12685}{{\ttfamily 2010.12685}}].

\bibitem{Figueroa:2020jkf}
D.~G. Figueroa, S.~Raatikainen, S.~Räsänen and E.~Tomberg,
  \emph{{Non-Gaussian Tail of the Curvature Perturbation in Stochastic
  Ultraslow-Roll Inflation: Implications for Primordial Black Hole
  Production}},
  \href{https://doi.org/10.1103/PhysRevLett.127.101302}{\emph{Physical Review
  Letters} {\bfseries 127}{\bfseries (10)} (2021) 101302}
  [\href{https://arxiv.org/abs/2012.06551}{{\ttfamily 2012.06551}}].

\bibitem{Figueroa:2021zah}
D.~G. Figueroa, S.~Raatikainen, S.~Räsänen and E.~Tomberg, \emph{Implications
  of stochastic effects for primordial black hole production in ultra-slow-roll
  inflation},
  \href{https://doi.org/10.1088/1475-7516/2022/05/027}{\emph{Journal of
  Cosmology and Astroparticle Physics} {\bfseries 2022}{\bfseries (05)} (2022)
  027} [\href{https://arxiv.org/abs/2111.07437}{{\ttfamily 2111.07437}}].

\bibitem{Mishra:2023lhe}
S.~S. Mishra, E.~J. Copeland and A.~M. Green, \emph{{Primordial black holes and
  stochastic inflation beyond slow roll: I -- noise matrix elements}},  (2023)
  [\href{https://arxiv.org/abs/2303.17375}{{\ttfamily 2303.17375}}].

\bibitem{Polarski1996}
D.~Polarski and A.~A. Starobinsky, \emph{Semiclassicality and decoherence of
  cosmological perturbations},
  \href{https://doi.org/10.1088/0264-9381/13/3/006}{\emph{Classical and Quantum
  Gravity} {\bfseries 13}{\bfseries (3)} (1996) 377}
  [\href{https://arxiv.org/abs/gr-qc/9504030}{{\ttfamily gr-qc/9504030}}].

\bibitem{Lesgourgues:1996jc}
J.~Lesgourgues, D.~Polarski and Starobinsk, \emph{Quantum-to-classical
  transition of cosmological perturbations for non-vacuum initial states},
  \href{https://doi.org/10.1016/S0550-3213(97)00224-1}{\emph{Nuclear Physics B}
  {\bfseries 497}{\bfseries (1--2)} (1997) 479}
  [\href{https://arxiv.org/abs/gr-qc/9611019}{{\ttfamily gr-qc/9611019}}].

\bibitem{Kiefer:2008ku}
C.~Kiefer and D.~Polarski, \emph{{Why do Cosmological Perturbations Look
  Classical to Us?}},
  \href{https://doi.org/10.1166/asl.2009.1023}{\emph{Advanced Science Letters}
  {\bfseries 2}{\bfseries (2)} (2009) 164}
  [\href{https://arxiv.org/abs/0810.0087}{{\ttfamily 0810.0087}}].

\bibitem{Martin:2015qta}
J.~Martin and V.~Vennin, \emph{{Quantum discord of cosmic inflation: Can we
  show that CMB anisotropies are of quantum-mechanical origin?}},
  \href{https://doi.org/10.1103/PhysRevD.93.023505}{\emph{Physical Review D}
  {\bfseries 93}{\bfseries (2)} (2016) 023505}
  [\href{https://arxiv.org/abs/1510.04038}{{\ttfamily 1510.04038}}].

\bibitem{Martin:2017zxs}
J.~Martin and V.~Vennin, \emph{{Obstructions to Bell CMB experiments}},
  \href{https://doi.org/10.1103/PhysRevD.96.063501}{\emph{Physical Review D}
  {\bfseries 96}{\bfseries (6)} (2017) 063501}
  [\href{https://arxiv.org/abs/1706.05001}{{\ttfamily 1706.05001}}].

\bibitem{Martin:2021znx}
J.~Martin, A.~Micheli and V.~Vennin, \emph{Discord and decoherence},
  \href{https://doi.org/10.1088/1475-7516/2022/04/051}{\emph{Journal of
  Cosmology and Astroparticle Physics} {\bfseries 2022}{\bfseries (04)} (2022)
  051} [\href{https://arxiv.org/abs/2112.05037}{{\ttfamily 2112.05037}}].

\bibitem{Cruces:2018cvq}
D.~Cruces, C.~Germani and T.~Prokopec, \emph{Failure of the stochastic approach
  to inflation beyond slow-roll},
  \href{https://doi.org/10.1088/1475-7516/2019/03/048}{\emph{Journal of
  Cosmology and Astroparticle Physics} {\bfseries 2019}{\bfseries (03)} (2019)
  048} [\href{https://arxiv.org/abs/1807.09057}{{\ttfamily 1807.09057}}].

\bibitem{Domenech:2023dxx}
G.~Domènech, G.~Vargas and T.~Vargas, \emph{An exact model for
  enhancing/suppressing primordial fluctuations},  (2023)
  [\href{https://arxiv.org/abs/2309.05750}{{\ttfamily 2309.05750}}].

\bibitem{Garcia-Bellido:2017mdw}
J.~García-Bellido and E.~Ruiz~Morales, \emph{Primordial black holes from
  single field models of inflation},
  \href{https://doi.org/10.1016/j.dark.2017.09.007}{\emph{Physics of the Dark
  Universe} {\bfseries 18} (2017) 47}
  [\href{https://arxiv.org/abs/1702.03901}{{\ttfamily 1702.03901}}].

\bibitem{Di:2017ndc}
H.~Di and Y.~Gong, \emph{Primordial black holes and second order gravitational
  waves from ultra-slow-roll inflation},
  \href{https://doi.org/10.1088/1475-7516/2018/07/007}{\emph{Journal of
  Cosmology and Astroparticle Physics} {\bfseries 2018}{\bfseries (07)} (2018)
  007} [\href{https://arxiv.org/abs/1707.09578}{{\ttfamily 1707.09578}}].

\bibitem{Ballesteros:2017fsr}
G.~Ballesteros and M.~Taoso, \emph{Primordial black hole dark matter from
  single field inflation},
  \href{https://doi.org/10.1103/PhysRevD.97.023501}{\emph{Physical Review D}
  {\bfseries 97}{\bfseries (2)} (2018) 023501}
  [\href{https://arxiv.org/abs/1709.05565}{{\ttfamily 1709.05565}}].

\bibitem{Geller:2022nkr}
S.~R. Geller, W.~Qin, E.~McDonough and D.~I. Kaiser, \emph{Primordial black
  holes from multifield inflation with nonminimal couplings},
  \href{https://doi.org/10.1103/PhysRevD.106.063535}{\emph{Physical Review D}
  {\bfseries 106}{\bfseries (6)} (2022) 063535}
  [\href{https://arxiv.org/abs/2205.04471}{{\ttfamily 2205.04471}}].

\bibitem{Rasanen:2018fom}
S.~Räsänen and E.~Tomberg, \emph{{Planck scale black hole dark matter from
  Higgs inflation}},
  \href{https://doi.org/10.1088/1475-7516/2019/01/038}{\emph{Journal of
  Cosmology and Astroparticle Physics} {\bfseries 2019}{\bfseries (01)} (2019)
  038} [\href{https://arxiv.org/abs/1810.12608}{{\ttfamily 1810.12608}}].

\bibitem{Stewart1993}
E.~D. Stewart and D.~H. Lyth, \emph{A more accurate analytic calculation of the
  spectrum of cosmological perturbations produced during inflation},
  \href{https://doi.org/10.1016/0370-2693(93)90379-V}{\emph{Physics Letters B}
  {\bfseries 302}{\bfseries (2--3)} (1993) 171}
  [\href{https://arxiv.org/abs/gr-qc/9302019}{{\ttfamily gr-qc/9302019}}].

\end{thebibliography}\endgroup

\end{document}